\documentclass[12pt,a4paper]{article}

\usepackage{epsfig}
\usepackage{amsmath,amsfonts,amssymb}
\usepackage{t1enc}
\usepackage{cite}

\providecommand{\openone}{\leavevmode\hbox{\small1\kern-3.8pt\normalsize1}}

\newcommand{\Esix}{\text{E}_6}
\newcommand{\fbin}{fb$^{-1}$}

\newcommand{\TT}{T \bar T}
\newcommand{\BB}{B \bar B}
\newcommand{\XX}{X \bar X}
\newcommand{\YY}{Y \bar Y}
\newcommand{\Ts}{T_\text{s}}
\newcommand{\Bs}{B_\text{s}}
\newcommand{\TBd}{TB_{\text{d}_1}}
\newcommand{\TBD}{TB_{\text{d}_2}}
\newcommand{\XTd}{XT_\text{d}}
\newcommand{\BYd}{BY_\text{d}}
\newcommand{\HZ}{V}
\newcommand{\TB}{(T \, B)}
\newcommand{\XT}{(X \, T)}
\newcommand{\BY}{(B \, Y)}

\newcommand{\gm}{\gamma^\mu}
\newcommand{\Wmp}{W_\mu^+}

\newcommand{\Zm}{Z_\mu}

\newcommand{\ptmiss}{p_T\!\!\!\!\!\!\!\!\not\,\,\,\,\,\,\,}

\begin{document}

\begin{center}
\begin{Large}
{\bf Identifying top partners at LHC}
\end{Large}

\vspace{0.5cm}
J. A. Aguilar--Saavedra  \\[0.2cm] 
{\it Departamento de F\'{\i}sica Te\'orica y del Cosmos and CAFPE, \\
Universidad de Granada, E-18071 Granada, Spain} \\[0.1cm]
\end{center}

\begin{abstract}
We systematically study the possible signals at LHC of new vector-like quarks mainly coupled to the third generation. We consider heavy quarks $T$, $B$, $X$, $Y$ of charges $2/3$, $-1/3$, $5/3$ and $-4/3$, respectively, in $\text{SU}(2)_L$ isosinglets $T_{L,R}$, $B_{L,R}$, or isodoublets $\TB_{L,R}$, $\XT_{L,R}$ or $\BY_{L,R}$. Analyses based on a fast detector simulation are presented for twelve different final states containing one, two, three or four charged leptons in several invariant mass regions, also considering various $b$ quark multiplicities. It is shown that with the combination of the different channels the new quarks can be identified and their charged and neutral decays established. The comparison among final states also shows that the single lepton one offers the best discovery potential at LHC. For heavy quark masses of 500 GeV, the $5\sigma$ discovery luminosities range from 0.16 \fbin\ for a $\XT_{L,R}$ doublet to 1.9 \fbin\ for a $B_{L,R}$ singlet.
\end{abstract}

\section{Introduction}
\label{sec:1}

Hadron colliders offer an excellent place to look for new quarks, as the top quark discovery~\cite{Abe:1994xt} and its recent observation in single production~\cite{Abazov:2009ii,Aaltonen:2009jj} at Tevatron evidence. In the near future, the operation of the Large Hadron Collider (LHC) will open a new, yet unexplored mass range for new physics searches, in particular for new quarks heavier than the top.
Despite the demanding environment, new quark searches at LHC will be relatively clean because they can be produced in pairs through strong interactions with a large cross section and, being rather heavy, their signals can be easily distinguished from the large background from top pair production and $W$ plus jets.

Although the possibility of a fourth standard model (SM) sequential generation has not yet been excluded~\cite{Alwall:2006bx,Kribs:2007nz,Holdom:2009rf} and partial wave unitarity allows for fourth generation masses up to 1 TeV~\cite{Chanowitz:1978mv},
new quarks heavier than the top are generally expected to be of vector-like nature if they exist. For example, extra-dimensional models with $t_R$ in the bulk~\cite{Mirabelli:1999ks,Chang:1999nh,Csaki:2004ay} predict a tower of charge $2/3$ isosinglets $T_{L,R}^{(n)}$, of which the lightest one can be light and have sizeable mixing with the third generation~\cite{delAguila:2000kb}. More recently, $\TB_{L,R}$ and $\XT_{L,R}$ isodoublets of hypercharges $1/6$, $7/6$ coupling to the third generation naturally emerge~\cite{Contino:2006qr,Carena:2006bn} in warped models implementing a custodial symmetry to protect the $Zbb$ coupling~\cite{Agashe:2006at}. Charge $-1/3$ isosinglets $B_{L,R}$ are predicted in grand unification theories based on $\Esix$, one of the most widely studied groups~\cite{Frampton:1999xi,Hewett:1988xc}, in which one such fermion per family appears in the {\bf 27} representation. 
Little Higgs models~\cite{ArkaniHamed:2001nc,ArkaniHamed:2002qy,Perelstein:2005ka} also 
introduce a new $T_{L,R}$ isosinglet partner of the top quark which ameliorates the quadratic divergences in the Higgs mass. 
In general, the new quarks predicted in these SM extensions are expected to couple mainly to the third generation. For generic Yukawa matrices and heavy quark mass terms, it has been shown~\cite{delAguila:1982fs} that the mixing of new vector-like quarks is of order $m/M$, where $m,M$ are the masses of SM and new quarks, respectively. Then, unless specific symmetries are imposed on the mass matrices,
the large mass hierarchy $m_t \gg m_{u,d}$, $m_b \gg m_{d,s}$ favours mixing with the third generation. Additionally, constraints on top couplings are weaker than for the rest of quarks~\cite{delAguila:1998tp,AguilarSaavedra:2002kr} so there is more room for mixing also from the experimental side. Note, however, that in some models it is possible to evade direct constraints with cancellations, and have large mixing with the first and second generations compatible with experimental data, see Ref.~\cite{Atre:2008iu}.

In case that any new physics is discovered at LHC, as it is hoped, it will be compulsory to determine its nature. For heavy quarks this means not only the observation of an event excess or even an invariant mass peak, but the determination of the quark charges and $\text{SU}(2)_L$ isospin, the investigation of the decay channels and the measurement of their mixing with the SM quarks. In this paper we address some of these issues. We study the pair production of vector-like singlets $T_{L,R}$, $B_{L,R}$ of charges $2/3$, $-1/3$ and doublets $\TB_{L,R}$, $\XT_{L,R}$, $\BY_{L,R}$ of hypercharges $1/6$, $7/6$, $-5/6$, respectively, with quarks $X$, $Y$ of charges $5/3$, $-4/3$.
(From now on we will drop the $L$, $R$ subscripts.) 
We will assume that the new quarks mainly couple to the third generation. 
Previous literature has also investigated some of these signals in specific final states. For example, pair production of $T$ singlets has been studied in the single lepton final state \cite{AguilarSaavedra:2005pv,AguilarSaavedra:2006gv,AguilarSaavedra:2006gw}, as well as pair production of charge $5/3$, $-1/3$ quarks in $\XT$, $\TB$ doublets producing like-sign dileptons~\cite{Contino:2008hi} and one charged
lepton~\cite{Dennis:2007tv}.\footnote{The discovery potential for $D$ singlets coupling to $u,d$ instead of the third generation has already been explored, for example in Refs.~\cite{Mehdiyev:2006tz,Sultansoy:2006cw,Mehdiyev:2007pf}.}
Here we will advance beyond previous work by analysing twelve multi-leptonic final states which give evidence for the several decay modes
\begin{align}
& T \to W^+ b \,, \quad T \to Zt \,,\quad T \to Ht \,, \displaybreak \notag \\
& B \to W^- t \,, \quad B \to Zb \,,\quad B \to Hb \,, \notag \\
& X \to W^+ t \,, \notag \\
& Y \to W^- b \,,
\label{ec:decall}
\end{align}
with the aim of model discrimination. It is well known since some time (see for example Ref.~\cite{delAguila:1989rq}) that the presence or absence of specific decay modes can characterise the new quarks eventually observed. Here we demonstrate how this could be done in practice. For example, $T$ quarks in a $\XT$ doublet have suppressed decay $T \to W^+ b$, so they are not seen in the $W^+ b \, W^- \bar b$ final state as $T$ singlets are.
But they have enhanced $T \to Ht$ decays, so if the Higgs boson is light (as preferred by electroweak precision data) they give a fairly large and clean $\TT \to Ht \, H \bar t \to H W^+ b \, H W^- \bar b$ signal with one charged lepton and six $b$ quarks. On the other hand, $\YY  \to W^- b \, W^+ \bar b$ cannot be distinguished from $\TT \to W^+ b \, W^- \bar b$ unless the $b$ jet charge is measured, which is very difficult and requires large statistics. But, apart from different signal branching ratios, $T$ quarks are cleanly identified by their characteristic $T \to Zt$ decay, which can be observed in the trilepton final state. $X$ and $B$ quarks can both decay into four $W$ final states, but in some models the latter also decays $B \to Zb$ producing a sharp peak in a $\ell^+ \ell^- b$ invariant mass distribution, which can be observed in dilepton and trilepton final states (here and in the following $\ell=e,\mu$). In summary, here it will be shown that the simultaneous study and comparison of several multi-leptonic final states, with the observation of invariant mass peaks in most cases, can establish the identity of the new quarks, if they are observed at LHC.

We remark that model discrimination is somewhat more demanding that evaluating the discovery potential of one's favourite model in some final state. From the technical point of view, it requires the complete signal generation with all decay channels. For $\TT$ and $\BB$ production there are in general nine decay modes according to Eqs.~(\ref{ec:decall}). When the decay of the $W$ and $Z$ bosons (up to four, depending on the channel) are included, a plethora of possible final states appears involving multi-lepton signals. These contributions are all included in our simulations, which take into account the effects of radiation, pile-up and hadronisation, performed by a parton shower Monte Carlo, and use a fast detector simulation. SM backgrounds have also to be generated and simulated, including those with huge cross sections such as $W$ and $Z$ production plus jets, which are computationally demanding.

Heavy quark pair production gives interesting signals in final states with one, two (like- and opposite-sign), three and four charged leptons. (Five and six lepton final states have too small branching ratios.) For model discrimination it is very convenient to classify signals not only by lepton multiplicity but by the number of $Z$ boson ``candidates'' present (same-flavour opposite-charge lepton pairs with an invariant mass consistent with $M_Z$). For example, the trilepton final state is divided into a sample of events having a $Z$ candidate (in which $\TT \to Zt \, W^- b$ and other signals involving $Z \to \ell^+ \ell^-$ would be found) and events without $Z$ candidates (to which $\XX \to W^+ t \, W^- \bar t$, for instance, would contribute). In some cases the number of $b$ jets present is also relevant. This gives a total of twelve interesting final states to be examined, and for which specific analyses are presented in this paper. But, even after this final state organisation in terms of charged lepton multiplicity and number of $Z$ candidates, there are final states where more than one type of quark give interesting signals. One of such cases is, for the trilepton final state
with a $Z$ candidate,
\begin{align}
& \TT \to Zt \, W^- \bar b \to Z W^+ b \, W^- \bar b &&
\quad Z \to \ell^+ \ell^- , WW \to \ell \nu q \bar q' \,,
\notag \\
& \BB \to Zb \, W^+ \bar t \to Z b \, W^+ W^- \bar b &&
\quad Z \to \ell^+ \ell^- , WW \to \ell \nu q \bar q'
\end{align}
(the charge conjugate modes are also understood). In these cases, a likelihood classification is performed to separate and identify the $\TT$ and $\BB$ signals, and reconstruct them accordingly. This approach is unavoidable, since in some models like the $\TB$ doublet both signals can be present.

Besides model discrimination, which is the main goal of this paper, the systematic study of all interesting final states offers several advantages. One of them is that the most sensitive ones can be identified. We find that the single lepton final state (with either two or four $b$-tagged jets) offers the best discovery potential for all the models studied. For quark masses of 500 GeV, $5\sigma$ significance could be achieved for integrated luminosities ranging from 0.16 \fbin\ for a $\XT$ doublet to 1.9 \fbin\ for a $B$ singlet. Our study also provides a guide of final state signatures to be searched in case that an event excess is identified in one of them. This complements previous work done for the characterisation and discrimination of seesaw models~\cite{delAguila:2008cj,delAguila:2008hw} and new heavy leptons~\cite{AguilarSaavedra:2009ik}.

The structure of the rest of this paper is the following. In section~\ref{sec:2} we introduce the models studied giving the relevant Lagrangian terms. In section~\ref{sec:3} we discuss the general features of heavy quark pair production at LHC, and some details associated to the signal and background generation. In sections~\ref{sec:4l}--\ref{sec:1l} the results for final states with four, three, two (like-sign and opposite-sign) and one charged lepton are presented, respectively. For the reader's convenience, the main results obtained  are summarised at the end of each section, so that in a first reading the details can be omitted. Section~\ref{sec:summ} is a general summary where we address model discrimination by comparing signals in different final states. Our conclusions are drawn in section~\ref{sec:concl}. The Feynman rules used in our Monte Carlo programs are given in the Appendix.

\section{Model overview}
\label{sec:2}

In this section we briefly review the electroweak interactions of the new quarks, which determine their decay modes and single production. Additional details can be found in many early references, for example
\cite{delAguila:1982fs,delAguila:1989rq,
Branco:1986my}.
The interactions of $\XT$ and $\BY$ doublets are also given in Refs.~\cite{delAguila:2000aa,delAguila:2000rc}.

\subsection{$T$ singlet}
\label{sec:2.1}

We denote the SM weak eigenstates as $q'_{Li}=(u'_{Li} \; d'_{Li})^T$, $u'_{Ri}$, $d'_{Ri}$, where Latin indices $i,j=1,2,3$ run over SM generations and Greek indices $\alpha,\beta=1,\dots,4$ over all quark fields. 
We use primes to distinguish them from mass eigenstates, where needed.
The addition of a $\text{SU}(2)_L$ isosinglet $u'_{L4}$, $u'_{R4}$ to the SM quark content does not modify the SM charged and neutral current interactions in the weak eigenstate basis. The new $u_{R4}'$ field has Yukawa couplings to the SM left-handed fields
(the Yukawa coupling matrix $\mathrm{Y}$ must not be confused with a charge $-4/3$ quark $Y$),
and a bare mass term can be written involving the new left-handed singlet $u'_{L4}$,\footnote{In full generality, the right-handed fields $u_{R\alpha}'$ can be redefined so that the bare mass term only involves  $u_{R4}'$. This change of basis also redefines the arbitrary matrix $\mathrm{Y}$ of Yukawa couplings.}
\begin{eqnarray}
\mathcal{L}_W & = & -\frac{g}{\sqrt 2} \, \bar u_{Li}' \gm d'_{Li} \, \Wmp + \text{H.c.}
\,, \notag \\
\mathcal{L}_Z & = & -\frac{g}{2 c_W} \left[ \bar u'_{Li} \gm u'_{Li} - 2 s_W^2 J_\text{EM}^\mu \right] \Zm \,, \notag \\
\mathcal{L}_\text{Y}  & = & - \mathrm{Y}_{i\beta}^u \; \bar q'_{Li} u'_{R\beta} \, \tilde \phi + \text{H.c.}
\,, \notag \\
\mathcal{L}_\text{bare}  & = & - M \bar u_{L4}' u_{R4}' + \text{H.c.}
\end{eqnarray}
In this and the rest of models,
the electromagnetic current $J_\text{EM}^\mu$ has the same expression as in the SM but summing over all quark fields. 
The Higgs doublet is
\begin{equation}
\phi = \left( \!\begin{array}{c} \phi^+ \\ \phi^0 \end{array} \!\right)
\to \frac{1}{\sqrt 2} \left( \!\begin{array}{c} 0 \\ v + H \end{array} \!\right) \,,\quad
\tilde \phi \equiv i\tau_2 \phi^*
\to \frac{1}{\sqrt 2} \left( \!\begin{array}{c} v + H \\ 0 \end{array} \!\right) \,,
\end{equation}
with $v=246$ GeV and $\tau$ the Pauli matrices. In the Lagrangians above we have omitted the terms in the down sector which are not affected by mixing. After the mass matrix diagonalisation
the $W$, $Z$ and $H$ interactions read
\begin{eqnarray}
\mathcal{L}_W & = & -\frac{g}{\sqrt 2} \, \bar u_{L \alpha} \gm \mathrm{V}_{\alpha j} d_{Lj} \, \Wmp + \text{H.c.} \,, \notag \\
\mathcal{L}_Z & = & -\frac{g}{2 c_W} \left[ \bar u_{L\alpha} \gm \mathrm{X}_{\alpha \beta} u_{L\beta} - 2 s_W^2 J_\text{EM}^\mu \right] \Zm \,, \notag \\
\mathcal{L}_H & = & - \frac{g}{2 M_W} \left[
\bar u_{L\alpha} \mathrm{X}_{\alpha \beta} \, m^u_\beta u_{R\beta} + \bar u_{R\alpha} m^u_\alpha \mathrm{X}_{\alpha \beta} u_{L\beta} \right] H \,,
\label{ec:Tint}
\end{eqnarray}
where $\mathrm{V}_{\alpha j}$ is the $4 \times 3$ generalisation of the Cabibbo-Kobayashi-Maskawa~\cite{Cabibbo:1963yz,Kobayashi:1973fv} (CKM) matrix, $\mathrm{X} = \mathrm{V} \mathrm{V}^\dagger$ a Hermitian $4 \times 4$ matrix (not to be confused with a charge $5/3$ quark $X$) and $m_\alpha^u$ the up-type quark masses. The electromagnetic current $J_\text{EM}^\mu$ obviously remains diagonal.
These equations, which result from a trivial change from weak to mass eigenstate basis, are exact and do not assume small mixing.
Notice the appearance of left-handed flavour-changing neutral (FCN) couplings among up-type quarks, due to the mixing of left-handed weak eigenstates of different isospin, which breaks the Glashow-Iliopoulos-Maiani~\cite{Glashow:1970gm} mechanism.
For a heavy quark $T \equiv u_4$ mixing with the top quark, and assuming small mixing, we have the approximate equality $\mathrm{X}_{Tt} \simeq \mathrm{V}_{Tb}$ among neutral and charged current couplings, replacing generation indices by quark labels. This is a very well known result: in the $T$ singlet model charged current mixing ($WTb$) automatically implies neutral current ($ZTt$) and scalar ($HTt$) interactions, all of the same strength up to multiplicative factors independent of mixing. The corresponding Feynman rules are given in the Appendix. These interactions determine the $T$ quark decays,
\begin{equation}
T \to W^+ b \,,\quad \quad T \to Zt \,,\quad \quad T \to Ht \,.
\end{equation}
This new eigenstate has a mass $m_T = M + O(v^2 \mathrm{Y}^2/M^2)$.

\subsection{$B$ singlet}
\label{sec:2.2}

The Lagrangian for a $B$ singlet is completely analogous to the one for a $T$ singlet, with few replacements. The relevant interactions in the weak eigenstate basis read
\begin{eqnarray}
\mathcal{L}_W & = & -\frac{g}{\sqrt 2} \, \bar u_{Li}' \gm d_{Li}' \, \Wmp + \text{H.c.}
\,, \notag \\
\mathcal{L}_Z & = & -\frac{g}{2 c_W} \left[- \bar d_{Li}' \gm d_{Li}' - 2 s_W^2 J_\text{EM}^\mu \right] \Zm \,, \notag \\
\mathcal{L}_\mathrm{Y} & = & - \mathrm{Y}_{i\beta}^d \; \bar q_{Li}' d_{R\beta}' \, \phi
+ \text{H.c.} \notag \\
\mathcal{L}_\text{bare}  & = & - M \bar d_{L4}' d_{R4}' + \text{H.c.}
\end{eqnarray}
After mass matrix diagonalisation, we have
\begin{eqnarray}
\mathcal{L}_W & = & -\frac{g}{\sqrt 2} \, \bar u_{Li} \gm \mathrm{V}_{i\beta} d_{L\beta} \, \Wmp + \text{H.c.} \,, \notag \\
\mathcal{L}_Z & = & -\frac{g}{2 c_W} \left[ - \bar d_{L\alpha} \gm \mathrm{X}_{\alpha \beta} d_{L\beta} - 2 s_W^2 J_\text{EM}^\mu \right] \Zm
\,, \notag \\
\mathcal{L}_H & = & - \frac{g}{2 M_W} \left[
\bar d_{L\alpha} \mathrm{X}_{\alpha \beta} \, m^d_\beta d_{R\beta} + \bar d_{R\alpha} m^d_\alpha \mathrm{X}_{\alpha \beta} d_{L\beta} \right] H \,.
\end{eqnarray}
The CKM matrix has dimension $3\times 4$, $\mathrm{X} = \mathrm{V}^\dagger \mathrm{V}$ in this case and $m_\alpha^d$ are the down-type quark masses. For $B$ mixing with the third generation we have $\mathrm{X}_{bB} \simeq \mathrm{V}_{tB}$, so that the new quark $B$ has $WtB$, $ZbB$ and $HbB$ interactions governed by a single mixing factor $\mathrm{V}_{tB}$, in analogy with the $T$ singlet model. The 
new quark $B$ has a mass $m_B \simeq M$, and its decays are
\begin{equation}
B \to W^- t \,,\quad \quad B \to Zb \,,\quad \quad B \to Hb \,.
\end{equation}

\subsection{$\TB$ doublet}
\label{sec:2.3}

With the addition of a vector-like doublet, the relevant Lagrangian in the weak interaction basis is
\begin{eqnarray}
\mathcal{L}_W & = & -\frac{g}{\sqrt 2} \left[
\bar u_{L\alpha}' \gm d_{L\alpha}' + \bar u_{R4}' \gm d_{R4}' \right] \Wmp
 + \text{H.c.} \,, \notag \\
\mathcal{L}_Z & = & -\frac{g}{2 c_W} \left[ \bar u_{L\alpha}' \gm u_{L\alpha}' + 
 \bar u_{R4}' \gm u_{R4}' - \bar d_{L\alpha}' \gm d_{L\alpha}' 
 - \bar d_{R4}' \gm d_{R4}' - 2 s_W^2 J_\text{EM}^\mu \right] \Zm \,, \notag \\
\mathcal{L}_\text{Y} & = &
  - \mathrm{Y}_{\alpha j}^u \; \bar q_{L\alpha}' u_{Rj}' \, \tilde \phi
  - \mathrm{Y}_{\alpha j}^d \; \bar q_{L\alpha}' d_{Rj}' \, \phi
+ \text{H.c.} \,, \notag \\
\mathcal{L}_\text{bare}  & = & - M \bar q_{L4}' q_{R4}' + \text{H.c.} \,,
\end{eqnarray}
with four SM-like left-handed doublets $q_{Li}'$ and one new right-handed doublet
$q_{R4}' = (u_{R4}' \, d_{R4}')^T$. The left-handed fields can be redefined so that the bare mass term only couples $q_{L4}'$.
In the mass eigenstate basis it is more transparent to write the Lagrangians at first order in the (small) light-heavy mixing,
\begin{eqnarray}
\mathcal{L}_W & = & -\frac{g}{\sqrt 2} \left[
  \bar u_{Li} \gm \mathrm{V}_{ij}^L d_{Lj} + \bar T_L \gm B_L 
+ \bar u_{R\alpha} \gm \mathrm{V}_{\alpha \beta}^R d_{R\beta}
 \right] \Wmp + \text{H.c.} \,, \notag \\
\mathcal{L}_Z & = & -\frac{g}{2 c_W} \left[
\bar u_{L\alpha} \gm u_{L\alpha}
+ \bar u_{R\alpha} \gm \mathrm{X}_{\alpha \beta}^u u_{R\beta} \right. \notag \\
& & \left. - \bar d_{L\alpha} \gm d_{L\alpha}
- \bar d_{R\alpha} \gm \mathrm{X}_{\alpha \beta}^d d_{R\beta}
- 2 s_W^2 J_\text{EM}^\mu \right] \Zm \,, \notag \\
\mathcal{L}_H & = & - \frac{g}{2 M_W} \left[
\bar u_{L\alpha} m_\alpha^u (\delta_{\alpha \beta}-\mathrm{X}^u_{\alpha \beta}) u_{R\beta}
+ \bar u_{R\alpha} (\delta_{\alpha \beta}-\mathrm{X}^u_{\alpha \beta}) m_\beta^u u_{L\beta}
\right. \notag \\
& & \left. 
+ \bar d_{L\alpha} m_\alpha^d (\delta_{\alpha \beta}-\mathrm{X}^d_{\alpha \beta}) d_{R\beta}
+ \bar d_{R\alpha} (\delta_{\alpha \beta}-\mathrm{X}^d_{\alpha \beta}) m_\beta^d d_{L\beta}
\right] H \,,
\end{eqnarray}
so that it is apparent that the mixing of the heavy quarks $T$, $B$ with SM quarks is only right-handed. The $4 \times 4$ matrix $\mathrm{V}^R$ is not unitary, and also determines the FCN interactions, because
$\mathrm{X}^u = \mathrm{V}^R \mathrm{V}^{R\dagger}$,  
$\mathrm{X}^d = \mathrm{V}^{R\dagger} \mathrm{V}^R$. Both $\mathrm{X}^u$ and $\mathrm{X}^d$ are Hermitian and non-diagonal, mediating FCN currents.
Then, charged current interactions of the new states with SM quarks imply FCN ones, which result from the mixing of right-handed weak eigenstates with different isospin. At first order we have $\mathrm{X}_{tT} \simeq \mathrm{V}_{tB}^R$, $\mathrm{X}_{Bb} \simeq \mathrm{V}_{Tb}^R$. The
new quarks are almost degenerate, with masses $m_T = m_B = M$, up to terms of order $v^2 \mathrm{Y}^2 / M^2$. One can distinguish three scenarios for the heavy quark decays, depending on the relative sizes of the charged current mixing of the new quarks. For $V_{Tb} \sim V_{tB}$ the decay modes, assuming that they couple to the third generation, are the same as for singlets,
\begin{align}
& T \to W^+ b \,,\quad \quad T \to Zt \,,\quad \quad T \to Ht \,, \notag \\
& B \to W^- t \,,\quad \quad B \to Zb \,,\quad \quad B \to Hb \,,
\end{align}
but with couplings of different chirality, which is reflected in some angular distributions.
For $V_{Tb} \ll V_{tB}$ ({\em i.e.} the top quark mixes with its partner much more than the bottom quark), the decays are
\begin{align}
& T \to Zt \,,\quad \quad T \to Ht \,, \notag \\
& B \to W^- t \,.
\end{align}
This scenario is the most natural one for generic Yukawa couplings due to the fact that the top quark is much heavier than the bottom quark, and is realised in some models \cite{Contino:2006nn}. Finally, a mixing $V_{Tb} \gg V_{tB}$ would give
\begin{align}
& T \to W^+ b \,, \notag \\
& B \to Zb \,,\quad \quad B \to Hb \,,
\end{align}
with signals similar to a hypercharge $-5/6$ doublet $\BY$ (see below). However, a mixing
$V_{Tb} \gg V_{tB}$ is not natural in view of the mass hierarchy $m_t \gg m_b$,
and is disfavoured by constraints on $b$ quark mixing.

\subsection{$\XT$ doublet}
\label{sec:2.4}

The interactions when a hypercharge $7/6$ doublet is added have some similarities and differences with the previous case. In the weak eigenstate basis we have
\begin{eqnarray}
\mathcal{L}_W & = & -\frac{g}{\sqrt 2} \left[ \bar u_{Li}' \gm d_{Li}'
 + \bar X_L \gm u_{L4}' + \bar X_R \gm u_{R4}'
  \right] \Wmp + \text{H.c.} \,, \notag \\
\mathcal{L}_Z & = & -\frac{g}{2 c_W} \left[ \bar u_{Li}' \gm u_{Li}'
 - \bar u_{L4}' \gm u_{L4}' - \bar u_{R4}' \gm u_{R4}'  + \bar X \gm X 
- 2 s_W^2 J_\text{EM}^\mu \right] \Zm \,, \notag \\
\mathcal{L}_\text{Y} & = &
  - \mathrm{Y}_{ij}^u \; \bar q_{Li}' u_{Rj}' \, \tilde \phi
  - \mathrm{Y}_{4j}^u \; (\bar X_L \; \bar u_{L4}') \, u_{Rj}' \, \phi
+ \text{H.c.} \,, \notag \\
\mathcal{L}_\text{bare}  & = & - M \left( \bar X_L \, \bar u_{L4}' \right)
\left( \! \begin{array}{c} X_R \\ u_{R4}' \end{array} \! \right) + \text{H.c.} \,,
\end{eqnarray}
where for the charge $5/3$ quark $X$ the weak interaction and mass eigenstates coincide.
We omit terms for the down sector which are unaffected by the presence of the new doublet.
In the mass eigenstate basis, at first order in the light-heavy mixing the Lagrangians read
\begin{eqnarray}
\mathcal{L}_W & = & -\frac{g}{\sqrt 2} \left[
  \bar u_{Li} \gm \mathrm{V}_{ij}^L d_{Lj} + \bar X_L \gm T_L
  + \bar X_R \gm \mathrm{V}_{4\beta}^R u_{R\beta}
 \right] \Wmp + \text{H.c.} \,, \notag \\
\mathcal{L}_Z & = & -\frac{g}{2 c_W} \left[
\bar u_{Li} \gm u_{Li} - \bar T_L \gm T_L
- \bar u_{R\alpha} \gm \mathrm{X}_{\alpha \beta} u_{R\beta} 
+ \bar X \gm X - 2 s_W^2 J_\text{EM}^\mu \right] \Zm \,, \notag \\
\mathcal{L}_H & = & - \frac{g}{2 M_W} \left[
\bar u_{L\alpha} m_\alpha^u (\delta_{\alpha \beta}-\mathrm{X}_{\alpha \beta}) u_{R\beta}
+ \bar u_{R\alpha} (\delta_{\alpha \beta}-\mathrm{X}_{\alpha \beta}) m_\beta^u u_{L\beta}
\right] H \,,
\end{eqnarray}
so that again the interactions of the new quarks $X$, $T$ with the SM ones are right-handed.
$\mathrm{V}^L$ is the usual CKM matrix.
 The $1 \times 4$ matrix $\mathrm{V}^R$ also determines the neutral mixing because $\mathrm{X} = \mathrm{V}^{R\dagger} \mathrm{V}^R$. Notice an important difference with the $T$ singlet and $\TB$ doublet: at first order the quark $T$ does not have charged current couplings to SM quarks but has neutral ones $ZTt$, $HTt$. For a mixing with the third generation we have $\mathrm{X}_{Tt} \simeq \mathrm{V}_{Xt}^R$. Obviously,
the charge $5/3$ quark $X$ only has charged current interactions with SM charge $2/3$ quarks. 
As in the $\TB$ doublet, the new mass eigenstates are almost degenerate, with masses $m_X \simeq m_T \simeq M$. Their allowed decays are
\begin{align}
& X \to W^+ t \,, \notag \\
& T \to Zt \,,\quad \quad T \to Ht \,. 
\end{align}

\subsection{$\BY$ doublet}
\label{sec:2.5}

Finally, the relevant Lagrangian for SM quarks plus a $\BY$ doublet is
\begin{eqnarray}
\mathcal{L}_W & = & -\frac{g}{\sqrt 2} \left[ \bar u_{Li}' \gm d_{Li}'
 + \bar d_{L4}' \gm Y_L + \bar d_{R4}' \gm Y_R
  \right] \Wmp + \text{H.c.} \,, \notag \\
\mathcal{L}_Z & = & -\frac{g}{2 c_W} \left[ - \bar d_{Li}' \gm d_{Li}'
 + \bar d_{L4}' \gm d_{L4}' + \bar d_{R4}' \gm d_{R4}' - \bar Y \gm Y 
- 2 s_W^2 J_\text{EM}^\mu \right] \Zm \,, \notag \\
\mathcal{L}_\text{Y} & = &
  - \mathrm{Y}_{ij}^d \; \bar q_{Li}' d_{Rj}' \, \phi
  - \mathrm{Y}_{4j}^d \; (\bar d_{L4}' \; \bar Y_L ) \, d_{Rj}' \, \tilde \phi
+ \text{H.c.} \,, \notag \\
\mathcal{L}_\text{bare}  & = & - M \left( \bar d_{L4}' \, \bar Y_L \right)
\left( \! \begin{array}{c} d_{R4}' \\ X_R \end{array} \! \right) + \text{H.c.}
\end{eqnarray}
At first order, the interactions in the mass eigenstate basis read
\begin{eqnarray}
\mathcal{L}_W & = & -\frac{g}{\sqrt 2} \left[
  \bar u_{Li} \gm \mathrm{V}_{ij}^L d_{Lj} + \bar B_L \gm Y_L
  + \bar d_{R\alpha} \gm \mathrm{V}_{\alpha 4}^R Y_R
 \right] \Wmp + \text{H.c.} \,, \notag \\
\mathcal{L}_Z & = & -\frac{g}{2 c_W} \left[
-\bar d_{Li} \gm d_{Li} + \bar B_L \gm B_L
+ \bar d_{R\alpha} \gm \mathrm{X}_{\alpha \beta} d_{R\beta} 
- \bar Y \gm Y - 2 s_W^2 J_\text{EM}^\mu \right] \Zm \,, \notag \\
\mathcal{L}_H & = & - \frac{g}{2 M_W} \left[
\bar d_{L\alpha} m_\alpha^d (\delta_{\alpha \beta}-\mathrm{X}_{\alpha \beta}) d_{R\beta}
+ \bar d_{R\alpha} (\delta_{\alpha \beta}-\mathrm{X}_{\alpha \beta}) m_\beta^d d_{L\beta}
\right] H \,.
\end{eqnarray}
The matrix $\mathrm{V}^R$ has dimension $4\times 1$ and $\mathrm{X} = \mathrm{V}^R \mathrm{V}^{R\dagger}$. At first order the quark $B$ does not have charged current couplings to SM quarks but has neutral ones.
(The charge $-4/3$ quark $Y$ has only charged current interactions with down-type SM quarks.)
For a mixing with the third generation we have $\mathrm{X}_{bB} \simeq \mathrm{V}_{bY}^R$.
The new quarks have masses $m_B \simeq m_Y \simeq M$, and their allowed decays are
\begin{align}
& B \to Zb \,,\quad \quad B \to Hb \,, \notag \\ 
& Y \to W^- b \,.
\end{align}
Notice that the $\bar Y \to W^+ \bar b$ decay is like $T \to W^+ b$ but with a $b$ antiquark instead of a quark. These decays can be distinguished using angular distributions but, except for small kinematical differences, the signatures of a $\BY$ doublet are similar to the ones of a $\TB$ doublet in which the $B$ quark mixes much more than the $T$ quark.

\section{Heavy quark production at LHC}
\label{sec:3}

\begin{figure}[t]
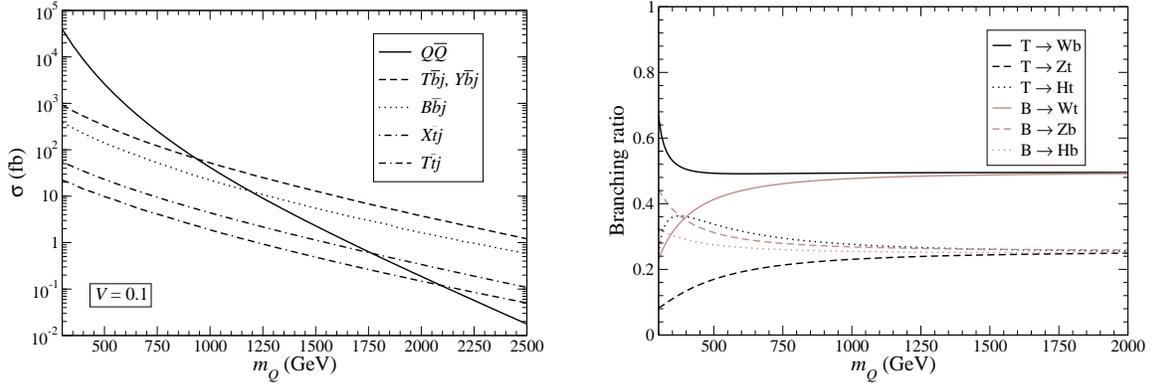

\begin{center}
\begin{tabular}{ccc}
\epsfig{file=Figs/mass-cross.eps,height=5.1cm,clip=} & \quad &
\epsfig{file=Figs/mass-BR.eps,height=5.1cm,clip=}
\end{tabular}
\caption{Left: Heavy quark production cross sections at LHC. Right: branching ratios for $T$ and $B$ decays.}
\label{fig:mass-cross}
\end{center}
\end{figure}

New heavy quarks can be produced in pairs via QCD interactions,
\begin{equation}
gg,q \bar q \to Q \bar Q \quad\quad (Q=T,B,X,Y) \,,
\end{equation}
in the same way as the top quark. The cross section only depends on the quark mass, and is plotted in Fig.~\ref{fig:mass-cross} (left).
For $T$ quark singlets the partial decay widths are
\begin{align}
\Gamma(T \to W^+ b) & = \frac{g^2}{64 \pi} |V_{Tb}|^2
  \frac{m_T}{M_W^2} \lambda(m_T,m_b,M_W)^{1/2} \nonumber \\
  & \times \left[ 1+\frac{M_W^2}{m_T^2}-2 \frac{m_b^2}{m_T^2}
  -2 \frac{M_W^4}{m_T^4}  +  \frac{M_W^4}{m_T^4} + \frac{M_W^2 m_b^2}{m_T^4}
  \right] \,, \nonumber \\
\Gamma(T \to Z t) & = \frac{g}{128 \pi c_W^2} |X_{Tt}|^2
  \frac{m_T}{M_Z^2} \lambda(m_T,m_t,M_Z)^{1/2} \nonumber \\
  & \times  \left[ 1 + \frac{M_Z^2}{m_T^2}
  - 2  \frac{m_t^2}{m_T^2} - 2  \frac{M_Z^4}{m_T^4}  + \frac{m_t^4}{m_T^4}
  + \frac{M_Z^2 m_t^2}{m_T^4} \right] \,, \nonumber \\
\Gamma(T \to H t) & = \frac{g^2}{128 \pi} |X_{Tt}|^2
 \frac{m_T}{M_W^2} \lambda(m_T,m_t,M_H)^{1/2} \nonumber \\
  & \times  \left[ 1 + 6 \frac{m_t^2}{m_T^2} - \frac{M_H^2}{m_T^2} 
  + \frac{m_t^4}{m_T^4} - \frac{m_t^2 M_H^2}{m_T^4} \right] \,,
\label{ec:Gamma}
\end{align}
being%
\begin{equation}
\lambda(x,y,z) \equiv (x^4 + y^4 + z^4 - 2 x^2 y^2 
- 2 x^2 z^2 - 2 y^2 z^2)
\end{equation}%
a kinematical function.
For a $B$ singlet, the expressions for $B \to W^- t$, $B \to Zb$, $B \to Hb$ can be obtained from Eqs.~(\ref{ec:Gamma}) by replacing the mixings $V_{Tb} \to V_{tB}$, $X_{Tt} \to X_{Bb}$ and the quark masses $m_T \to m_B$, $m_t \to m_b$, $m_b \to m_t$. The branching ratios as a function of the heavy quark mass are presented in Fig.~\ref{fig:mass-cross} (right), fixing $M_H = 115$ GeV. For $\TB$ doublets the analytical expressions of the widths are the same as for the singlets, although the relation beween the neutral and charged current mixings differs. For equal mixings $V_{Tb} \simeq V_{tB}$ the branching ratios are the same as for singlets, while for $V_{Tb} \ll V_{tB}$ the decays
$T \to W^+ b$, $B \to Zb$, $B \to Hb$ are absent, so that $\mathrm{Br}(T \to Zt) \simeq \mathrm{Br}(T \to Ht) \simeq 0.5$, $\mathrm{Br}(B \to W^- t) = 1$.
For $T$, $B$ quarks in $\XT$ and $\BY$ doublets the charged decay modes are absent, and thus the partial widths for the other modes are roughly one half. For $X \to W^+ t$ and $Y \to W^- b$ the widths are as for $B \to W^- t$, $T \to W^+ b$ replacing the mixings by $\mathrm{V}_{Xt}^R$ and $\mathrm{V}_{bY}^R$, respectively, as well as the quark masses. These are the only decay modes for $X$, $Y$ quarks.

Electroweak single heavy quark production is also possible at LHC, for example in the $t$-channel processes
\begin{align}
& gq \to T \bar b q' \,, \quad\quad gq \to T \bar t q' \,,  \notag \\
& gq \to B \bar b q \,, \quad\quad  gq \to X \bar t q' \,,  \notag \\
& gq \to Y \bar b q' \,.
\label{ec:single}
\end{align}
For $T \bar b j$, $X \bar t j$ and $Y \bar b j$ production ($j=q,q'$ denotes a light jet) the processes involve a $t$-channel $W$ boson, while $B \bar b j$ and $T \bar t j$ production exchange a $Z$ boson. This latter process has a much smaller cross section than $T \bar b j$ but is the only possibility for the $T$ quark in a $\XT$ doublet. 
The cross sections for the processes in Eqs.~(\ref{ec:single}) are also plotted in Fig.~\ref{fig:mass-cross}, for reference mixings $V,X=0.1$ with the third generation and including heavy quark and antiquark production. For the $2 \to 2$ processes $bq \to Tj$, $bq \to Bj$, etc. the cross sections are very close to the ones for their $2 \to 3$ counterparts $T \bar bj$, $B \bar bj$, etc.
Next-to-leading order corrections \cite{Campbell:2009gj,Berger:2009qy} are not included.
In this work we do not consider single production as a means for model discrimination.
Nevertheless, depending on the heavy quark mass and mixing, some single production processes can be important, as it can be observed in Fig.~\ref{fig:mass-cross}.
In any case, it is important to remark that single production processes are crucial to 
measure the heavy quark mixing with SM quarks.

The heavy quark signals studied in this paper have been calculated by implementing 
pair ($\TT$, $\BB$, $\XX$ and $\YY$) and single ($Tj$, $T \bar b j$, $T \bar t j$, $Bj$, $B \bar b j$, $X \bar t j$, $Yj$, $Y \bar b j$) production in the generator {\tt Protos}~\cite{AguilarSaavedra:2008gt}, for the six models considered. All the decay channels in Eqs.~(\ref{ec:decall}) are included, with the subsequent $W$ and $Z$ boson decays in all channels. The Higgs boson decay, which does not carry any spin information, is left to the parton shower Monte Carlo.
A complete signal evaluation is necessary for a study like the one presented here,  which surveys final states from one to four leptons, and various $b$ quark multiplicities in some cases. But also because sometimes charged leptons are missed by the detector, {\em e.g.} in $Z \to \ell^+ \ell^-$, resulting in contributions with fewer detected charged leptons than were generated at the partonic level.
Matrix elements are calculated using {\tt HELAS} \cite{helas}, to take finite width and spin effects into account, and integration in phase space is done by {\tt Vegas} \cite{vegas}. The output is given in a suitable form to be interfaced to the parton shower Monte Carlo {\tt Pythia} 6.4\cite{Sjostrand:2006za} to add initial and final state radiation (ISR, FSR) and pile-up, and perform hadronisation. 

In this work we restrict our detailed simulations to heavy quark pair production,
assuming heavy quark masses of 500 GeV and $m_t = 175$ GeV, $M_H = 115$ GeV. Cross sections and branching ratios are independent of the heavy-light mixing for $T$, $B$ singlets and
$\XT$, $\BY$ doublets, and a mixing $V = 0.1$ is assumed for definiteness. For the $\TB$ doublet we study two scenarios: (1) equal mixing $V_{Tb} = V_{tB} = 0.1$, in which the signals are quite similar to the ones of two $T$, $B$, singlets; (2) doublet mixing mainly with the top quark, $V_{Tb} = 0$, $V_{tB} = 0.1$. The signals for a doublet mixing mainly with the bottom are practically the same (except for the exchange of $b$ quarks and antiquarks and small kinematical differences) as for the $\BY$ doublet, and are not presented for brevity.
These six models are identified by the labels $\Ts$, $\Bs$, $\TBd$, $\TBD$, $\XTd$ and $\BYd$ in tables and figures.
Signals are generated with statistics of 300 fb$^{-1}$ and rescaled to a reference luminosity of 30 fb$^{-1}$, in order to reduce statistical fluctuations.
The factorisation and renormalisation scales used equal the heavy quark mass.
We use the fast simulation {\tt AcerDET}~\cite{RichterWas:2002ch} which is a generic LHC detector simulation, neither of ATLAS nor of CMS, with standard settings.
In particular, the lepton isolation criteria require a separation $\Delta R > 0.4$ from other clusters and a maximum energy deposition $\Sigma E_T = 10$ GeV in a cone of $\Delta R = 0.2$ around the reconstructed electron or muon. Jets are reconstructed using a cone algorithm with $\Delta R = 0.4$. In this analysis we only focus on central jets with pseudo-rapidity $|\eta| < 2.5$. Forward jets with $2.5 < |\eta| < 5$ can also be present but are not considered for signal reconstruction nor for background rejection.
For central jets, a simple $b$ tagging is performed with probabilities of 60\% for $b$ jets, 10\% for charm and 1\% for light jets. We remark that the inclusion of radiation and hadronisation effects, as well as a detector simulation, is essential for our study. In an ideal situation in which the number of jets matches the number of partons in the hard process, the combinatorics to reconstruct the signals is relatively simple. In a real experiment, however, the presence of several more jets than were present at the partonic level, the radiation and the presence of mistags make it much more difficult to reconstruct and identify signals than it would be apparent with a toy parton-level simulation. An explicit example of these difficulties will be found in the single lepton channel in section~\ref{sec:1l}, where we will show that $\TT$ and $\BB$ signals can sometimes be very alike, despite the very different decay chains involved.

An adequate background calculation is another essential ingredient for our evaluations.
For multi-lepton signals, especially trileptons and like-sign dileptons, $t \bar t nj$ (where $nj$ stands for $n$ additional jets at the partonic level) is one of the largest and most dangerous backgrounds, due to its large cross section and the fact that $b$ quark decays sometimes produce isolated charged leptons. This background simply cannot be estimated with a parton-level calculation. Another important effect to be taken into account is the correct matching between the ``soft'' radiation generated by the parton shower Monte Carlo and the ``hard'' jets generated by the matrix element generator.
In order to have predictions for SM backgrounds as accurate as possible we use {\tt Alpgen}~\cite{Mangano:2002ea} to generate hard events which are interfaced to {\tt Pythia}
using the MLM prescription \cite{mlm} to perform the matching avoiding double counting.
The processes generated are collected in Table~\ref{tab:allbkg}, where we also give the equivalent luminosity generated (30 fb$^{-1}$ in most cases) and the number of events after matching. The additional SM processes $b \bar b nj$ and $c \bar c nj$, which were previously shown to be negligible after selection cuts for multi-lepton states~\cite{delAguila:2008cj} are ignored in this work. (They are very likely to be negligible in the single lepton channel too, after the transverse energy and invariant mass cuts.)

\begin{table}[t]
\begin{center}
\begin{small}
\begin{tabular}{llcc}
Process & Decay & $L$ & Events \\
\hline
$t \bar t nj$, $n=0,\dots,6$   & semileptonic           & 30 fb$^{-1}$   & 6.1 M \\
$t \bar t nj$, $n=0,\dots,6$   & dileptonic             & 30 fb$^{-1}$   & 1.5 M \\
$tj$                          & $W \to l \nu$       & 30 fb$^{-1}$   & 0.9 M  \\
$t\bar b$                     & $W \to l \nu$       & 30 fb$^{-1}$   & 54 K  \\
$tW$                          & all                    & 30 fb$^{-1}$   & 1.6 M   \\
$t \bar t t \bar t$           & all                    & 30 fb$^{-1}$   & 160   \\
$t \bar t b \bar b$           & all                    & 30 fb$^{-1}$   & 34 K   \\
$Wnj$, $n=0,1,2$              & $W \to l \nu$       & 3 fb$^{-1}$    & 167 M \\
$Wnj$, $n=3,\dots,6$           & $W \to l \nu$       & 30 fb$^{-1}$    & 10 M \\
$W b \bar b nj$, $n=0,\dots,4$ & $W \to l \nu$       & 30 fb$^{-1}$   & 520 K \\
$W c \bar c nj$, $n=0,\dots,4$ & $W \to l \nu$       & 30 fb$^{-1}$   & 550 K \\
$W t \bar t nj$, $n=0,\dots,4$ & $W \to l \nu$       & 30 fb$^{-1}$   & 5.1 K \\
$Z/\gamma\, nj$, $n=0,1,2$, $m_{ll} < 120$ GeV
                              & $Z \to l^+ l^-$  & 3 fb$^{-1}$    & 16.5 M \\
$Z/\gamma\, nj$, $n=3,\dots,6$, $m_{ll} < 120$ GeV
                              & $Z \to l^+ l^-$  & 30 fb$^{-1}$    & 1.1 M \\
$Z/\gamma\, nj$, $n=0,\dots,6$, $m_{ll} > 120$ GeV
                              & $Z \to l^+ l^-$  & 30 fb$^{-1}$   & 1.7 M \\
$Z b \bar b nj$, $n=0,\dots,4$ & $Z \to l^+ l^-$  & 30 fb$^{-1}$   & 200 K \\
$Z c \bar c nj$, $n=0,\dots,4$ & $Z \to l^+ l^-$  & 30 fb$^{-1}$   & 180 M \\
$Z t \bar t nj$, $n=0,\dots,4$ & $Z \to l^+ l^-$  & 30 fb$^{-1}$   & 1.9 K \\
$WWnj$, $n=0,\dots,3$          & $W \to l \nu$       & 30 fb$^{-1}$   & 290 K \\
$WZnj$, $n=0,\dots,3$          & $W \to l \nu$, $Z \to l^+ l^-$
                                                       & 30 fb$^{-1}$   & 37.7 K \\
$ZZnj$,  $n=0,\dots,3$          & $Z \to l^+ l^-$  & 30 fb$^{-1}$   & 3.7 K \\
$WWWnj$, $n=0,\dots,3$         & $2W \to l \nu$      & 30 fb$^{-1}$   & 1.5 K \\
$WWZnj$, $n=0,\dots,3$         & all                    & 30 fb$^{-1}$   & 4.9 K \\
$WZZnj$, $n=0,\dots,3$         & all                    & 30 fb$^{-1}$   & 1.5 K
\end{tabular}
\end{small}
\caption{Background processes considered in the simulations. The second column indicates the decay modes included (where $l=e,\mu,\tau$), and the third column the luminosity equivalent generated. The last column corresponds to the number of events after matching, with K and M standing for $10^3$ and $10^6$ events, respectively.}
\label{tab:allbkg}
\end{center}
\end{table}

The procedure used for estimating the statistical significance of a signal is considered case by case. To claim discovery we require both (i) a statistical significance larger tan $5\sigma$; (ii) at least 10 signal events. 
In the absence of any systematic uncertainty on the background, the statistical significance would be $\mathcal{S}_0 \equiv S/\sqrt B$, where $S$ and $B$ are the number of signal and background events, or its analogous from the $P$-number for small backgrounds where Poisson statistics must be applied. Nevertheless, there are systematic uncertainties in the background evaluation from several sources: the theoretical calculation, parton distribution functions (PDFs), the collider luminosity, pile-up, ISR and FSR, etc. as well as some specific uncertainties related to the detector like the energy scale and $b$ tagging efficiency. Such uncertainties have little relevance in the cleanest channels, where the discovery luminosity is controlled by the requirement of at least 10 signal events, being the significance far above $5\sigma$. For the channels in which the background normalisation can be important, we consider whether the signal manifests as a clear peak in a distribution. In such case it would be possible in principle to normalise the background directly from data, and extract the peak significance. Otherwise, we include a 20\% background uncertainty in the significance summed in quadrature, using as estimator $\mathcal{S}_{20} \equiv S/\sqrt{B+(0.2 B)^2}$.

\section{Final state $\ell^+ \ell^+ \ell^- \ell^-$}
\label{sec:4l}

\begin{figure}[htb]
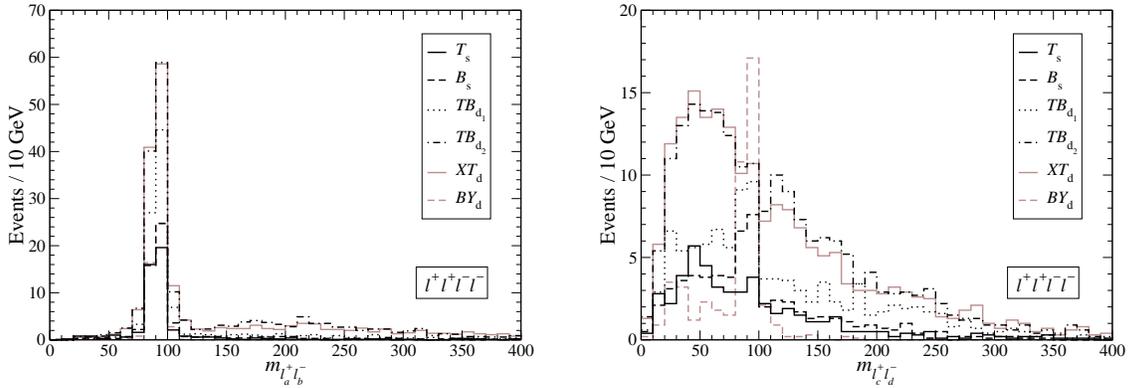

\begin{center}
\begin{tabular}{ccc}
\epsfig{file=Figs/mZ1-4Q0.eps,height=5.1cm,clip=}  & \quad &
\epsfig{file=Figs/mZ2-4Q0.eps,height=5.1cm,clip=}
\end{tabular} 
\caption{$\ell_a^+ \ell_b^-$, $\ell_c^+ \ell_d^-$ invariant mass distributions for the six models in the $\ell^+ \ell^+ \ell^- \ell^-$ final state (see the text).
The luminosity is 30 fb$^{-1}$.}
\label{fig:mrec-4Q0}
\end{center}
\end{figure}

We begin our survey of the relevant final states with the one containing four leptons, which is the cleanest and less demanding one. The heavy quark signal reconstruction is not possible in most cases, but a simple event counting in several four-lepton subsamples already provides an useful test of the heavy quark signals. 
Having a small branching ratio in general, four leptons can be produced in several cascade decays of heavy quark pairs, for example
\begin{align}
& T \bar T \to Zt \, W^- \bar b \to Z W^+b \, W^- \bar b
&& \quad Z \to \ell^+ \ell^- , W \to \ell \nu \,, \nonumber \\
& T \bar T \to Zt \, \HZ \bar t \to Z W^+b \, \HZ W^- \bar b
&& \quad Z \to \ell^+ \ell^- , W \to \ell \nu , \HZ \to q \bar q/\nu \bar \nu \,, \nonumber \\
& B \bar B \to Zb \, Z \bar b 
&& \quad Z \to \ell^+ \ell^- \,, \nonumber \\
& B \bar B \to Zb \, W^+ \bar t \to Z b W^+ W^- \bar b
&& \quad Z \to \ell^+ \ell^- , W \to \ell \nu \,, \nonumber \\
& B \bar B \to W^- t \, W^+ \bar t \to W^- W^+ b \, W^+ W^- \bar b
&& \quad W \to \ell \nu \,, \nonumber \\
& X \bar X \to W^+ t \, W^- \bar t \to W^+ W^+ b \, W^- W^- \bar b 
&& \quad W \to \ell \nu \,, 
\label{ec:ch4Q0} 
\end{align}
with $\HZ=Z,H$. The charge conjugate channels are implicitly included as well.
The SM background is mainly constituted by $ZZnj$, $t \bar t nj$ and $Z t \bar t nj$. The first one can be suppressed simply by requiring the presence of at least one $b$-tagged jet, which hardly affects the signals which have two or more $b$ quarks. Thus, 
for signal pre-selection we demand (i) four leptons summing a zero total charge, two of them with transverse momenta $p_T > 30$ GeV and the other two with $p_T > 10$ GeV; (ii) at least one $b$-tagged jet with $p_T > 20$ GeV. We then develop three different analyses with disjoint event samples, aiming to separate the different signal sources of four leptons ($ZZ$, $ZWW$ or $WWWW$ leptonic decays). The criterion for the subdivision is the number of same-flavour, opposite-charge lepton pairs with an invariant mass consistent with $M_Z$ within some given interval, and the samples are labelled as `$ZZ$', `$Z$' and `no $Z$', respectively. The invariant mass distribution of opposite-sign pairs can be studied by choosing pairs $\ell_a^+ \ell_b^-$ as follows:
\begin{enumerate}
\item If the charged leptons can be combined to form two $Z$ candidates (there are two possibilities to construct two opposite-sign pairs), we label these pairs as $\ell_a^+ \ell_b^-$, $\ell_c^+ \ell_d^-$, ordered by transverse momentum.
\item If not, we still look for a $Z$ candidate combining opposite-sign pairs (there are four possible combinations). If found, we label this pair as $\ell_a^+ \ell_b^-$
and the remaining leptons as $\ell_c^+$, $\ell_d^-$.
\item If no $Z$ candidates can be found, we construct pairs $\ell_a^+ \ell_b^-$,
$\ell_c^+ \ell_d^-$ ordered by transverse momentum.
\end{enumerate}
The interval chosen to accept a $Z$ boson candidate is $M_Z \pm 15$ GeV, which provides a good balance between signal efficiency (for true $Z$ boson decays) and rejection of non-resonant  $W^+ W^-$ decays giving opposite-charge leptons.
The $\ell_a^+ \ell_b^-$ and $\ell_c^+ \ell_d^-$ invariant mass distributions are presented in Fig.~\ref{fig:mrec-4Q0} for the six models, which are identified by the labels
$\Ts$, $\Bs$, $\TBd$, $\TBD$ (corresponding to the two mixing scenarios defined in section~\ref{sec:2.3}),
$\XTd$ and $\BYd$.
These plots illustrate the relative size of the different signal contributions. Most signal events have at least one $Z$ boson candidate: the $WWWW$ decays correspond to $\ell_a^+ \ell_b^-$ outside the $Z$ peak (left plot). Events with two $Z$ candidates are the ones with $\ell_c^+ \ell_d^-$ at the $Z$ peak (right plot). The distribution of signal and background events in the three samples at pre-selection is given in Table~\ref{tab:nsnb-4Q0}.

\begin{table}[htb]
\begin{center}
\begin{small}
\begin{tabular}{ccccccccccc}
               & Total & $ZZ$ & $Z$  & no $Z$ & \quad &          & Total & $ZZ$ & $Z$ & no $Z$
\\[1mm]
$\TT$ ($\Ts$)  & 50.0  & 4.7  & 33.3 & 12.0   & & $\BB$ ($\Bs$)  & 58.9 & 12.3 & 32.2 & 14.4 \\
$\TT$ ($\TBd$) & 52.4  & 3.9  & 35.2 & 13.3   & & $\BB$ ($\TBd$) & 54.3 & 12.4 & 28.3 & 13.6 \\
$\TT$ ($\TBD$/$\XTd$) & 114.8 & 12.1 & 77.5 & 25.2 & & $\BB$ ($\TBD$) & 86.3 & 1.2 & 19.7 & 65.4 \\
$\XX$ ($\XTd$) & 81.9  & 1.0  & 21.2 & 59.7   & & $\BB$ ($\BYd$) & 46.7 & 29.7 & 14.7 & 2.3\\
               &       &      &      &        & & $\YY$ ($\BYd$) & 0.0  & 0.0  & 0.0  & 0.0  \\
\hline
$t \bar t nj$  & 7     & 0    & 3    & 4      & & $Z t \bar tnj$ & 15   & 0    &  15  & 0 \\
$Z b \bar bnj$ & 1     & 0    & 1    & 0      & & $ZZnj$         & 2    & 2    &  0   & 0
\end{tabular}
\end{small}
\end{center}
\caption{Number of events in the $\ell^+ \ell^+ \ell^- \ell^-$ final state for
the signals and main backgrounds with a luminosity of 30 fb$^{-1}$, at pre-selection level.}
\label{tab:nsnb-4Q0}
\end{table}

\subsection{Final state $\ell^+ \ell^+ \ell^- \ell^-$ ($ZZ$)}

\begin{figure}[ht]
\begin{center}
\epsfig{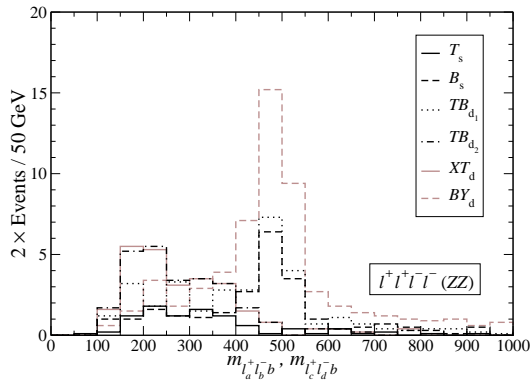}  
\caption{$\ell_a^+ \ell_b^- b$, $\ell_c^+ \ell_d^- b$ invariant mass distribution for the six models in the $\ell^+ \ell^+ \ell^- \ell^-$ ($ZZ$) final state, with two entries per event. The luminosity is 30 fb$^{-1}$.}
\label{fig:mrec-4Q0-ZZ}
\end{center}
\end{figure}

In this sample we do not impose any further requirement for event selection because the background is already tiny. The numbers of signal and background events can be read from Table~\ref{tab:nsnb-4Q0}.
We observe that this final state is most useful for the model with a $\BY$ doublet where the decays $B \to Zb$ are enhanced. The presence of the heavy quark $B$ can be established by constructing a plot  with the invariant mass of the $b$-tagged jet and each of the two reconstructed $Z$ bosons
(two entries per event).
This is shown in Fig.~\ref{fig:mrec-4Q0-ZZ} for the six models considered, summing the contribution of the two quarks in the case of the doublets. The background, only two events, is not included. Notice that the bumps around 200 GeV in the $\TBD$ and $\XTd$ models cannot be mistaken by a charge $-1/3$ quark even with low statistics: for such a mass the heavy quark production cross section would be more than 100 times larger.

We give in Table~\ref{tab:sig-4Q0-ZZ} the luminosity required to have a $5\sigma$ discovery,
including all signal contributions within a specific model. The background normalisation uncertainty has little relevance in these cases, because the background itself is very small and the discovery luminosity is mainly determined by the minimum of 10 signal events.
We also include in this table whether a mass peak can be reconstructed, although in this case the peak observation and reconstruction clearly requires more luminosity than $5\sigma$ discovery, due to the small statistics. We point out that, since the heavy resonance is observed to decay into a $Z$ boson and a $b$ quark, it can be identified as a heavy $B$ quark. This, however, can also be done in the opposite-sign dilepton final state with six times better statistics.
For the models with $T$ quarks, mass peaks could in principle be reconstructed with a high integrated luminosity, but this is far more interesting to do in the trilepton channel where statistics are larger.

\begin{table}[ht]
\begin{center}
\begin{tabular}{ccccccc}
       & $L$      & Rec.  & \quad  &        & $L$       & Rec.  \\[1mm]
$\Ts$  & --       & no    &        & $\TBD$ & 23 \fbin  & no    \\
$\Bs$  & 24 \fbin & $m_B$ &        & $\XTd$ & 23 \fbin  & no    \\
$\TBd$ & 18 \fbin & $m_B$ &        & $\BYd$ & 10 \fbin  & $m_B$
\end{tabular}
\end{center}
\caption{Luminosity $L$ required to have a $5\sigma$ discovery in the $\ell^+ \ell^+ \ell^- \ell^-$ ($ZZ$) final state.  A dash indicates no signal or a luminosity larger than 100 \fbin. We also indicate whether a mass peak can be reconstructed in this final state.}
\label{tab:sig-4Q0-ZZ}
\end{table}

\subsection{Final state $\ell^+ \ell^+ \ell^- \ell^-$ ($Z$)}

Events with only one $Z$ boson candidate are selected in this sample. Additionally, the presence of two ($b$-tagged or not) extra jets with $p_T > 20$ GeV is required to reduce the $Z t \bar t nj$ background, hardly affecting the signals. The number of signal and background events at pre-selection and selection is collected in Table~\ref{tab:nsnb-4Q0-Z}.
Notice that for $X \bar X$ production, where $Z$ bosons are not produced in the decay,
in some cases a pair of charged leptons from $W^+ W^-$ decays accidentally have an invariant mass in the interval selected. Nevertheless, this non-resonant contribution is 5 times smaller than the one from pair production of its $T$ partner. The same comment applies to $B \bar B$ production in the $\TBD$ model.

\begin{table}[htb]
\begin{center}
\begin{tabular}{cccccccc}
               & Pre. & Sel. & \quad &          & Pre. & Sel. \\[1mm]
$\TT$ ($\Ts$)  & 33.3 & 29.5 & & $\BB$ ($\Bs$)  & 32.2 & 25.1 \\
$\TT$ ($\TBd$) & 35.2 & 31.5 & & $\BB$ ($\TBd$) & 28.3 & 21.4 \\
$\TT$ ($\TBD$/$\XTd$) & 77.5 & 74.9 & & $\BB$ ($\TBD$) & 19.7 & 14.0 \\
$\XX$ ($\XTd$) & 21.2 & 15.5 & & $\BB$ ($\BYd$) & 14.7 & 12.5 \\
               &      &      & & $\YY$ ($\BYd$) & 0.0  & 0.0   \\
\hline
$t \bar t nj$  & 3    & 0    & & $Zt \bar tnj$  & 15   & 8   \\
$Z b \bar bnj$ & 1    & 0    & & $ZZnj$         & 0    & 0     \\
\end{tabular}
\end{center}
\caption{Number of events in the $\ell^+ \ell^+ \ell^- \ell^-$ ($Z$) sample for
the signals and main backgrounds with a luminosity of 30 \fbin.}
\label{tab:nsnb-4Q0-Z}
\end{table}

The reconstruction in this final state is very difficult due to the presence of two final state neutrinos, each resulting from the decay of one heavy quark. Thus, we restrict our analysis of this sample to the $Z$ boson identification and a simple counting of events, which can already be an useful test of the different models. In Table~\ref{tab:sig-4Q0-Z} we collect the discovery luminosities for the six models studied.
We observe that those with $T$ quarks give important signals, especially the ones with enhanced branching ratio for $T \to Zt$, and the discovery luminosities are relatively small.
In these interesting cases the background normalisation uncertainty is not important because the signals are much larger.

\begin{table}[t]
\begin{center}
\begin{tabular}{ccccccc}
       & $L$      & Rec. & \quad &        & $L$       & Rec. \\[1mm]
$\Ts$  & 11 \fbin & no   &       & $\TBD$ & 3.4 \fbin & no \\
$\Bs$  & 14 \fbin & no   &       & $\XTd$ & 3.3 \fbin & no \\
$\TBd$ & 5.7 \fbin & no  &       & $\BYd$ & 50 \fbin        & no
\end{tabular}
\end{center}
\caption{Luminosity $L$ required to have a $5\sigma$ discovery in the $\ell^+ \ell^- \ell^+ \ell^-$ ($Z$) final state.
We also indicate whether a mass peak can be reconstructed in this final state.}
\label{tab:sig-4Q0-Z}
\end{table}

\subsection{Final state $\ell^+ \ell^+ \ell^- \ell^-$ (no $Z$)}

This sample contains the signal and background events for which all opposite-sign pairs have invariant masses $|m_{\ell_i^+ \ell_j^-} - M_Z| > 15$ GeV. We do not apply any further event selection criteria since the background at pre-selection is already rather small. The number of signal and background events can be read in Table~\ref{tab:nsnb-4Q0}. The most important signals are from $X \bar X$ production, for which the decay $X \to W^+ t \to W^+ W^+ b$ has branching ratio unity, and $B \bar B$ production in the $\TBD$ model, with unit branching ratio for
$B \to W^- t \to W^- W^+ b$. The latter decay approximately has a branching ratio of 0.25 for the $B$ singlet and $\TBd$ doublet models, and is absent for the $\BY$ doublet.
$T \bar T$ production, which in the four-lepton final state at least involves one $Z$ leptonic decay, gives a small contribution which is only due to the finite $Z$ width and energy resolution of the detector. 

\begin{table}[h]
\begin{center}
\begin{tabular}{ccccccc}
       & $L$      & Rec. & \quad &        & $L$       & Rec. \\[1mm]
$\Ts$  & 35 \fbin & no   &       & $\TBD$ & 3.3 \fbin & no \\
$\Bs$  & 25 \fbin & no   &       & $\XTd$ & 3.5 \fbin & no \\
$\TBd$ & 11 \fbin & no   &       & $\BYd$ & --        & no
\end{tabular}
\end{center}
\caption{Luminosity $L$ required to have a $5\sigma$ discovery in the $\ell^+ \ell^+ \ell^- \ell^-$ (no $Z$) final state. A dash indicates no signal or a luminosity larger than 100 \fbin.
We also indicate whether a mass peak can be reconstructed in this final state.}
\label{tab:sig-4Q0-noZ}
\end{table}

We collect in Table~\ref{tab:sig-4Q0-noZ} the luminosity required for $5\sigma$ discovery of the six models considered in this work.
The reconstruction in this final state is virtually impossible because four neutrinos are present in the final state and, in fact, all like-sign and opposite-sign dilepton distributions seem very similar. Nevertheless, as in the previous sample, the number of events itself is a very good check of the different models.
For the most interesting signals (with $\TB$ and $\XT$ doublets) the background normalisation is not important, while for the other cases the luminosities given are a little optimistic.

\subsection{Summary}

Four lepton final states have seldom been considered in the context of heavy quark searches, perhaps because they are less relevant for the traditionally most popular models with $T$ or $B$ singlets. Nevertheless, for the $\XT$ and $\BY$ doublets and the $\TBD$ model the multi-lepton signals are larger in general: either for the decays $T \to Zt$, $B \to Zb$ (which have branching ratios two times larger than in the singlet case), or from the decays $X \to W^+ t$, $B \to W^- t$ (with unit branching ratio). Thus, the four-lepton final state can be interesting for this class of models. One has to note here that the sensitivity to heavy quark signals in other final states is much better, and discovery luminosities one order of magnitude smaller. Still, four lepton signals would be visible with a moderate luminosity and should be explored to test the models.

It is very convenient to divide the four lepton final state in three different subsets (`$ZZ$', `$Z$' and `no $Z$') depending on the number of $Z$ boson candidates (2, 1 and 0, respectively) present. This subdivision allows for some model discrimination from event counting in this final state alone, for example:
\begin{itemize}
\item If a signal is simultaneously observed in the `$Z$' and `no $Z$' samples with a similar luminosity, but not in the `$ZZ$' one, it points towards a $\XT$ doublet or a $\TB$ doublet predominantly mixing  with the top quark ($\TBD$ model).
\item If, conversely, a signal is observed exclusively in the `$ZZ$' sample, it corresponds to a $\BY$ doublet. The presence of the heavy $B$ quark can also be established by the observation of a peak in the $Zb$ invariant mass distribution. However, this can also be done in the opposite-sign dilepton final state with six times better statistics.
\end{itemize}
Finally, it is worth mentioning that
the four lepton final state is also a possible signal of heavy charged lepton in several models~\cite{AguilarSaavedra:2009ik}, but in that case the invariant mass of three charged leptons displays a very clear and sharp peak at the heavy charged lepton mass $m_E$, and $b$ quarks are not produced. Four leptons are also produced in the decay of doubly charged scalars produced in pairs (for a detailed analysis see Ref.~\cite{delAguila:2008cj}) but for the scalar triplet signals are clearly distinguishable by the presence of narrow peaks in the like-sign dilepton invariant mass distributions.

\section{Final state $\ell^\pm \ell^\pm \ell^\mp$}
\label{sec:3l}

The trilepton final state offers a good balance between signal branching ratio in $\TT$, $\BB$ and $\XX$ production, and SM background. Three leptons can result from several heavy quark pair cascade decays, either involving the leptonic decay of a $Z$ and a $W$ boson, as for example in
\begin{align}
& T \bar T \to Zt \, W^- \bar b \to Z W^+b W^- \bar b
&& \quad Z \to \ell^+ \ell^- , WW \to \ell \nu q \bar q' \,, \nonumber \\
& T \bar T \to Zt \, \HZ \bar t \to Z W^+b \, \HZ W^- \bar b
&& \quad Z \to \ell^+ \ell^- , WW \to \ell \nu q \bar q' , \HZ \to q \bar q/\nu \bar \nu \,, \nonumber \\
& B \bar B \to Z b \, W^+ \bar t \to Z b \, W^+ W^- \bar b
&& \quad Z \to \ell^+ \ell^- , WW \to \ell \nu q \bar q' \,, 
\label{ec:ch3Q1Z}
\end{align}
with $\HZ=Z,H$, or of three $W$ bosons,
\begin{align}
& B \bar B \to W^- t \, W^+ \bar t \to W^- W^+ b \, W^+ W^- \bar b
&& \quad 3W \to \ell \nu , 1W \to q \bar q' \,, \nonumber \\
& X \bar X \to W^+ t \, W^- \bar t \to W^+ W^+ b \, W^- W^- \bar b 
&& \quad 3W \to \ell \nu , 1W \to q \bar q' \,.
\label{ec:ch3Q1noZ}
\end{align}
The charge conjugate channels are implicitly included in all cases. All these production and decay channels are interesting and a first signal discrimination can be made, as in the previous section, by the presence or not of $Z$ boson candidates in the final state. 
In the sample with $Z$ candidates it is necessary to go further and try to separate the three channels in Eqs.~(\ref{ec:ch3Q1Z}). An obvious reason motivating this separation is that for $\TB$ doublets both $\TT$ and $\BB$ pairs can be produced and the three processes in Eqs.~(\ref{ec:ch3Q1Z}) are present in general. Then, it is quite desirable to separate the signals of $T$ and $B$ quarks, identifying their production and decay channels. The discrimination is possible with a probabilistic analysis which classifies the events into the three processes in Eqs.~(\ref{ec:ch3Q1Z}) with a good efficiency.

The main SM backgrounds to trilepton signals are from $WZnj$ and $t \bar t nj$ production, both roughly of the same size. The latter is originated when the two $W$ bosons decay leptonically and one $b$ quark gives a third isolated lepton but, as in the like-sign dilepton final state examined in the next section, it can be significantly reduced by asking that the two like-sign leptons have high transverse momenta. Thus, for event pre-selection we require the presence of three charged leptons (summing a total charge $\pm 1$), the like-sign pair having $p_T > 30$ GeV and the third lepton with $p_T > 10$ GeV. As mentioned above, we divide the trilepton sample into two disjoint ones. The first one contains events where a $Z$ boson candidate can be identified, that is, when two same-flavour opposite-charge leptons have an invariant mass consistent with $M_Z$. The other sample contains events without $Z$ candidates. The interval  in which a lepton pair is accepted as a $Z$ candidate is chosen to be of 15 GeV around $M_Z$. We can compare the signal contributions to the two samples by plotting the invariant mass of two opposite-charge leptons $\ell_a^+$, $\ell_b^-$, chosen in the following way:
\begin{enumerate}
\item If there is a $Z$ candidate, we label the corresponding leptons as $\ell_a^+$, $\ell_b^-$. In case that there are two $Z$ candidates, which can accidentally happen, the leptons with largest transverse momenta are chosen. 
\item If there are no $Z$ candidates, we choose $\ell_a^+$, $\ell_b^-$ with the largest transverse momenta.
\end{enumerate}
\begin{figure}[htb]
\begin{center}
\epsfig{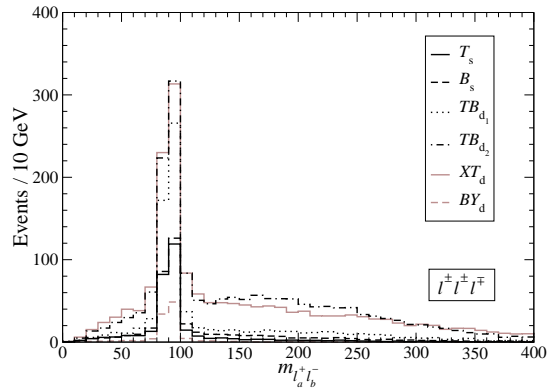}
\caption{$\ell_a^+ \ell_b^-$ invariant mass distributions for the six models in the $\ell^\pm \ell^\pm \ell^\mp$ final state (see the text for the definition of $\ell_a^+$ and $\ell_b^-$).
The luminosity is 30 fb$^{-1}$.}
\label{fig:mZrec-3Q1}
\end{center}
\end{figure}
\begin{table}[h]
\begin{center}
\begin{tabular}{cccccccccccc}
               & Total  & $Z$   & no $Z$ & \quad &          & Total & $Z$ & no $Z$ \\[1mm]
$\TT$ ($\Ts$)  & 320.7  & 212.4 & 108.3  & & $\BB$ ($\Bs$)  & 421.9 & 227.9 & 194.0 \\
$\TT$ ($\TBd$) & 349.0  & 229.9 & 119.1  & & $\BB$ ($\TBd$) & 484.5 & 237.0 & 247.5 \\
$\TT$ ($\TBD$/$\XTd$) & 654.6  & 435.8 & 218.8  & & $\BB$ ($\TBD$) & 1174.4 & 144.0 & 1030.4 \\
$\XX$ ($\XTd$) & 1181.8 & 143.9 & 1037.9 & & $\BB$ ($\BYd$) & 106.3 & 88.3 & 18.0 \\
               &        &       &        & & $\YY$ ($\BYd$) & 0.5   & 0.1  & 0.4 \\
\hline
$t \bar t nj$  & 464    & 114   & 350    & & $WZnj$         & 4258  & 4196 & 62  \\
$W t \bar tnj$ & 78     & 11    & 67     & & $ZZnj$         & 424   &  417 & 7  \\
$Z t \bar tnj$ & 189    & 169   & 20      \\
\end{tabular}
\end{center}
\caption{Number of events in the $\ell^\pm \ell^\pm \ell^\mp $ final state for
the signals and main backgrounds with a luminosity of 30 fb$^{-1}$, at pre-selection level.}
\label{tab:nsnb-3Q1}
\end{table}
The resulting distribution is shown in Fig.~\ref{fig:mZrec-3Q1}. 
We observe that there is a large off-peak signal from $B \bar B$ and $X \bar X$ decays in the $\TBD$ and $\XTd$ models, respectively. The number of events at pre-selection in each sample is given in Table~\ref{tab:nsnb-3Q1}.
A sizeable fraction of events from $\TT$ decays, which in the trilepton channel always involve a $Z$ boson, are classified in the `no $Z$' set, while around 10\% of the $\BB$ and $\XX$ events in which $Z$ bosons are not present are accepted in the `$Z$' sample. The rate of wrong assignments can be reduced at the cost of losing signal efficiency, by strenghtening the classification criteria. For example, $Z$ candidates could be accepted only in the interval $M_Z \pm 10$ GeV and  events in the `no $Z$' subsample could be rejected if opposite-charge pairs have an invariant mass in the range $M_Z \pm 20$ GeV. This fine tuning of the analysis makes more sense with a full detector simulation, and is not necessary for model discrimination, anyway.

\subsection{Final state $\ell^\pm \ell^\pm \ell^\mp$ ($Z$)}

This final state receives important contributions from $\TT$ and $\BB$ production in the channels of Eqs.~(\ref{ec:ch3Q1Z}). We will first perform an analysis with fewer selection criteria to suppress the background and obtain the heavy quark discovery potential for this final state. Then, we will address the identification of a heavy quark signal eventually observed, strengtheining our requirements on signal and background events and using a likelihood function which assigns them to each of the decay channels in Eqs.~(\ref{ec:ch3Q1Z}). After that,
events will be reconstructed accordingly to their classification and to the kinematics assumed in each case.

\subsubsection{Discovery potential}

\begin{figure}[p]
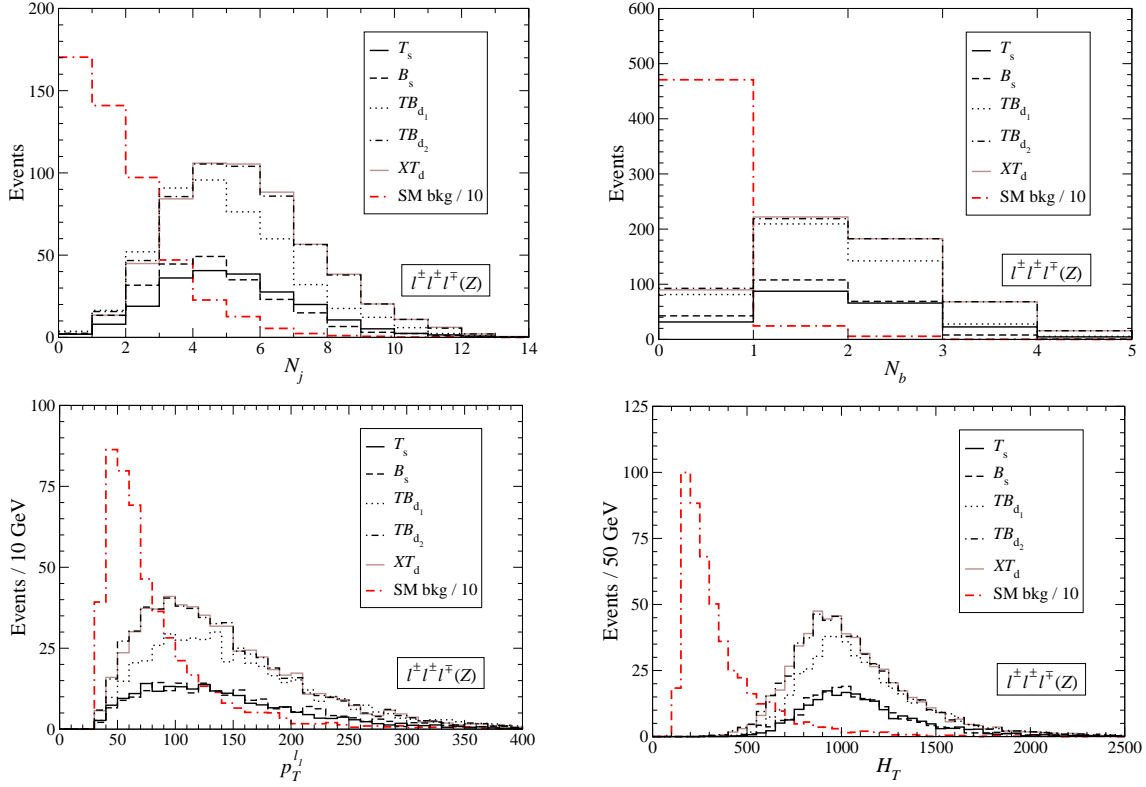

\begin{center}
\begin{tabular}{ccc}
\epsfig{file=Figs/mult-3Q1-Z.eps,height=5.1cm,clip=} & \quad &
\epsfig{file=Figs/bmult-3Q1-Z.eps,height=5.1cm,clip=} \\
\epsfig{file=Figs/ptlep1-3Q1-Z.eps,height=5.1cm,clip=} & \quad &
\epsfig{file=Figs/HT-3Q1-Z.eps,height=5.1cm,clip=}
\end{tabular}
\caption{Kinematical distributions of variables used in selection and recontruction criteria
for the $\ell^\pm \ell^\pm \ell^\mp$ ($Z$) final state: light jet multiplicity $N_j$, $b$ jet multiplicity $N_b$, transverse momentum of the leading lepton and total transverse energy. The luminosity is 30 fb$^{-1}$.}
\label{fig:dist-3Q1-Z}
\end{center}
\end{figure}

For events with one $Z$ candidate we ask (i) at least two light jets with $p_T > 20$ GeV; (ii) one $b$-tagged jet also with $p_T > 20$ GeV; (iii) transverse momentum $p_T > 50$ GeV for the leading charged lepton $\ell_1$; (iv) transverse energy $H_T > 500$ GeV. The kinematical distributions of these variables at pre-selection are presented in Fig.~\ref{fig:dist-3Q1-Z} for the relevant signals and the SM background. 
In particular, requiring a $b$-tagged jet hardly affects the signals but practically eliminates the $WZnj$ background which does not have $b$ quarks. The cuts on transverse energy and leading charged lepton momentum are quite general to look for new heavy quarks and are not optimised for the input masses used in our calculation. Notice also that the $H_T$ distribution for the signals clearly indicates that one or more heavy particles with masses summing around 1 TeV are produced. This data will be crucial later when we address the disentanglement and reconstruction of different signal channels.

\begin{table}[p]
\begin{center}
\begin{tabular}{cccccccccc}
               & Pre.  & Sel.  & Rec.  & \quad &          & Pre. & Sel.   & Rec. \\[1mm]
$\TT$ ($\Ts$)  & 212.4 & 162.7 & 82.9  & & $\BB$ ($\Bs$)  & 227.9 & 162.4 & 65.3 \\
$\TT$ ($\TBd$) & 229.9 & 181.3 & 87.1  & & $\BB$ ($\TBd$) & 237.0 & 174.3 & 72.6 \\
$\TT$ ($\TBD$/$\XTd$) & 435.8 & 356.1 & 211.0 & & $\BB$ ($\TBD$) & 144.0 & 94.9 & 34.2 \\
$\XX$ ($\XTd$) & 143.9 & 99.9  & 35.7  & & $\BB$ ($\BYd$) & 88.3  & 58.6  & 21.2 \\
               &       &       &       & & $\YY$ ($\BYd$) & 0.1   & 0.0   & 0.0 \\
\hline
$t \bar t nj$  & 114   & 1     & 0     & & $WZnj$         & 4196  & 24    & 0 \\
$W t \bar tnj$ & 11    & 3     & 1     & & $ZZnj$         & 417   & 1     & 0 \\
$Z t \bar tnj$ & 169   & 89    & 32     \\
\end{tabular}
\end{center}
\caption{Number of events at the pre-selection, selection and reconstruction levels in the $\ell^\pm \ell^\pm \ell^\mp$ ($Z$) sample for
the signals and main backgrounds with a luminosity of 30 fb$^{-1}$.}
\label{tab:nsnb-3Q1-Z}
\end{table}

The number of signal and background events at the selection level is given in Table~\ref{tab:nsnb-3Q1-Z}, also including the values at pre-selection for better comparison.
As it might be expected, the most important background after cuts is $Z t \bar t nj$, which has a $Z$ boson, two $b$ quarks, two $W$ bosons and large transverse energy, and is then quite similar to the signals. More aggresive cuts will of course reduce this and the other backgrounds but we refrain ourselves from performing such optimisations. The discovery luminosities are given in Table~\ref{tab:sig-3Q1-Z}, summing all signal contributions within a given model. We observe that this clean channel offers a good potential to discover $T$ and $B$ quarks in singlet or doublet representations. For the $\BY$ doublet the discovery luminosity may be optimistic because it does not take into account the background normalisation uncertainty, which may be important in this case where the signal is small. 

\begin{table}[ht]
\begin{center}
\begin{tabular}{ccccccc}
       & $L$      & Rec. & \quad &        & $L$       & Rec. \\[1mm]
$\Ts$  & 3.4 \fbin  & $m_T$   &       & $\TBD$ & 0.73 \fbin & $m_T$ \\
$\Bs$  & 3.4 \fbin & $m_B$   &       & $\XTd$ & 0.72 \fbin & $m_T$ \\
$\TBd$ & 1.1 \fbin & $m_T$, $m_B$    &       & $\BYd$ & 26 \fbin    & $m_B$
\end{tabular}
\end{center}
\caption{Luminosity $L$ required to have a $5\sigma$ discovery in the $\ell^\pm \ell^\pm \ell^\mp$ ($Z$) final state.
We also indicate whether a mass peak can be reconstructed in this final state.}
\label{tab:sig-3Q1-Z}
\end{table}

\subsubsection{Heavy quark reconstruction}
\label{sec:3l-Z-2}

The broad sensitivity to $\TT$ and $\BB$ signals of this final state implies that if a positive excess is observed, identifying its nature will require a more elaborate analysis. Indeed, the decay modes in Eqs.~(\ref{ec:ch3Q1Z}) give signals only differing by the number of jets and the location of the resonant peaks. The identification can be done efficiently, however, by using a likelihood method which gives the probability that a given event corresponds to each of the decay modes. 
We build probability distribution functions (p.d.f.) for three signal classes: ($a$) $\TT \to ZtWb$; ($b$) $\TT \to Zt \HZ t$, with $\HZ=H,Z$ decaying hadronically or invisibly; ($c$) $\BB \to Zb Wt$. We generate high-statistics samples different from the ones used for the final analysis. We do not include a separate class for the background, because the likelihood function is only used to identify signals and not to reject the background which is rather small. Nevertheless, the discriminant analysis also affects the background and, for instance, if we restrict ourselves to events
classified as resulting from $\BB$ production, a sizeable part of the background is classified as $\TT$-like and thus rejected.
Note also that an essential parameter for building the kinematical distributions for the signals is the heavy quark mass. If a heavy quark signal is observed at LHC, the approximate value of the heavy quark mass can be estimated from the transverse energy distribution for the signal,\footnote{In the case of the doublets the new states are expected to be nearly degenerate, simplifying their approximate determination from this distribution.}  and then a probabilistic analysis can be performed to separate signal contributions and reconstruct the decay chain event by event.

\begin{figure}[t]
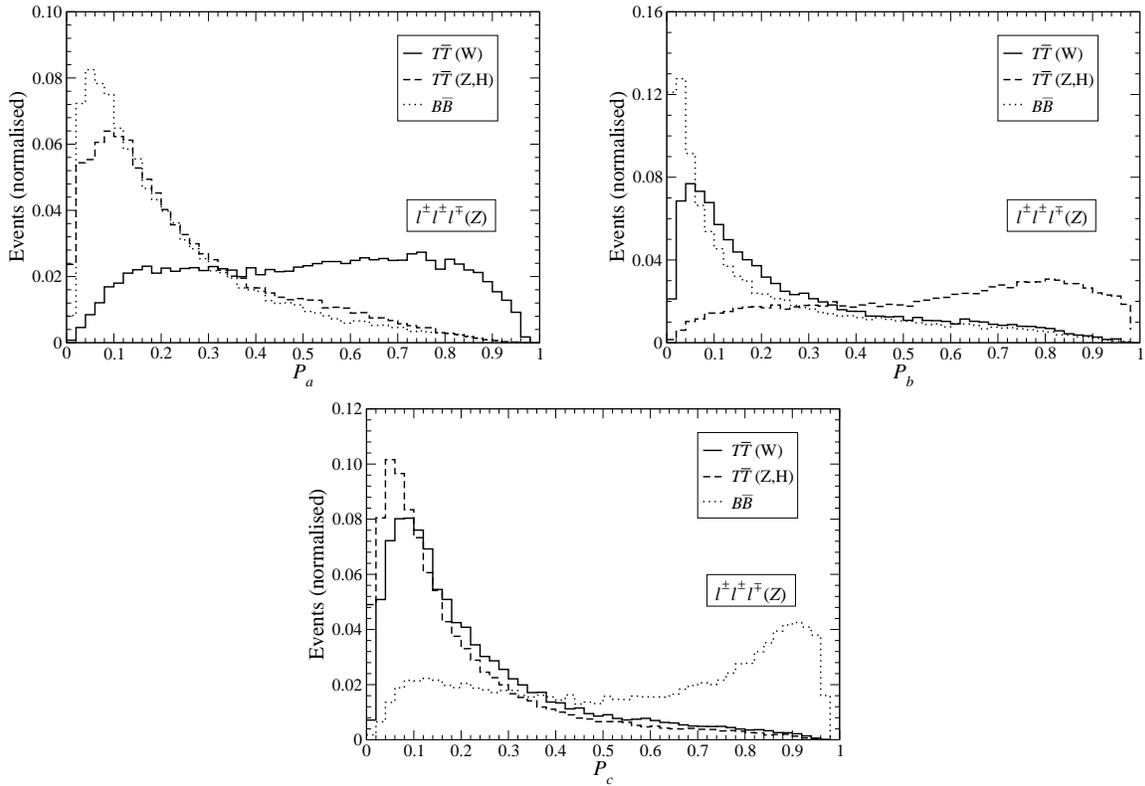

\begin{center}
\begin{tabular}{ccc}
\epsfig{file=Figs/D-Pa-3Q1-Z.eps,height=5.1cm,clip=} & \quad &
\epsfig{file=Figs/D-Pb-3Q1-Z.eps,height=5.1cm,clip=} \\
\multicolumn{3}{c}{\epsfig{file=Figs/D-Pc-3Q1-Z.eps,height=5.1cm,clip=}}
\end{tabular}
\caption{Probability distribution functions for events in the reference samples.}
\label{fig:lik2-3Q1-Z}
\end{center}
\end{figure}

For the signal discrimination and reconstruction we demand, in addition to the selection criteria already specified, the presence of at least two $b$-tagged jets. The number of signal and background events with this last requirement is given in Table~\ref{tab:nsnb-3Q1-Z}. We begin by finding two $W$ bosons decaying hadronically and leptonically. 
The former is approximately reconstructed at this stage by selecting among the light jets with largest $p_T$ (up to a maximum of four) the two ones which give an invariant mass closest to $M_W$. The latter is approximately reconstructed from the charged lepton not included in the $Z$ candidate and the missing energy, with the procedure explained below, and selecting the solution giving the smallest neutrino energy.
The variables used in the likelihood function are:
\begin{figure}[htb]
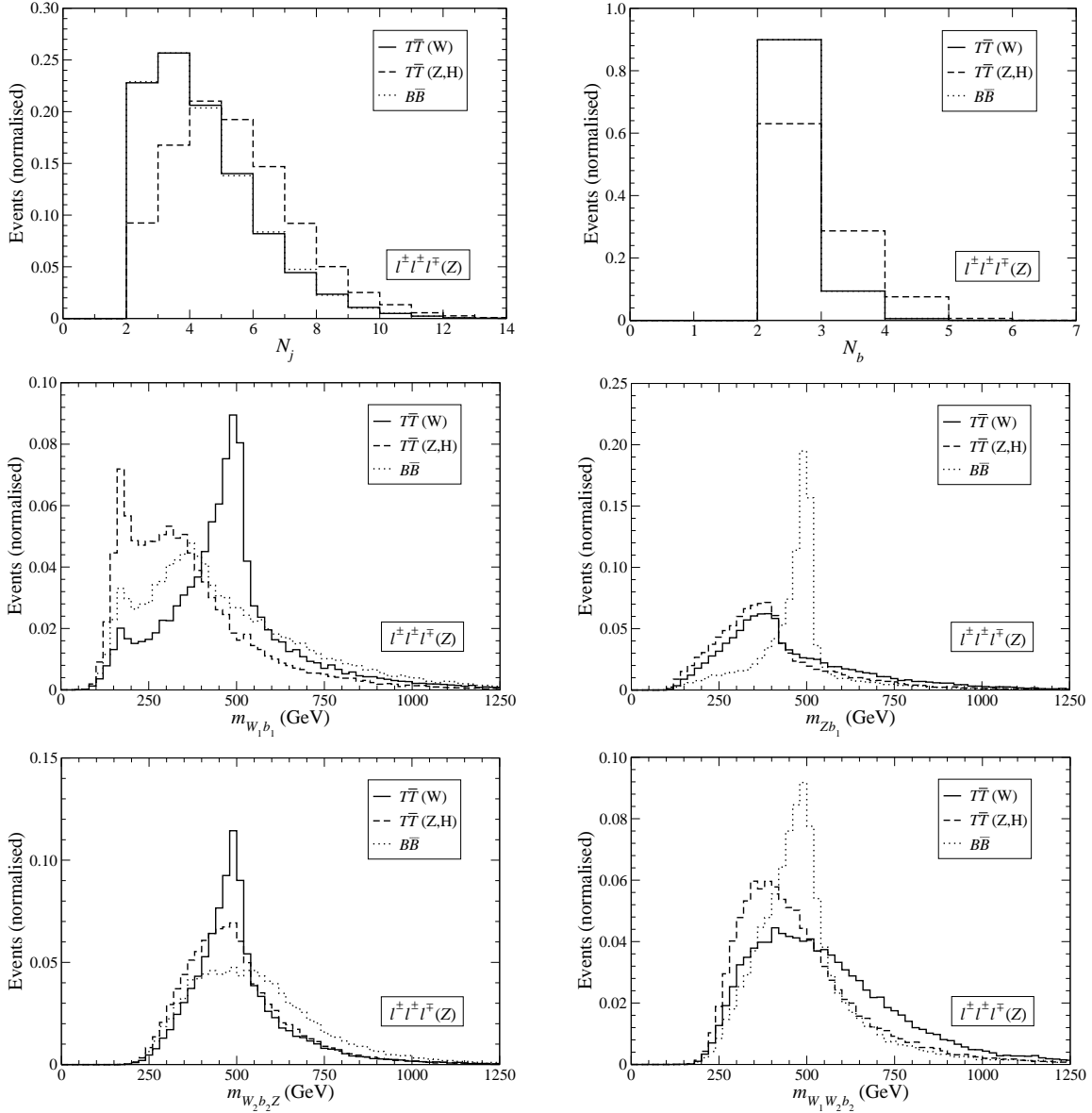

\begin{center}
\begin{tabular}{ccc}
\epsfig{file=Figs/D-mult-3Q1-Z.eps,height=5.1cm,clip=} & \quad &
\epsfig{file=Figs/D-bmult-3Q1-Z.eps,height=5.1cm,clip=} \\
\epsfig{file=Figs/D-mW1b1-3Q1-Z.eps,height=5.1cm,clip=} & \quad &
\epsfig{file=Figs/D-mZb1-3Q1-Z.eps,height=5.1cm,clip=} \\
\epsfig{file=Figs/D-mW2b2Z-3Q1-Z.eps,height=5.1cm,clip=} & \quad &
\epsfig{file=Figs/D-mW1W2b2-3Q1-Z.eps,height=5.1cm,clip=} \\
\end{tabular}
\caption{Kinematical variables used to classify the three heavy quark signals in the
$\ell^\pm \ell^\pm \ell^\mp$ ($Z$) final state.}
\label{fig:lik-3Q1-Z}
\end{center}
\end{figure}
\begin{itemize}
\item The light jet and $b$ jet multiplicities.
\item The invariant mass of the reconstructed $W$ boson (decaying hadronically or leptonically) with larger transverse momentum, labelled as $W_1$, plus the $b$ quark with largest transverse momentum, $b_1$. For $T \to Wb$ decays the $W$ boson as well as the $b$ quark are expected to have larger $p_T$, and we observe in Fig.~\ref{fig:lik-3Q1-Z} that this is often the case.
\item The invariant mass of the reconstructed $Z$ boson and the $b$ quark with highest $p_T$, which for the $\BB$ signal is most times the one resulting from $B \to Zb$, as we observe in the distribution of Fig.~\ref{fig:lik-3Q1-Z}.
\item The invariant mass of the reconstructed $W$ with smaller $p_T$ ($W_2$), the $Z$ boson and the $b$ quark with smaller transverse momentum ($b_2$).
\item The invariant mass of the two $W$ bosons and the $b$ quark with smallest $p_T$, which for the $\BB$ signal are the ones from $B \to Wt \to WWb$ in most cases.
\end{itemize}
The likelihood function evaluated on the three class samples gives the probability distributions in Fig.~\ref{fig:lik2-3Q1-Z}, where  $P_a$, $P_b$, $P_c$ are the probabilities that events correspond to each of the three likelihood classes in Eq.~(\ref{ec:ch3Q1Z}). 
Events are assigned to the class ($a$, $b$ or $c$) which has the highest probability $P_a$, $P_b$ or $P_c$, respectively. Table~\ref{tab:lik-3Q1-Z} shows the performance of the likelihood function on the reference samples.
Events in a class $x$ are correctly classified if $P_x > P_y,P_z$, where $y$, $z$ are the other classes. The probabilities for correct assignments are in the range $0.61-0.69$, which suffice to achieve a good reconstruction of the heavy resonances. We now describe the procedure followed in each case.

\begin{table}[t]
\begin{center}
\begin{tabular}{cccc}
Class & $P_a > P_b,P_c$ & $P_b > P_a,P_c$ & $P_c > P_a,P_b$ \\
\hline
($a$) & 0.61 & 0.24 & 0.15 \\
($b$) & 0.19 & 0.69 & 0.12 \\
($c$) & 0.15 & 0.20 & 0.65
\end{tabular}
\end{center}
\caption{Performance of the likelihood function on the $\ell^\pm \ell^\pm \ell^\mp$ event reference samples: fractions of events in each sample and their classification. Events in a class $x$ are correctly classified if $P_x > P_y,P_z$, where $y$, $z$ are the other classes.}
\label{tab:lik-3Q1-Z}
\end{table}

{\em Class} ($a$): $\TT \to Zt W \bar b \to Z Wb Wb$. Events which are identified as resulting from this decay chain are reconstructed using this procedure:
\begin{enumerate}
\item The $Z$ boson momentum is obtained from the opposite-sign lepton pair $Z$ candidate. 
\item Two light jets are selected to form the hadronic $W$, labelled as $W_H$.
If there are only two light jets these are automatically chosen; if there are more than two, only up to four (ordered by decreasing $p_T$) are considered. 
\item The leptonic $W$ (labelled as $W_L$) is obtained from the charged lepton $\ell$ not included in the $Z$ candidate and the missing energy, identifying $(p_\nu)_T = \ptmiss$, requiring $(p_{\ell}+p_\nu)^2 = M_W^2$ and solving for the longitudinal component of the neutrino momentum. If no real solution exists, the neutrino transverse momentum is decreased in steps of 1\% and the procedure is repeated. If no solution is still found after 100 iterations, the discriminant of the quadratic equation is set to zero.
Both solutions for the neutrino momentum are kept, and the one giving best reconstructed masses is selected.
\item Two $b$ jets are selected among the ones present, to be paired with $W_H$ and $W_L$, respectively.
\item The top quark is reconstructed from one of the $Wb$ pairs, and its parent heavy quark $T_1$ from the top quark and the $Z$ boson.
\item The other heavy quark $T_2$ is reconstructed from the remaining $Wb$ pair.
\item Among all choices for $b$ and light jets and all possible pairings, the combination minimising the quantity
\begin{small}
\begin{equation}
\frac{(m_{W_H}^\text{rec}-M_W)^2}{\sigma_W^2} + 
\frac{(m_{W_L}^\text{rec}-M_W)^2}{\sigma_W^2} + 
\frac{(m_t^\text{rec}-m_t)^2}{\sigma_t^2} +
\frac{(m_{T_1}^\text{rec}-m_{T_2}^\text{rec})^2}{\sigma_T^2}
\end{equation}
\end{small}%
is selected, with $\sigma_W = 10$ GeV, $\sigma_t = 14$ GeV~\cite{Aad:2009wy}, $\sigma_T = 20$ GeV. Notice that we include the leptonic $W$ boson reconstructed mass in the minimisation. Since the quadratic equation is forced to have a solution in all cases, sometimes the reconstructed mass is not the $W$ mass.
\end{enumerate}

{\em Class} ($b$): $\TT \to Zt \HZ \bar t \to Z Wb \HZ Wb$. For events identified as resulting from this decay chain which have at least six jets (otherwise they they are dropped) we proceed through the same steps $1-4$ as in class $(a)$, and subsequently:
\begin{enumerate}\setcounter{enumi}{4}
\item The hadronic and leptonic tops $t_H$, $t_L$ are obtained from the two $Wb$ pairs.
\item One heavy quark $T_1$ is reconstructed from one top and the $Z$ boson.
The other heavy quark is obtained from the other top and two jets chosen among the ones present ($b$-tagged or not).
\item The combination minimising
\begin{small}
\begin{equation}
\frac{(m_{W_H}^\text{rec}-M_W)^2}{\sigma_W^2} + 
\frac{(m_{W_L}^\text{rec}-M_W)^2}{\sigma_W^2} + 
\frac{(m_{t_H}^\text{rec}-m_t)^2}{\sigma_t^2} +
\frac{(m_{t_L}^\text{rec}-m_t)^2}{\sigma_t^2} +
\frac{(m_{T_1}^\text{rec}-m_{T_2}^\text{rec})^2}{\sigma_T^2}
\end{equation}
\end{small}%
is finally selected.
\end{enumerate}

{\em Class} ($c$): $\BB \to Zb Wt \to Zb WWb$. The reconstruction of this channel proceeds through  
the same steps $1-3$ as in the previous two channels, and then:
\begin{enumerate}\setcounter{enumi}{3}
\item Two $b$ jets are selected among the ones present, and one of them is paired with the $Z$ boson to reconstruct a heavy quark $B_1$.
\item The second $b$ jet is associated with one of the $W$ bosons to reconstruct a top quark, and then with the other $W$ boson to reconstruct the second  heavy quark $B_2$.
\item The combination minimising
\begin{small}
\begin{equation}
\frac{(m_{W_H}^\text{rec}-M_W)^2}{\sigma_W^2} + 
\frac{(m_{W_L}^\text{rec}-M_W)^2}{\sigma_W^2} + 
\frac{(m_{t}^\text{rec}-m_t)^2}{\sigma_t^2} +
\frac{(m_{B_1}^\text{rec}-m_{B_2}^\text{rec})^2}{\sigma_B^2}
\end{equation}
\end{small}%
is finally selected, with $\sigma_B = 20$ GeV.
\end{enumerate}

\begin{figure}[htb]
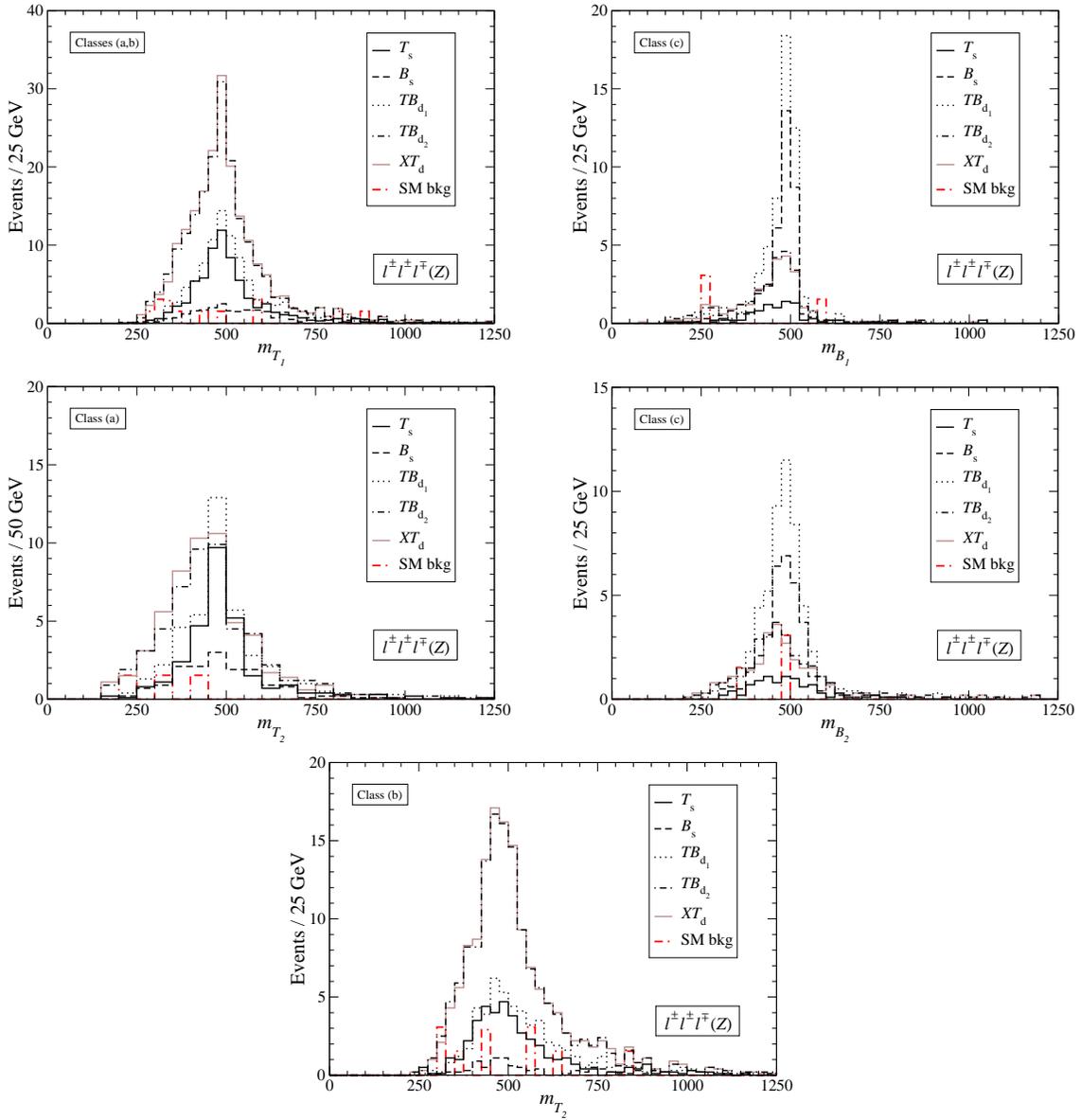

\begin{center}
\begin{tabular}{ccc}
\epsfig{file=Figs/mtZ-3Q1-Z.eps,height=5.1cm,clip=} & \quad &
\epsfig{file=Figs/mbZ-3Q1-Z.eps,height=5.1cm,clip=} \\
\epsfig{file=Figs/mbW-3Q1-Z.eps,height=5.1cm,clip=} & \quad &
\epsfig{file=Figs/mtW-3Q1-Z.eps,height=5.1cm,clip=} \\
\multicolumn{3}{c}{\epsfig{file=Figs/mtH-3Q1-Z.eps,height=5.1cm,clip=}}
\end{tabular}
\caption{Reconstructed heavy quark masses in the $\ell^\pm \ell^\pm \ell^\mp$ ($Z$) final state.}
\label{fig:mrec-3Q1-Z}
\end{center}
\end{figure}

\begin{table}[htb]
\begin{center}
\begin{small}
\begin{tabular}{cccccccccccc}
               & Total & ($a$) & ($b$) & $(c)$  & \quad &          & Total & ($a$) & ($b$) & $(c)$\\[1mm]
$\TT$ ($\Ts$)  & 82.9  & 29.1 & 40.7  & 9.3  && $\BB$ ($\Bs$)  & 65.3 & 15.3 & 9.7  & 36.5 \\
$\TT$ ($\TBd$) & 87.1  & 29.3 & 41.7  & 11.4 && $\BB$ ($\TBd$) & 72.6 & 11.4 & 12.8 & 45.2 \\
$\TT$ ($\TBD$/$\XTd$) & 211.0 & 41.0 & 133.0 & 24.2 && $\BB$ ($\TBD$) & 34.2 & 12.3 & 11.8 & 3.2 \\
$\XX$ ($\XTd$) & 35.7  & 14.2 & 12.6  & 2.3  && $\BB$ ($\BYd$) & 21.2 & 7.0  & 2.7  & 11.3 \\
               &       &      &       &      && $\YY$ ($\BYd$) & 0.0  & 0.0  & 0.0  & 0.0 \\
\hline
$Z t \bar tnj$ & 32    & 5    & 12    & 5   \\
\end{tabular}
\end{small}
\end{center}
\caption{Number of signal and background events in the $\ell^\pm \ell^\pm \ell^\mp$ ($Z$) final state at the reconstruction level assigned to each event class. The luminosity is 30 fb$^{-1}$.}
\label{tab:nsnb-3Q1-Z-C}
\end{table}

We present our results in Fig.~\ref{fig:mrec-3Q1-Z}, including all signal contributions in a given model, as well as the SM background, and discuss them in turn.
We do not include $W$ and top reconstructed masses, which show good peaks at the true masses with the optimised method used.
The separate contributions of each process are given in Table~\ref{tab:nsnb-3Q1-Z-C}, skipping several backgrounds which are practically removed at the last stage of event selection. (The total number of events includes in each case those in class $(b)$ which are later rejected by the reconstruction algorithm because they do not have at least 6 jets.)
The first plot (up, left) shows the reconstructed $T_1$ mass for events assigned to classes $(a,b)$. This heavy quark is the one decaying $T \to Zt$, with $t$ decaying either hadronically or semileptonically. The reconstruction of this peak in the $Zt$ invariant mass distribution implies that $T$ has charge $2/3$, and also shows the vector-like nature of $T$. The counterpart for $B$ quarks is shown in the second plot (up, right), with the reconstructed mass of  $B_1$, which is the quark decaying $B \to Zb$. The reconstruction of a peak in the $Zb$ invariant mass distribution shows that $B$ has charge $-1/3$ and is vector-like.

The other resonant peaks also give information regarding heavy quark decays. We show in the third plot (middle, left) the reconstructed $T_2$ mass for events in class $(a)$, which corresponds to the decay $T \to Wb$, with $W$ decaying either leptonically or hadronically. For the $T$ singlet and $\TB$ doublet in scenario 1, where this decay takes place, the peaks are sharp, and they might be observed with sufficient luminosity. We point out that the presence of events with $W$ decaying leptonically (about one half of the total) clearly indicate the $T \to Wb$ decay, but this can also be established in the single lepton final state with much larger statistics. The other models with $T$ quarks in which the decay $T \to Wb$ does not happen still have a fraction of events incorrectly assigned to this class. In these models the $Wb$ invariant mass distribution, which should peak at $m_t$, is broader and shifted towards larger values because the reconstruction procedure enforces equal masses for both heavy quarks.  The fourth plot (middle, right) represents the $B_2$ invariant mass distribution for events in class ($c$), from the decay $B \to Wt \to WWb$. This plot also shows the presence of a resonance decaying into a top quark and a $W$ boson, the latter reconstructed either from two jets or from a charged lepton plus missing energy. This peak establishes the decay $B \to Wt$, which is absent in the $\BY$ doublet. Finally, the fifth plot (down) shows the reconstructed $T_2$ mass in class $(b)$, corresponding to the decay $T \to \HZ t$, with $\HZ$ decaying into two jets. This distribution shows the presence of a resonance but does not help establish its nature, because the identity of $\HZ$ is not determined. 

A few remarks are in order. It is clear that detecting the presence of a resonant peak and drawing conclusions about the nature of the heavy quark requires a significant amount of statistics, and a compromise should be taken between having good reconstructed peaks (imposing quality cuts on class identification as well as on reconstructed $W$ boson and top quark masses, for example) and having a sufficient number of events. Here we have made no quality cuts in order to keep the signals as large as possible. But even with this conservative approach the contributions of the three cascade decays in Eqs.~(\ref{ec:ch3Q1Z}) can be disentangled,  and invariant mass peaks can be reconstructed so that, if sufficient luminosity is collected, the decays $T \to Zt$, $T \to Wb$, $B \to Zb$, $B \to Wt$ can be established. 

\begin{figure}[t]
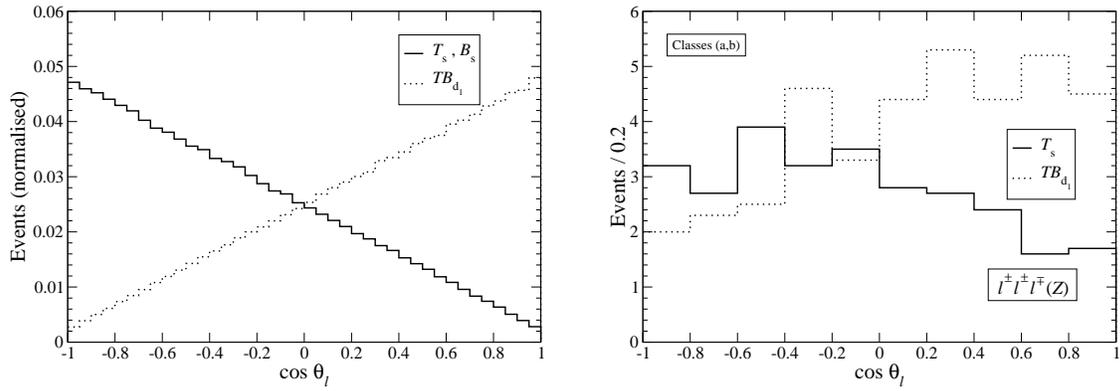

\begin{center}
\begin{tabular}{ccc}
\epsfig{file=Figs/cos_l-th.eps,height=5.1cm,clip=} & \quad &
\epsfig{file=Figs/cos_l-3Q1-Z-ab.eps,height=5.1cm,clip=} 
\end{tabular}
\caption{Left: Charged lepton distribution in the top quark rest frame for $T \to Zt$ and $B \to W^- t$ decays. Right: distribution for the $T$ singlet and $\TB$ doublet after simulation.}
\label{fig:cosl-3Q1-Z}
\end{center}
\end{figure}

Finally, we address the discrimination of $T$ singlets and $T$ quarks of a $\TB$ doublet in scenario 1, using angular distributions. (In this scenario the $T$ decay branching ratios are the same as for $T$ singlets.) In $T \to Zt$ decays, as well as in $B \to W^- t$, the top quarks are produced with a high polarisation $P = \pm 0.91$ in the helicity axis (negative for the singlets and positive for the doublets), and the opposite polarisation for antiquarks.
This allows to determine the chirality of the $WTb$ and $WtB$ couplings by looking at the charged lepton distribution in the top quark rest frame for the event subset in which the top decays leptonically. We show in Fig.~\ref{fig:cosl-3Q1-Z} (left) the theoretical distributions as computed with the Monte Carlo generator, which are the same for $T$ and $B$ quarks, since the decays $T \to Zt$ and $B \to W^- t$ involve a coupling with the same chirality, left-handed for singlets and right-handed for doublets. 
On the right panel we show the reconstructed distribution for the $T$ singlet and $\TBd$ model, including in the latter case the $B$ contribution which is flat and slightly smooths the slope of the distribution. It is clear that large statistics are required to discriminate both cases, but the differences are visible already without the need of unfolding detector effects.
The forward-backward asymmetries computed from the reconstructed distributions are $A_\text{FB} = -0.19$ for $\Ts$ and $A_\text{FB} = 0.24$ for $\TBd$, so that with 30 \fbin\ (corresponding to 27.7 and 38.6 events in each case) the statistical difference would amount to $2.4\sigma$. A complete analysis unfolding the detector effects and with an appropriate calculation of systematic uncertainties (see for example Ref.~\cite{AguilarSaavedra:2007rs} for a similar analysis)
is beyond the scope of this work. For $B \to W^- t$ decays the results are completely analogous but with smaller statistics. We also note that for these large heavy quark masses the $Z$ bosons produced in 
$T \to Zt$, $B \to Zb$ are mostly longitudinal, and the angular distribution of the $\ell^+ \ell^-$ pair from $Z$ decay is almost indistinguishable for $T$, $B$ singlets and doublets already at the generator level.

\subsection{Final state $\ell^\pm \ell^\pm \ell^\mp$ (no $Z$)}

\begin{figure}[t]
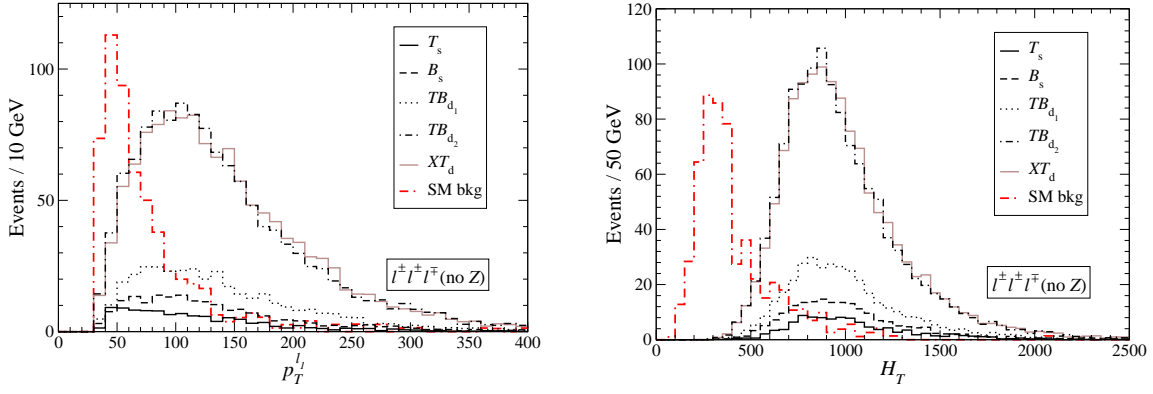

\begin{center}
\begin{tabular}{ccc}
\epsfig{file=Figs/ptlep1-3Q1-noZ.eps,height=5.1cm,clip=} & \quad &
\epsfig{file=Figs/HT-3Q1-noZ.eps,height=5.1cm,clip=}
\end{tabular}
\caption{Left: transverse momentum distribution of the leading like-sign lepton. Right: total transverse energy. The luminosity is 30 fb$^{-1}$.}
\label{fig:dist-3Q1-noZ}
\end{center}
\end{figure}

In the sample without $Z$ candidates we ask for event selection that (i) the leading like-sign lepton $\ell_1$ has transverse momentum $p_T > 50$ GeV; (ii) the total transverse energy $H_T$ is larger than 500 GeV. Notice again that these cuts are not optimised to reduce the background but are quite general to search for new heavy states. 
The kinematical distributions of the two variables at pre-selection are shown in Fig.~\ref{fig:dist-3Q1-noZ} for the SM background and all models except the $\BY$ doublet model which has a very small signal.
As in other final states,
the $H_T$ distribution clearly indicates in all cases that one or more heavy particles, summing a mass around 1 TeV, are produced. 
The number of events after this selection is given in Table~\ref{tab:nsnb-3Q1-noZ}, including also the numbers at pre-selection for comparison. We do not require $b$-tagged jets at this stage because it does not improve the background rejection, since most of the backgrounds have two top quarks.
\begin{table}[t]
\begin{center}
\begin{tabular}{cccccccccc}
               & Pre.   & Sel.  & Rec.  & \quad &          & Pre. & Sel.   & Rec. \\[1mm]
$\TT$ ($\Ts$)  & 108.3  & 96.6  & 11.6  & & $\BB$ ($\Bs$)  & 194.0 & 181.1 & 25.1 \\
$\TT$ ($\TBd$) & 119.1  & 111.3 & 15.2  & & $\BB$ ($\TBd$) & 247.5 & 235.8 & 39.9 \\
$\TT$ ($\TBD$/$\XTd$) & 218.8  & 200.0 & 33.6  & & $\BB$ ($\TBD$) & 1030.4 & 977.8 & 177.0 \\
$\XX$ ($\XTd$) & 1037.9 & 988.9 & 187.1 & & $\BB$ ($\BYd$) & 18.0 & 16.9  & 1.1 \\
               &        &       &       & & $\YY$ ($\BYd$) & 0.4   & 0.3   & 0.0 \\
\hline
$t \bar t nj$  & 350    & 41    & 3     & & $WZnj$         & 62    & 16    & 0 \\
$W t \bar tnj$ & 67     & 37    & 3     & & $ZZnj$         & 7     & 1     & 0 \\
$Z t \bar tnj$ & 20     & 15    & 5   \\
\end{tabular}
\end{center}
\caption{Number of events at the pre-selection, selection and reconstruction levels in the $\ell^\pm \ell^\pm \ell^\mp$ (no $Z$) final state for the signals and main backgrounds with a luminosity of 30 fb$^{-1}$.}
\label{tab:nsnb-3Q1-noZ}
\end{table}
The $5\sigma$ discovery luminosities in this final state for the six models, including the contribution of both members in the case of the doublets, are given in Table~\ref{tab:sig-3Q1-noZ}. 
The luminosity indicated for the $T$ singlet may be optimistic because the background normalisation uncertainty (unimportant for the other models which have much larger signals) may be relevant.

\begin{figure}[t]
\begin{center}
\epsfig{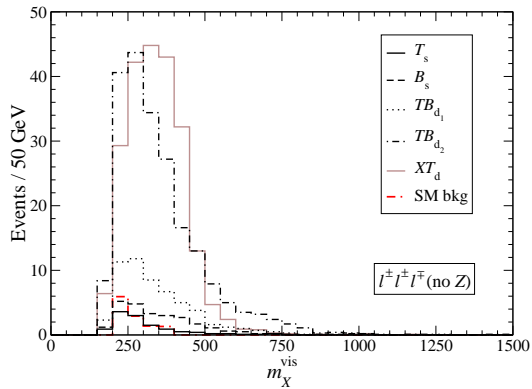}
\caption{Visible reconstructed mass $m_X^\text{vis}$ distribution of one of the heavy quarks (see the text). The luminosity is 30 \fbin.}
\label{fig:mrec-3Q1-noZ}
\end{center}
\end{figure}

In this final state the heavy quark masses cannot be directly reconstructed because each heavy quark has among its decay products an invisible neutrino, and there are three neutrinos in total. 
In $\XX$ production we can still have information about the mass of one of the heavy quarks, the one decaying $X \to W t \to WWb$, with $WW \to \ell \nu q \bar q'$, by reconstructing its decay products except the missing neutrino.\footnote{Notice that the transverse mass is not useful in this case because there are two more neutrinos from the other heavy quark decay.} For the reconstruction we demand the presence of at least one $b$-tagged jet and two light jets with $p_T > 20$ GeV, and select the hadronic $X$ decay products as follows:
\begin{table}[t]
\begin{center}
\begin{tabular}{ccccccc}
       & $L$      & Rec. & \quad &        & $L$       & Rec. \\[1mm]
$\Ts$  & 11 \fbin  & no   &       & $\TBD$ & 0.25 \fbin & no \\
$\Bs$  & 3.5 \fbin & no   &       & $\XTd$ & 0.25 \fbin & no \\
$\TBd$ & 1.1 \fbin & no   &       & $\BYd$ & --        & no
\end{tabular}
\end{center}
\caption{Luminosity $L$ required to have a $5\sigma$ discovery in the $\ell^\pm \ell^\pm \ell^\mp$ (no $Z$) final state. A dash indicates no signal or a luminosity larger than 100 \fbin.
We also indicate whether a mass peak can be reconstructed in this final state.}
\label{tab:sig-3Q1-noZ}
\end{table}
\begin{enumerate}
\item We select a $b$ jet among the ones present.
\item We select a pair of light jets $j_1$, $j_2$ among the three ones with highest $p_T$ (in case there are only two, we select these.
\item We choose the combination minimising the quantity
\begin{small}
\begin{equation}
\frac{(m_{j_1 j_2}-M_W)^2}{\sigma_W^2} + 
\frac{(m_{j_1 j_2 b}-m_t)^2}{\sigma_t^2} \,,
\end{equation}
\end{small}%
with $\sigma_W = 10$ GeV, $\sigma_t = 14$ GeV.
\end{enumerate}
The ``visible'' component of the heavy quark mass $m_X^\text{vis}$ is then reconstructed as the invariant mass of these jets plus the opposite-sign lepton, of the three ones present. However,
in the decay $X \to W t \to WWb$, only half of the time the two jets and $b$ quark will correspond to the top decay. We then set a cut
\begin{equation}
140~\text{GeV} < m_{j_1 j_2 b} < 210~\text{GeV}
\label{ec:cut-3Q1-noZ}
\end{equation}
to ensure that the event topology is consistent with the decay chain assumed. The number of events after the additional reconstruction conditions, including the cut in Eq.~(\ref{ec:cut-3Q1-noZ}), can be found in Table~\ref{tab:nsnb-3Q1-noZ}.
The $m_X^\text{vis}$ distribution is shown in Fig.~\ref{fig:mrec-3Q1-noZ}. For $\XX$ production we observe an endpoint around 500 GeV, which is not present for the other signals nor the SM background. Hence, if a signal is observed, template methods may be used to measure $m_X$ in this decay. Notice that a similar procedure to reconstruct the mass in $\BB$ decays is more difficult due to combinatorics, because the two $W$ bosons from the heavy quark decay have opposite sign.
Finally, it is worth remarking that the size of the signal itself would already give a strong hint that $\XX$ or $\BB$ pairs are produced, as it is apparent from the comparison of the numbers of events in Table~\ref{tab:nsnb-3Q1-noZ}.

\subsection{Summary}

In this section it has been shown that the trilepton final state has very good sensitivity to $\TT$, $\BB$ and $\XX$ production. Pair production of $T$, $B$ and $X$ quarks
gives final states with three leptons with branching ratios not too small, and trilepton backgrounds can be significantly reduced with not very strong selection criteria, which keep most of the signals. The discovery potential found is similar to the dilepton final states, which have larger branching ratios but also larger backgrounds, but worse than in the single lepton one. However, the main interest of the trilepton final state is not heavy quark discovery but model discrimination, via the observation or not of several quark decays in this unique channel.
Signals in the trilepton final state involve either the leptonic decay of a $Z$ and a $W$ boson, or of three $W$ bosons. Hence, we have split the sample into two subsamples, one in which a $Z$ candidate can be found (labelled as `$Z$') and the other one in which no such candidates can be identified (labelled as `no $Z$')

The $\ell^\pm \ell^\pm \ell^\mp$ ($Z$) final state is very interesting because it is very sensitive to $\TT$ and $\BB$ production, in the decay channels of Eqs.~(\ref{ec:ch3Q1Z}). $T$ and $B$ singlets with a mass of 500 GeV can both be discovered with a luminosity of 3.4 \fbin. Discovery of a $\TB$ doublet requires 1.1 \fbin\ (0.73 \fbin) in scenario 1 (2), and a $\XT$ doublet 0.72 \fbin\, being the main contribution to the signal from the $T$ quark in the latter two cases. But this broad sensitivity brings an additional difficulty for the discrimination of signals: the three signal channels in Eqs.~(\ref{ec:ch3Q1Z}) are interesting and it is necessary to identify event by event to which one it corresponds. In this sense, this final state is also very adequate because the kinematics of the decay chain can be fully reconstructed  as there is only one light neutrino. With this purpose we have developed a likelihood analysis to discriminate among the 
three signal channels and then reconstruct the events accordingly.
We have shown that the contributions of the three cascade decays in Eqs.~(\ref{ec:ch3Q1Z}) can be disentangled, invariant mass peaks can be reconstructed, and the decays $T \to Zt$, $T \to W^+ b$, $B \to Zb$, $B \to W^-t$ can be established if sufficient luminosity is collected. This clean channel, in which the combinatorics is  moderate, is also a good candidate to determine the chiralities of $T$, $B$ quarks with the analysis of
angular distributions in the semileptonic decay of the top quarks produced in
$T \to Zt$, $B \to W^- t$. For example, a simple analysis presented shows that for 30 \fbin\ the differences in a forward-backward asymmetry between a $T$ quark singlet and a $\TB$ doublet in scenario 1 would amount to $2.4\sigma$. In $T \to Ht$ decays (seen for instance in the single lepton channel) the statistics is larger but the top quark polarisation is smaller and the reconstruction less clean than here.

The final state without $Z$ candidates is also very interesting because of its excellent sensitivity to $\XX$ and $\BB$ production, very similar to the one in the like-sign dilepton channel: 3.5 \fbin\ for $B$ singlets and 0.25 \fbin\ for both $\XT$ doublets and the $\TBD$ model, with the main contribution resulting from $\XX$ and $\BB$ production, respectively. Although the masses cannot be fully reconstructed in this final state, in $\XX$ production the $X$ mass can be still determined from the endpoint of an invariant mass distribution. The presence of a signal for $B$ quarks gives indirect evidence for the $B \to W^- t$ decay, which is absent for the $B$ quark in a $\BY$ doublet.

To conclude this section, it is worth mentioning that the trilepton final state is also very sensitive to heavy Dirac or Majorana neutrinos in singlet, doublet or triplet representations~\cite{AguilarSaavedra:2009ik}. Those signals can be distinguished from heavy quark production because in that case the heavy neutrino can be observed as a peak in the invariant mass distribution of two opposite-charge leptons plus missing energy, with an additional peak in the distribution of the remaining lepton plus two jets. Scalar triplets also give trilepton signals (see for example Ref.~\cite{delAguila:2008cj}) but the like-sign dilepton invariant mass distribution displays a very sharp peak in scalar triplet production, which is of course absent in the case of heavy quarks.
\section{Final state $\ell^\pm \ell^\pm$}
\label{sec:2lik}

\begin{figure}[htb]
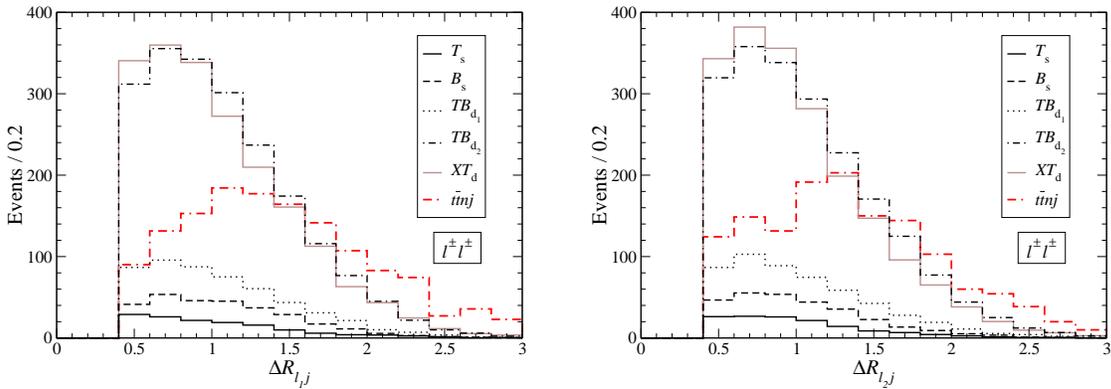

\begin{center}
\begin{tabular}{ccc}
\epsfig{file=Figs/dRl1j-2Q2.eps,height=5.1cm,clip=} & \quad &
\epsfig{file=Figs/dRl2j-2Q2.eps,height=5.1cm,clip=}
\end{tabular}
\caption{Lego-plot separation between the leptons and the closest jet for several signals and the $t \bar t nj$ background, at the pre-selection level (these variables are not used for event selection). The luminosity is 30 fb$^{-1}$.}
\label{fig:dist2-2Q2}
\end{center}
\end{figure}

This conspicuous signal can be produced in decays of $B$ and $X$ quark pairs when two same-sign $W$ bosons decay leptonically,
\begin{align}
& B \bar B \to W^- t \, W^+ \bar t \to W^- W^+ b \, W^+ W^- \bar b
&& \quad W^\pm \to \ell^\pm \nu , W^\mp \to q \bar q' \,, \nonumber \\
& X \bar X \to W^+ t \, W^- \bar t \to W^+ W^+ b \, W^- W^- \bar b 
&& \quad W^\pm \to \ell^\pm \nu , W^\mp \to q \bar q' \,,
\label{ec:ch2Q2} 
\end{align}
and also in decays of $T$ quark pairs involving $Z$ bosons 
\begin{align}
& T \bar T \to Zt \, W^- \bar b \to Z W^+b W^- \bar b
&& \quad Z \to \ell^+ \ell^- , WW \to \ell \nu q \bar q' 
\label{ec:ch2Q2b} 
\end{align}
when the opposite-charge lepton from the $Z$ decay is missed by the detector (the charge conjugate channel is also included). Like-sign dilepton signals are relatively clean, their largest SM background being $t \bar t nj$ in the semileptonic channel, where one of the two like-sign leptons results from a $b$ quark decay.
A very large source of $\ell^\pm \ell^\pm$ events but for low lepton transverse momenta is $b \bar b nj$, with a cross section of 1.4 $\mu$b (for a detailed discussion of like-sign dilepton backgrounds see Ref.~\cite{delAguila:2007em}). For example, requiring only $p_T > 15$ GeV for the charged leptons the number of like-sign dilepton events from $t \bar t nj$, $b \bar b nj$ is around 25000 and 150000, respectively, for a luminosity of 30 \fbin~\cite{delAguila:2007em}.
In order to reduce such backgrounds, we demand for event pre-selection (i) the presence of two like-sign leptons with transverse momentum $p_T > 30$ GeV; (ii) the absence of non-isolated muons. The first condition practically eliminates $b \bar b nj$ while the latter reduces $WZnj$, which gives this final state when the opposite-charge lepton from $Z$ decay is missed by the detector. The number of signal and background events at pre-selection can be read in Table~\ref{tab:nsnb-2Q2}. 
\begin{table}[htb]
\begin{center}
\begin{tabular}{cccccccccccc}
               & Pre.   & Sel.   & Rec.  & \quad &          & Pre.  & Sel.  & Rec. \\[1mm]
$\TT$ ($\Ts$)  & 139.6  & 79.3   & 65.9  & & $\BB$ ($\Bs$)  & 291.5 & 170.1 & 137.9 \\
$\TT$ ($\TBd$) & 156.3  & 90.4   & 74.8  & & $\BB$ ($\TBd$) & 368.6 & 223.2 & 172.8 \\
$\TT$ ($\TBD$/$\XTd$) & 263.8  & 177.8  & 149.1 & & $\BB$ ($\TBD$) & 1737.4 & 1122.8 & 890.0 \\
$\XX$ ($\XTd$) & 1684.7 & 1138.6 & 900.4 & & $\BB$ ($\BYd$) & 15.8  & 5.6   & 45.0 \\
               &        &        &       & & $\YY$ ($\BYd$) & 2.0   & 0.7   & 0.2 \\
\hline
$t \bar t nj$  & 1413   & 43     & 23    & & $WWnj$         & 245   & 7     & 0  \\
$W t \bar tnj$ & 184    & 47     & 34    & & $WZnj$         & 1056  & 9     & 1  \\
$Z t \bar tnj$ & 28     & 9      & 5     & & $WWWnj$        & 110   & 11    & 3 \\
\end{tabular}
\end{center}
\caption{Number of events at the pre-selection, selection and reconstruction levels in the $\ell^\pm \ell^\pm$ final state for the signals and main backgrounds with a luminosity of 30 fb$^{-1}$.}
\label{tab:nsnb-2Q2}
\end{table}
We note that it is sometimes claimed in the literature (without actually providing a proof) that SM backgrounds with charged leptons from $b$ decays, namely $t \bar t nj$ and $b \bar b nj$, can be removed or suppressed to negligible levels by isolation criteria.
However, recent analyses for supersymmetry searches performed with a full detector simulation~\cite{Aad:2009wy} arrive at the opposite conclusion. This can already be seen at the level of a fast detector simulation.
We show in Fig.~\ref{fig:dist2-2Q2} the minimum $\Delta R$ distance between the two charged leptons and the closest jet, for various signals and the $t \bar t nj$ background. We observe that these variables, which in general do not bring a tremendous improvement in the signal to background ratio, are even inadequate in this case when the signals have many hard jets.

For the like-sign dilepton final state we first perform a ``discovery'' analysis with conditions aiming only to improve the signal significance by reducing the background. Then, we impose additional requirements (which reduce the signal statistical significance) to try to reconstruct the event kinematics and detect heavy quark mass peaks. These two analyses are presented in turn.

\subsection{Discovery potential}

\begin{figure}[t]
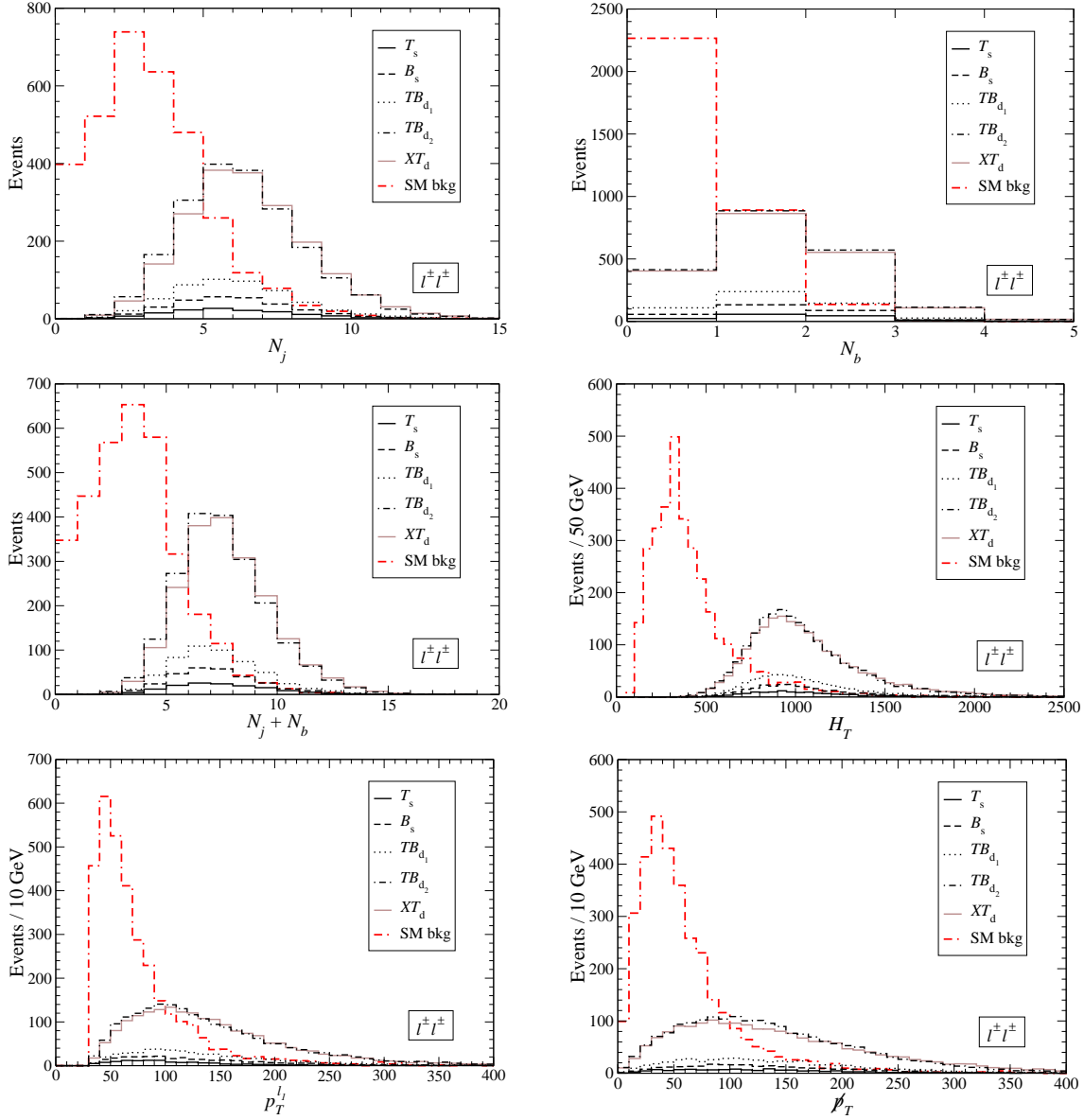

\begin{center}
\begin{tabular}{ccc}
\epsfig{file=Figs/lmult-2Q2.eps,height=5.1cm,clip=} & \quad &
\epsfig{file=Figs/bmult-2Q2.eps,height=5.1cm,clip=} \\
\epsfig{file=Figs/mult-2Q2.eps,height=5.1cm,clip=} & \quad &
\epsfig{file=Figs/HT-2Q2.eps,height=5.1cm,clip=} \\
\epsfig{file=Figs/ptlep1-2Q2.eps,height=5.1cm,clip=}  & \quad &
\epsfig{file=Figs/ptmiss-2Q2.eps,height=5.1cm,clip=}
\end{tabular}
\caption{Light and $b$-tagged jet multiplicity, total jet multiplicity, total transverse energy, transverse momentum of the leading lepton and missing energy. The luminosity is 30 fb$^{-1}$.}
\label{fig:dist1-2Q2}
\end{center}
\end{figure}

To evaluate the discovery potential for heavy quark signals we require for event selection
(i) the presence of at least six jets, $b$ tagged or not, with $p_T > 20$ GeV; 
(ii) transverse momentum $p_T > 50$ GeV for the leading charged lepton $\ell_1$;
(iii) missing energy $\ptmiss > 50$ GeV;
(iv) transverse energy larger than 500 GeV.
The kinematical distributions of these variables at pre-selection are presented in Fig.~\ref{fig:dist1-2Q2}. We also show for completeness the separate multiplicity distributions of light and $b$-tagged jets. Notice that the maximum in the transverse energy distribution for the signals indicates that one or more heavy particles with a total mass around 1 TeV is produced.
The number of signal and background events after our selection cuts is listed in Table~\ref{tab:nsnb-2Q2}. We observe that $t \bar t nj$ still amounts to one third of the total like-sign dilepton background after being reduced by the cuts.
The corresponding discovery luminosities for each model can be found in Table~\ref{tab:sig-2Q2},  also indicating whether a mass peak can be reconstructed (see the next subsection). It is noticeable that not only $\XX$ and $\BB$ decays give like-sign dileptons but also $\TT$ decays, although the discovery luminosity in the $T$ singlet model is much larger than for the rest. Finally, we note that for the $B$ singlets and , $\TB$, $\XT$ doublets the signals are much larger than the background and thus the uncertainty in the latter is not crucial for the evaluation of the discovery potential.

\begin{table}[ht]
\begin{center}
\begin{tabular}{ccccccc}
       & $L$       & Rec. & \quad &        & $L$       & Rec. \\[1mm]
$\Ts$  & 17 \fbin  & no   &       & $\TBD$ & 0.23 \fbin & no \\
$\Bs$  & 4.1 \fbin & no   &       & $\XTd$ & 0.23 \fbin & $m_X$ \\
$\TBd$ & 1.5 \fbin & no   &       & $\BYd$ & --        & no
\end{tabular}
\end{center}
\caption{Luminosity $L$ required to have a $5\sigma$ discovery in the $\ell^\pm \ell^\pm$ final state. A dash indicates no signal or a luminosity larger than 100 \fbin.
We also indicate whether a mass peak can be reconstructed in this final state.}
\label{tab:sig-2Q2}
\end{table}

\subsection{Heavy quark reconstruction}
\label{sec:6.2}

\begin{figure}[t]
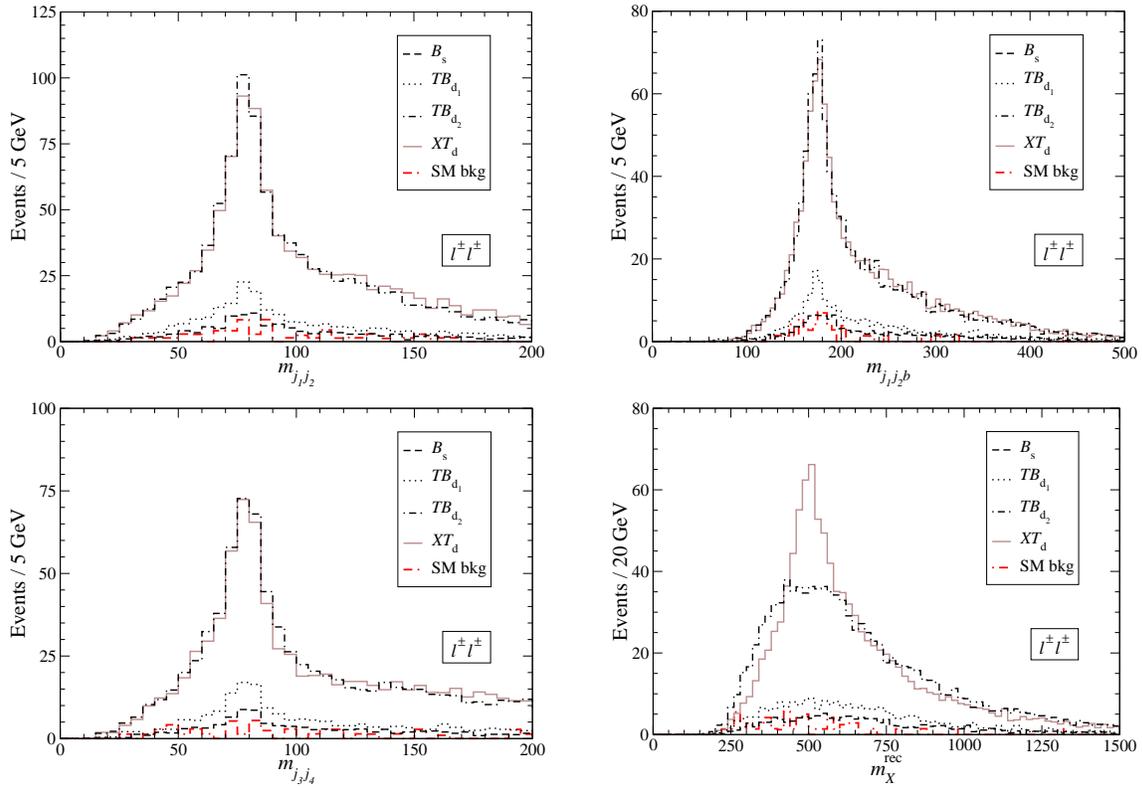

\begin{center}
\begin{tabular}{ccc}
\epsfig{file=Figs/mw1-2Q2.eps,height=5.1cm,clip=} & \quad &
\epsfig{file=Figs/mt-2Q2.eps,height=5.1cm,clip=} \\
\epsfig{file=Figs/mw2-2Q2.eps,height=5.1cm,clip=} & \quad &
\epsfig{file=Figs/mQ-2Q2.eps,height=5.1cm,clip=}
\end{tabular}
\caption{Reconstructed masses of the two $W$ bosons, the top and the heavy quark. The luminosity is 30 \fbin.}
\label{fig:mrec-2Q2}
\end{center}
\end{figure}

In $\XX$ production the invariant mass of the quark decaying hadronically can be reconstructed from its decay products: a $b$ quark and four jets from $W$ decays. In order to do so, we restrict ourselves to events with at least one $b$-tagged jet and four light (non-tagged) jets.
The number of signal and background events after these additional reconstruction criteria is given in Table~\ref{tab:nsnb-2Q2}.
The reconstruction is performed as follows:
\begin{enumerate}
\item A $b$-tagged jet is selected among the ones present.
\item The four highest $p_T$ light jets are grouped in two pairs $j_1 j_2$, $j_3 j_4$ trying to reconstruct two $W$ bosons, the first one from the top quark decay and the second one from $X \to Wt$.
\item The $b$ jet is associated to the first light jet pair $j_1 j_2$ to reconstruct a top quark.
\item Among all the possible choices for the $b$ jet and light jet combinations, the one minimising the quantity
\begin{equation}
\frac{(m_{j_1 j_2}-M_W)^2}{\sigma_W^2} + 
\frac{(m_{j_3 j_4}-M_W)^2}{\sigma_W^2} + 
\frac{(m_{j_1 j_2 b}-m_t)^2}{\sigma_t^2}
\end{equation}
is chosen, taking $\sigma_W = 10$ GeV, $\sigma_t = 14$ GeV.
\end{enumerate}
The reconstructed heavy quark mass $m_X$ is then defined as the invariant mass of the $b$-tagged and four light jets. These distributions are presented in Fig.~\ref{fig:mrec-2Q2}. The long tails in the distributions of $W$ and top reconstructed masses are mainly caused by wrong assignments. In particular, when only one $b$-tagged jet is present in the event, half of the times it corresponds to the $b$ quark from the other heavy quark $X$ in which the $W$ bosons decay leptonically. Still, the heavy quark mass peak is clearly observed without the need of quality cuts on $W$ and top reconstructed masses (which of course sharpen the $m_X$ peak). The rest of signals and the SM background do not exhibit any resonant structure, which shows that the above procedure does not introduce any bias. Note that a similar mass reconstruction cannot be achieved for $\BB$ or $\TT$ decays, due to the missing neutrino from each of the heavy quark decays.

\begin{figure}[t]
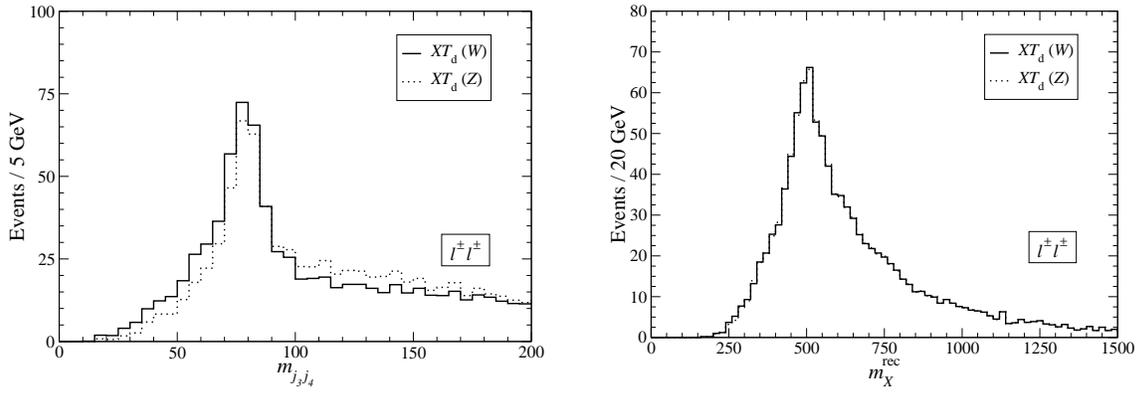

\begin{center}
\begin{tabular}{ccc}
\epsfig{file=Figs/mw2alt-2Q2.eps,height=5.1cm,clip=} & \quad &
\epsfig{file=Figs/mQalt-2Q2.eps,height=5.1cm,clip=}
\end{tabular}
\caption{Comparison between the reconstructed masses of the extra boson and the heavy quark, using an alternative procedure (see the text). The luminosity is 30 \fbin.}
\label{fig:mrec2-2Q2}
\end{center}
\end{figure}

Although the cross section and mass reconstruction are consistent with the production of a $\XX$ pair we still can ask ourselves to which extent we can conclude that $\XX$ pairs are produced and not other possibility consistent with charge conservation and giving the same final state. We can establish $\XX$ production in two steps. First,
we ensure that the extra boson from the heavy quark decaying hadronically, reconstructed from two jets $j_3$ and $j_4$, is a $W$ boson and not a $Z$. It is not surprising that the $j_3 j_4$ invariant mass distribution in Fig.~\ref{fig:mrec-2Q2} has a peak around $M_W$, since this is imposed in the reconstruction procedure. The question is then what would happen if, instead of choosing the pair of jets which best reconstruct a $W$ boson, we chose the pair which best reconstruct a $Z$. The comparison between both situations is shown in Fig.~\ref{fig:mrec2-2Q2}, including the two signals in the $\XT$ doublet ($\XX$ and $\TT$). We observe that if we select the pair of jets with $m_{j_3 j_4}$ closest to $M_Z$ the distribution is slightly shifted but the peak is maintained at $M_W$, and the heavy quark reconstruction is unaffected. Then, up to a more detailed study with a full detector simulation to confirm these results, it seems that the identity of the gauge boson from the heavy quark decay can be established. This leaves us with two options for the heavy quark: a $B$ (charge $-1/3$) or $X$ (charge $5/3$) quark. (The opposite charges for antiquarks are understood.)

\begin{figure}[htb]
\begin{center}
\epsfig{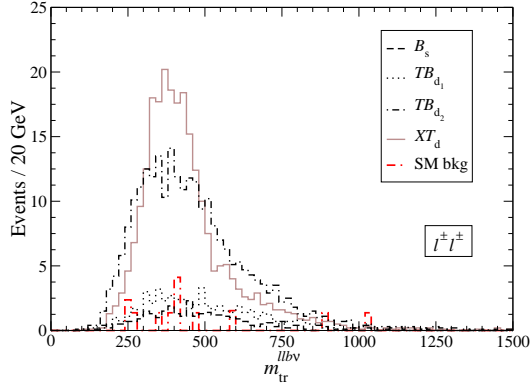}
\caption{Transverse mass distribution of the two charged leptons, a $b$ jet and the missing energy. The luminosity is 30 \fbin.}
\label{fig:mrec3-2Q2}
\end{center}
\end{figure}

Then, we examine the transverse mass distribution of the rest of particles in the event, restricting ourselves to events with two $b$ tags for simplicity and imposing the quality cuts
\begin{align}
40~\text{GeV} < m_{j_1 j_2} < 120~\text{GeV} \,, \notag \\
125~\text{GeV} < m_{j_1 j_2 b} < 225~\text{GeV}
\end{align}
on the reconstruction (see Fig.~\ref{fig:mrec-2Q2}).
We define the transverse mass as in Ref.~\cite{Contino:2008hi},
\begin{equation}
m_\text{tr}^2 = (E_T^{\ell \ell b} + \ptmiss)^2 - (p_T^{\ell \ell b \nu})^2 \,,
\end{equation}
with $(E_T^{\ell \ell b})^2 = (p_T^{\ell \ell b})^2 + m_{\ell \ell b}^2$, $p_T^\nu \equiv \ptmiss$ and the transverse momenta of different particles summed vectorially. This distribution, shown in Fig.~\ref{fig:mrec3-2Q2} for the relevant signals, has an edge around $m_X$ for the $\XT$ doublet signal, showing that the like-sign charged leptons and the $b$ quark result from the decay of a 500 GeV resonance. Then, charge conservation and the absence of significant additional jet activity (which could be identified as additional partons produced in the hard process) implies the possible charge assignments $(Q_h,Q_l) = \pm (5/3,-5/3),\pm (5/3,-7/3), \pm (1/3,-5/3)$ for the heavy quark decaying 
hadronically and leptonically, respectively. Of these three possibilities, the only one consistent with a small mixing of the third SM generation and the new quarks with the first two SM generations (so that $b$ quarks are produced in $b \bar b$ pairs) is the first one, corresponding to $\XX$ production.

\subsection{Summary}

We have shown in this section that the like-sign dilepton final state has an excellent discovery potential for $\XX$ production: a $(X \, T)$ doublet of 500 GeV could be discovered with only 0.23 \fbin. This discovery potential is only matched by the trilepton and single lepton final states (but in the latter the quark observed is the $T$ partner).
Moreover, a heavy quark mass peak can be found in the invariant mass distribution of a reconstructed top quark and two extra jets, resulting from the hadronic decay of a $W$ boson. Despite the fact that heavy quark charges cannot be directly measured (unless the $b$ jet charge is measured, which is very difficult at LHC), the detailed analysis of the event kinematics can eventually establish that the signal corresponds to $\XX$ production if this indeed is the case.

For $\BB$ production the signals are also interesting and the discovery potential is also very good:  4.1 \fbin\ for $B$ singlets and 1.1 \fbin, 0.23 \fbin\ for $(T \, B)$ doublets in scenarios 1 and 2, respectively. For $\BB$ production the heavy quark mass peaks cannot be reconstructed because each heavy quark has among its decay products an invisible neutrino. But the presence of a signal distinguishes a $B$ singlet or $\TB$ doublet, which have decays $B \to W^- t$, from the $B$ quark in a $\BY$ doublet which does not.
This final state has some sensitivity to $\TT$ production, with one heavy quark decaying $T \to Zt$, $Z \to \ell^+ \ell^-$ and one of these leptons missed by the detector. The discovery potential is worse than in other final states, however, and the heavy quark masses cannot be reconstructed.

We note that
our results exhibit some differences with respect to previous work~\cite{Contino:2008hi}, due to two different sources:
\begin{itemize}
\item We include a fast detector simulation and pile-up, and in order to reduce SM backgrounds we must apply tighter event selection criteria. For example, the event selection in Ref.~\cite{Contino:2008hi} demands five jets with pseudo-rapidity $|\eta| < 5$. This is not sensible in the presence of pile-up, so we restrict our analysis to the central region of the calorimeter, $|\eta| < 2.5$. In fact, we have applied the cuts in Ref.~\cite{Contino:2008hi} to our simulation, obtaining a similar signal efficiency but a background 5 times larger (see Table~\ref{tab:comp-contino}). In particular $t \bar t nj$, not considered there, amounts to 52 events, about one half of the total SM contribution.
Therefore, reducing the background requires stronger selection criteria which obviously reduce the signal as well.
\item As already indicated in section \ref{sec:3}, our $5\sigma$ discovery criterion is that (i) the signal statistical significance (possibly evaluated with Poisson statistics) is larger than $5\sigma$; (ii) the number of events is larger than 10. This second condition, which determines the limits on the $\XX$ and $\BB$ signals for this final state, is not included in Ref.~\cite{Contino:2008hi}. Then, even for equal number of signal and background events, their limits are better.
\end{itemize}
Our mass reconstruction method is also more involved and adapted to the more realistic conditions and the higher jet multiplicities found in our analysis.

\begin{table}[t]
\begin{center}
\begin{tabular}{ccc}
& Ref.~\cite{Contino:2008hi} & Our analysis \\[1mm]
$\XX$ ($\XTd$) & 440  & 470.6 \\
$\BB$ ($\TBD$) & 424  & 470.7 \\
Background          & 23   & 116  
\end{tabular}
\end{center}
\caption{Number of signal and background events for 10 \fbin\ using the selection criteria in Ref.~~\cite{Contino:2008hi}.}
\label{tab:comp-contino}
\end{table}

We finally comment on some other models giving the same final state. Like-sign dilepton signals without significant missing energy are characteristic of the presence of a heavy Majorana neutrinos~\cite{Keung:1983uu}, either in singlet or triplet $\text{SU}(2)_L$ representations (for a detailed comparison see Ref.~\cite{AguilarSaavedra:2009ik}). Those models can be easily distinguished from heavy quark production because in the heavy neutrino case (i) the missing energy is very small; (ii) two heavy resonances can be reconstructed, each one consisting of a charged lepton and two jets. On the other hand, like-sign dileptons with large missing energy are characteristic of heavy Dirac neutrinos in triplet $\text{SU}(2)_L$ representations~\cite{AguilarSaavedra:2009ik}, but in this case a resonance can be reconstructed with one charged lepton and two jets. Scalar triplet production also gives this final state but with the like-sign dilepton invariant mass displaying a sharp peak at the doubly charged scalar mass (see for example Ref.~\cite{delAguila:2008cj}).

\section{Final state $\ell^+ \ell^-$}
\label{sec:2opp}

\begin{figure}[htb]
\begin{center}
\epsfig{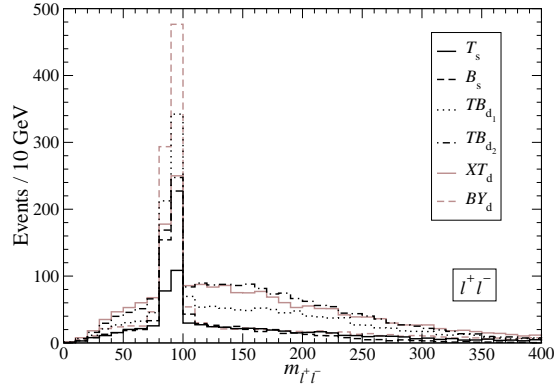}
\caption{$\ell^+ \ell^-$ invariant mass distributions for the six models in the $\ell^+ \ell^-$ final state. The luminosity is 30 fb$^{-1}$.}
\label{fig:mZrec-2Q0}
\end{center}
\end{figure}

This final state has large SM backgrounds which make it more difficult to observe positive signals by simply counting events with few selection criteria, as it is possible in 
the cleaner final states, and demand either a signal reconstruction to observe invariant mass peaks or an efficient background reduction. We ask for pre-selection the presence of (i) two opposite-charged leptons with $p_T > 30$ GeV; (ii) two $b$-tagged jets with $p_T > 20$ GeV. Dilepton signals result from many signal decay channels, for example
\begin{align}
& T \bar T \to Zt \, W^- \bar b \to Z W^+b W^- \bar b
&& \quad Z \to \ell^+ \ell^- , W \to q \bar q' \,, \nonumber \\
& T \bar T \to Zt \, \HZ \bar t \to Z W^+b \, \HZ W^- \bar b
&& \quad Z \to \ell^+ \ell^- , W \to q \bar q' , \HZ \to q \bar q/\nu \bar \nu \,, \nonumber \\
& B \bar B \to Z b \, W^+ \bar t \to Z b \, W^+ W^- \bar b
&& \quad Z \to \ell^+ \ell^- , W \to q \bar q' \,, 
\label{ec:ch2Q0Z}
\end{align}
involving a $Z \to \ell^+ \ell^-$ decay (here $\HZ=Z,H$), or
\begin{align}
& T \bar T \to W^+ b \, W^- \bar b 
&& \quad W \to \ell \nu\,, \nonumber \\
& T \bar T \to W^+ b \, \HZ \bar t \to W^+ b \, \HZ W^- \bar b 
&& \quad W \to \ell \nu , \HZ \to q \bar q/\nu \bar \nu\,, \nonumber \\
& B \bar B \to W^- t \, W^+ \bar t \to W^- W^+ b \, W^+ W^- \bar b
&& \quad 2W \to \ell \nu , 2W \to q \bar q' \,, \nonumber \\
& X \bar X \to W^+ t \, W^- \bar t \to W^+ W^+ b \, W^- W^- \bar b 
&& \quad 2W \to \ell \nu , 2W \to q \bar q' \,, \nonumber \\
& Y \bar Y \to W^- b \, W^+ \bar b 
&& \quad W \to \ell \nu\,, 
\label{ec:ch2Q0noZ}
\end{align}
with two leptonic $W$ decays. 
As it has been done in other final states, we separate the sample into two ones, one for events with a $Z$ candidate (when the two charged leptons have the same flavour and invariant mass
$|m_{\ell^+ \ell^-} - M_Z| < 15$ GeV), and the other one with the rest of events, which do not fulfill one of these conditions. Backgrounds are also separated by this division: the ones involving $Z$ production like $Znj$ and $Z b \bar bnj$ mainly contribute to the former while $t \bar t nj$ contributes to the latter. 
The total number of signal and background events at pre-selection level in both samples is given in Table~\ref{tab:nsnb-2Q0}, and the dilepton mass distribution for the signals in Fig.~\ref{fig:mZrec-2Q0}.

\begin{table}[htb]
\begin{center}
\begin{tabular}{cccccccccccc}
                 & Total  & $Z$   & no $Z$ & \quad &          & Total & $Z$   & no $Z$ \\[1mm]
$\TT$ ($\Ts$)    & 715.9  & 179.6 & 536.3  & & $\BB$ ($\Bs$)  & 819.8 & 393.4 & 426.4  \\
$\TT$ ($\TBd$)   & 799.4  & 174.1 & 625.3  & & $\BB$ ($\TBd$) & 907.5 & 388.1 & 519.4 \\
$\TT$ ($\TBD$/$\XTd$)   & 1007.7 & 341.6 & 666.1  & & $\BB$ ($\TBD$) & 1105.4 & 55.4 & 1050.0 \\
$\XX$ ($\XTd$)   & 1147.4 & 60.7  & 1086.7 & & $\BB$ ($\BYd$) & 902.5 & 780.3 & 122.2 \\
                 &        &       &        & & $\YY$ ($\BYd$) & 570.4 & 25.8  & 544.6 \\
\hline
$t \bar t nj$    & 68493  & 7464  & 61029  & & $Z^*/\gamma^*nj$ & 5245   & 4875  & 370 \\
$tW$             & 2135   & 212   & 1923   & & $Zb\bar b nj$  & 10132 & 9807  & 325 \\
$t\bar tb\bar b$ & 347    & 38    & 309    & & $Zc \bar c nj$ & 931   & 883   & 48 \\
$Wt \bar t nj$   & 63     & 4     & 59     & & $Zt \bar t nj$ & 106   & 88    & 18 
\end{tabular}
\end{center}
\caption{Number of events in the $\ell^+ \ell^-$ final state for
the signals and main backgrounds with a luminosity of 30 fb$^{-1}$, at pre-selection level.}
\label{tab:nsnb-2Q0}
\end{table}

\subsection{Final state $\ell^+ \ell^-$ ($Z$)}

In the sample with $|m_{\ell^+ \ell^-}-M_Z| < 15$ GeV we first perform a generic analysis sensitive to $T$ and $B$ quarks, to obtain the discovery potential in this sample. Then, we perform a specific one aiming to detect the decay $B \to Hb$ (and thus the Higgs boson) in $\BB \to ZbH \bar b$ decays. This final state is interesting for Higgs boson discovery in models with doublets $(B \, Y)$ where the only decays of the $B$ quark are $B \to Zb$, $B \to Hb$ and the decay $B \to Hb$ cannot be reconstructed in the single lepton channel.

\subsubsection{Discovery potential}

\begin{figure}[htb]
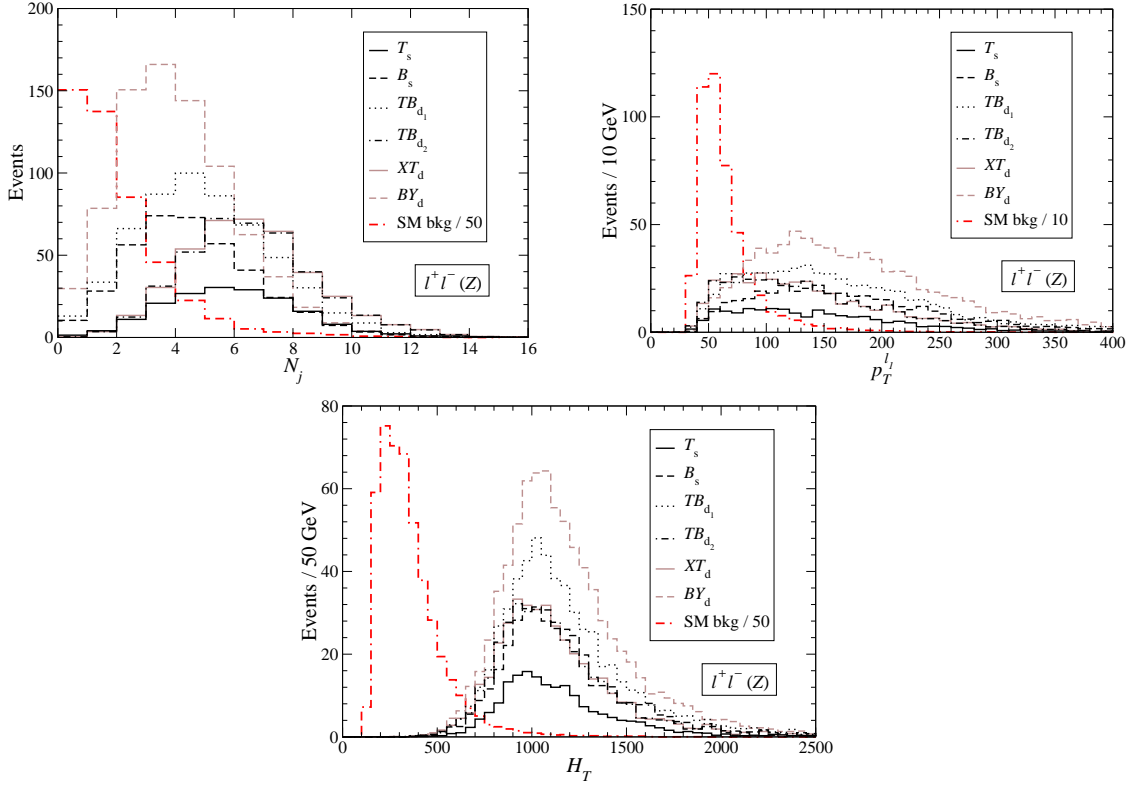

\begin{center}
\begin{tabular}{ccc}
\epsfig{file=Figs/mult-2Q0-Z.eps,height=5.1cm,clip=} & \quad &
\epsfig{file=Figs/ptlep1-2Q0-Z.eps,height=5.1cm,clip=} \\
\multicolumn{3}{c}{\epsfig{file=Figs/HT-2Q0-Z.eps,height=5.1cm,clip=}}
\end{tabular}
\caption{Kinematical distributions of variables used in selection criteria
for the $\ell^+ \ell^-$ ($Z$) final state: light jet multiplicity, transverse momentum of the leading lepton and total transverse energy. The luminosity is 30 fb$^{-1}$.}
\label{fig:dist-2Q0-Z}
\end{center}
\end{figure}

Here we demand for event selection (i) at least four jets with $p_T > 20$ GeV; (ii) transverse momentum $p_T > 50$ GeV for the leading charged lepton $\ell_1$; (iii) transverse energy $H_T > 500$ GeV. The kinematical distributions of the three variables are presented in Fig.~\ref{fig:dist-2Q0-Z}.
Most signal channels, in particular those in Eqs.~(\ref{ec:ch2Q0Z}), have at least four jets at the partonic level.
One exception is $\BB \to Zb \HZ b$ with $\HZ \to q \bar q/\nu \bar \nu$, in which additional jets are only produced by radiation or fragmentation. Still, this signal is sizeable after the multiplicity cut.
The reader may also notice that the background could be further reduced by requiring for example $H_T > 1$ TeV. However, it is not clear whether this would indeed improve the signal observability.
If the signal cannot be seen as a clear peak (or bump) over the background lineshape, its observation requires a simple event counting in which, if the background is large as it is our case, the background normalisation uncertainty plays an important role. On the other hand, if the signal displays a peak the background can be in principle normalised from off-peak measurements and its uncertainty will be smaller. The selection cuts made here represent a (conservative) compromise between having a manageable background and still observe the signal peak structure.

\begin{figure}[t]
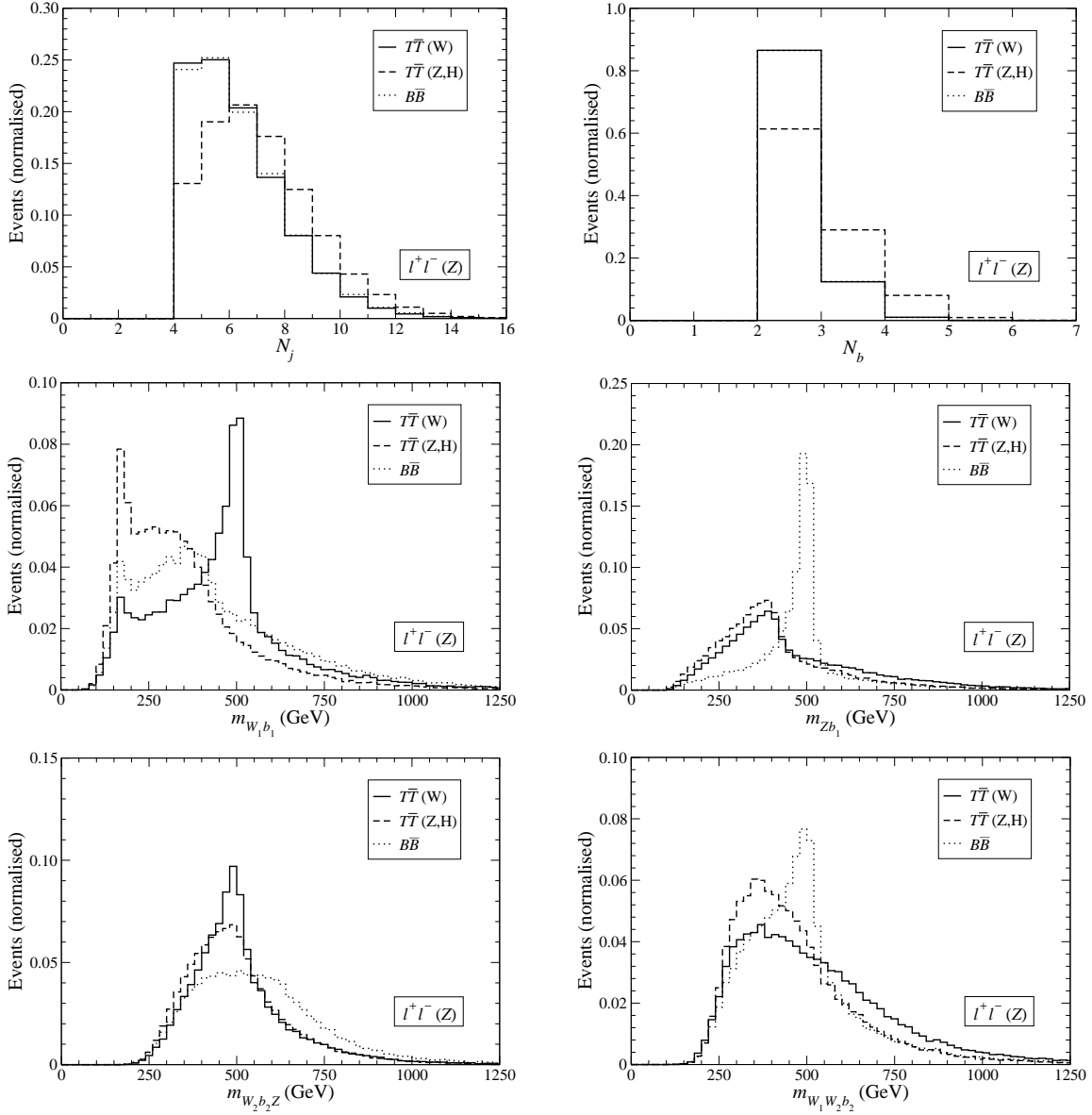

\begin{center}
\begin{tabular}{ccc}
\epsfig{file=Figs/D-mult-2Q0-Z.eps,height=5.1cm,clip=} & \quad &
\epsfig{file=Figs/D-bmult-2Q0-Z.eps,height=5.1cm,clip=} \\
\epsfig{file=Figs/D-mW1b1-2Q0-Z.eps,height=5.1cm,clip=} & \quad &
\epsfig{file=Figs/D-mZb1-2Q0-Z.eps,height=5.1cm,clip=} \\
\epsfig{file=Figs/D-mW2b2Z-2Q0-Z.eps,height=5.1cm,clip=} & \quad &
\epsfig{file=Figs/D-mW1W2b2-2Q0-Z.eps,height=5.1cm,clip=} \\
\end{tabular}
\caption{Kinematical variables used to classify the three heavy quark signals in the
$\ell^+ \ell^-$ ($Z$) final state.}
\label{fig:lik-2Q0-Z}
\end{center}
\end{figure}

As it was done for the trilepton channel, we build here a likelihood function to discriminate among the three signal channels in Eqs.~(\ref{ec:ch2Q0Z}), building probability functions for three signal classes: ($a$) $\TT \to ZtWb$; ($b$) $\TT \to Zt \HZ t$; ($c$) $\BB \to Zb Wt$. We generate high-statistics samples different from the ones used for the final analysis. We choose not to include a separate class for the background, because that would strongly bias it towards signal-like distributions and
jeopardise the observation of reconstructed peaks. At any rate, the
discriminant analysis implemented here rejects a large fraction of the background (which is classified as $\TT$-like) when we concentrate ourselves on the $\BB$ signal.
To build the discriminant variables we use an approximate reconstruction of the two $W$ bosons decaying hadronically, choosing among the light jets (up to a maximum of 6) the four ones which best reconstruct two $W$ bosons. Then, we use the same variables as in the trilepton channel but this time with two hadronic $W$ bosons $W_1$, $W_2$, ordered by transverse momenta as well as the $b$-tagged jets $b_1$, $b_2$. The resulting distributions are presented in Fig.~\ref{fig:lik-2Q0-Z}.
\begin{figure}[t]
\begin{center}
\begin{tabular}{ccc}
\epsfig{file=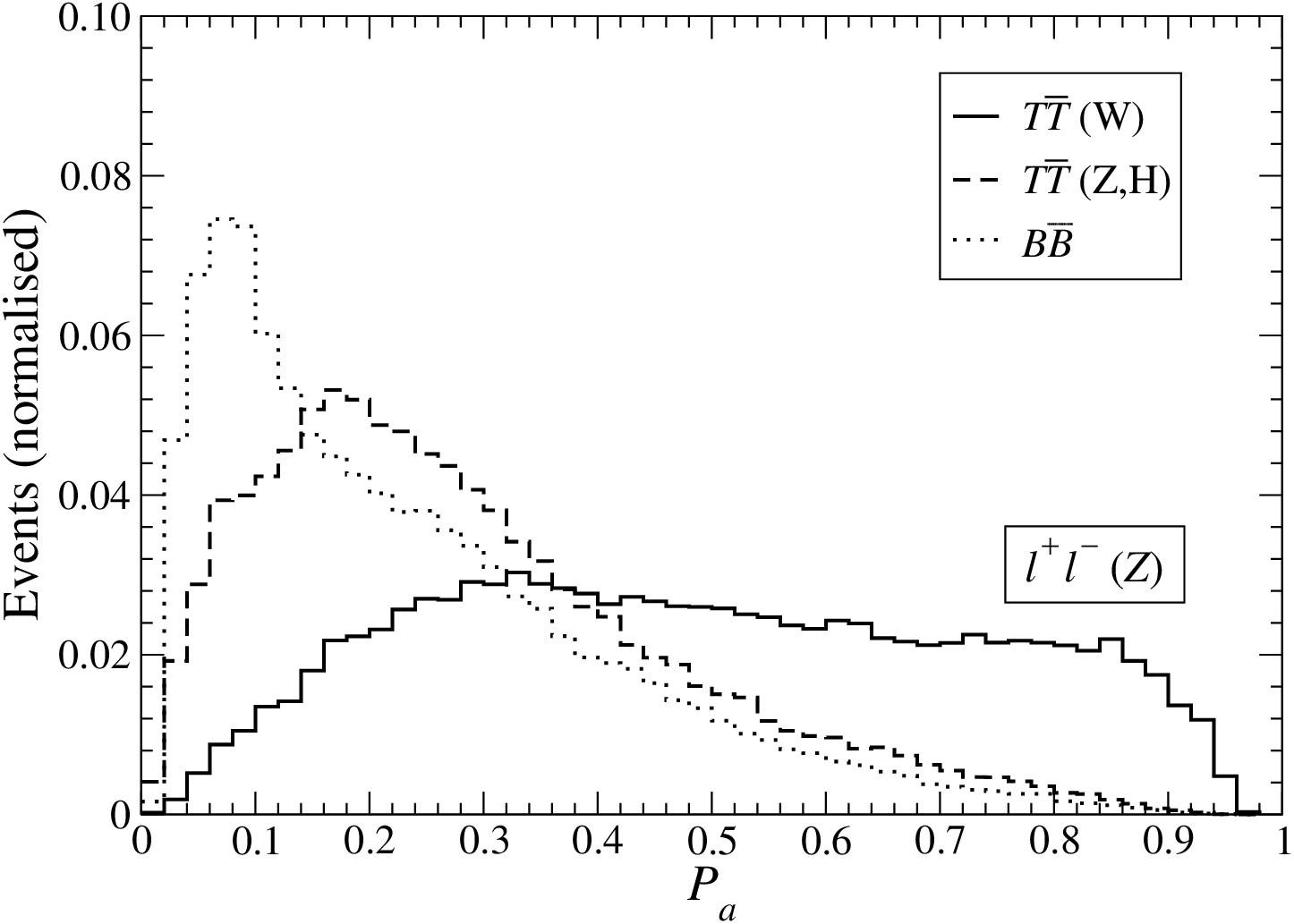,height=5.1cm,clip=} & \quad &
\epsfig{file=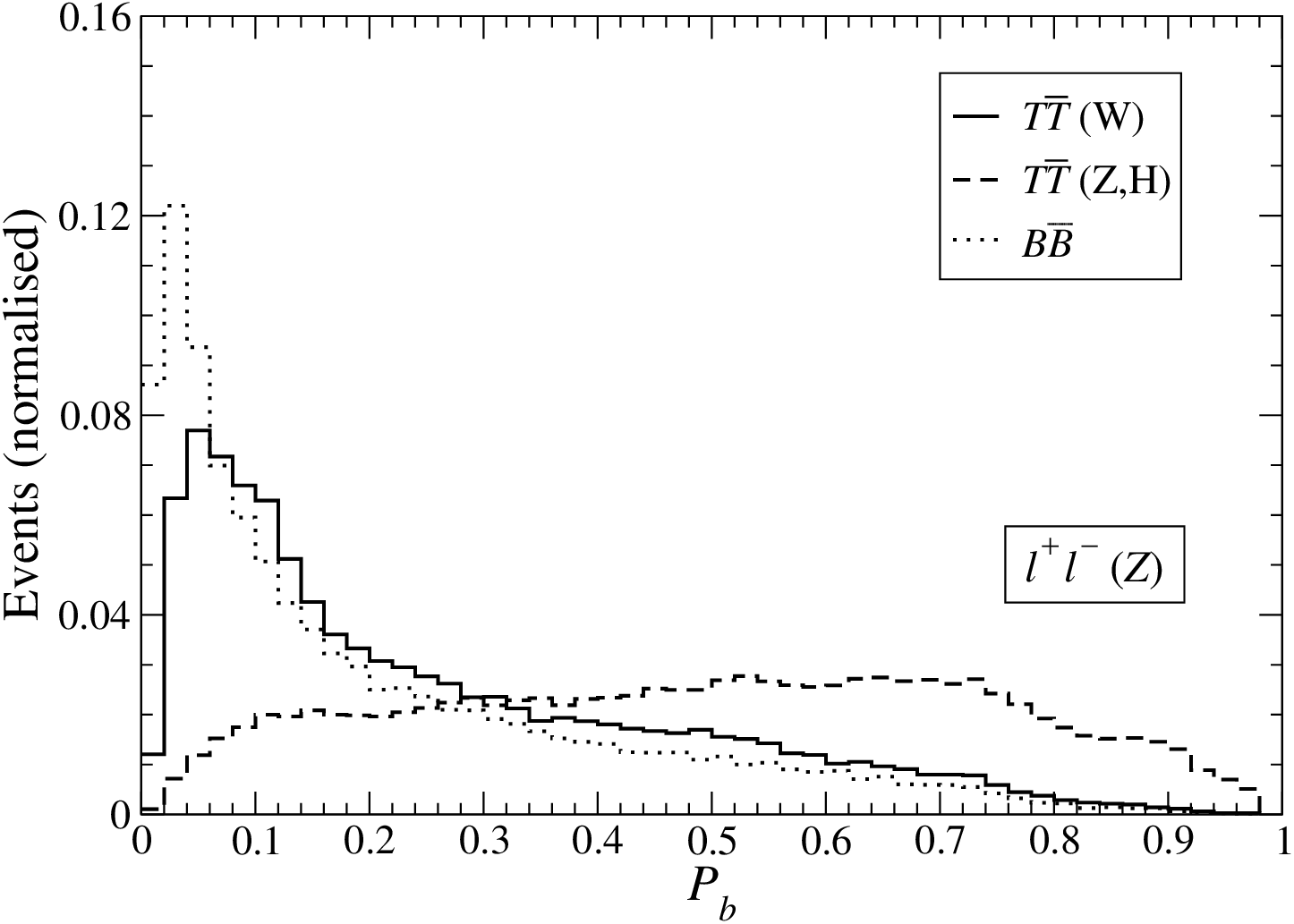,height=5.1cm,clip=} \\
\multicolumn{3}{c}{\epsfig{file=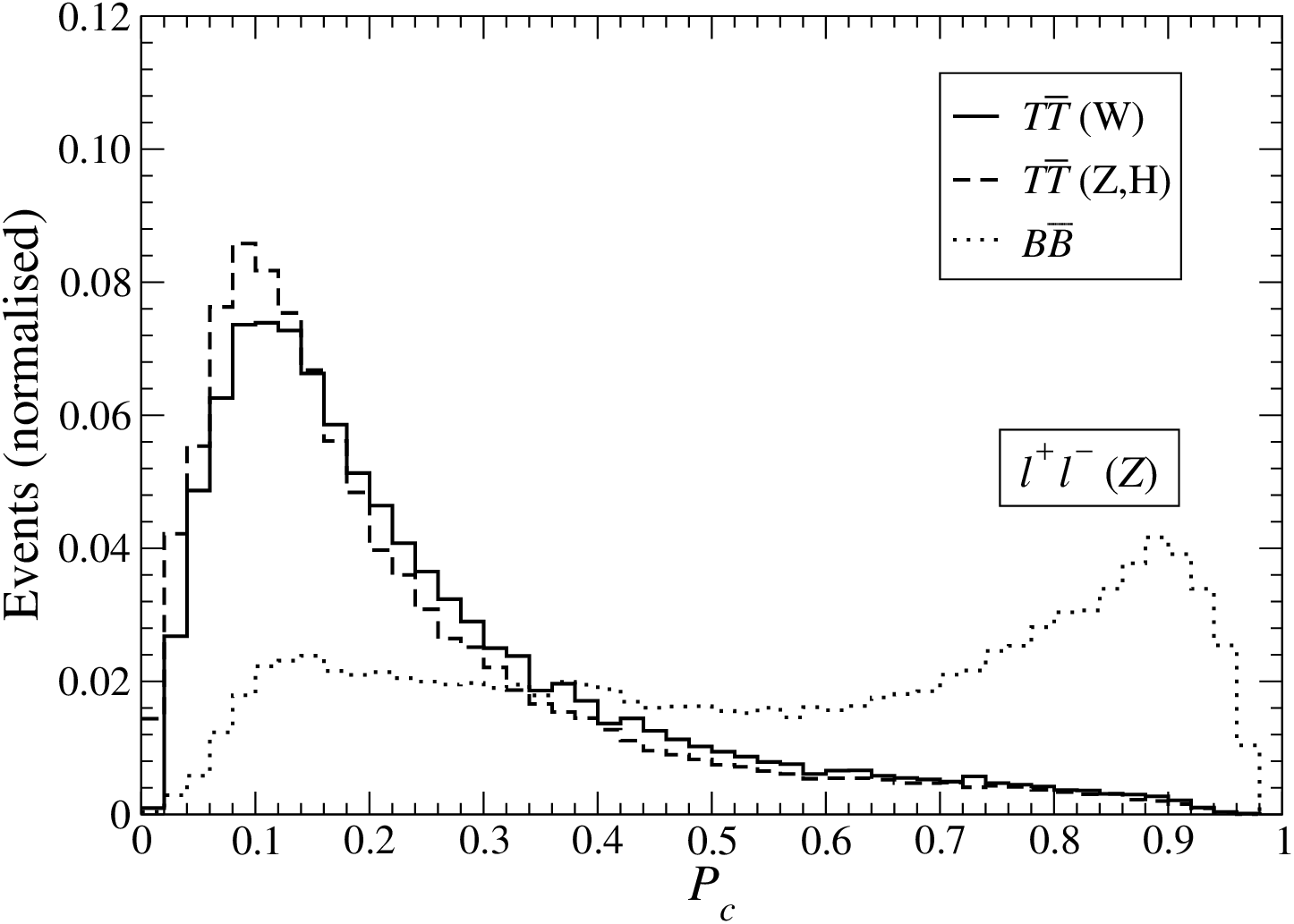,height=5.1cm,clip=}}
\end{tabular}
\caption{Probability distribution functions for events in the 
reference samples.}
\label{fig:lik2-2Q0-Z}
\end{center}
\end{figure}
It is seen that the 
discriminating power is practically the same as in the trilepton channel, as it can be better observed in Fig.~\ref{fig:lik2-2Q0-Z} which represents the likelihood function evaluated on the three class samples, giving the probabilities $P_a$, $P_b$, $P_c$ that the events correspond to each class.
Table~\ref{tab:lik-2Q0-Z} shows the performance of the likelihood function on the reference samples.
Events in a class $x$ are correctly classified if $P_x > P_y,P_z$, where $y$, $z$ are the other classes.

\begin{table}[htb]
\begin{center}
\begin{tabular}{cccc}
Class & $P_a > P_b,P_c$ & $P_b > P_a,P_c$ & $P_c > P_a,P_b$ \\
\hline
($a$) & 0.59 & 0.25 & 0.16 \\
($b$) & 0.23 & 0.62 & 0.15 \\
($c$) & 0.17 & 0.18 & 0.65
\end{tabular}
\end{center}
\caption{Performance of the likelihood function on the $\ell^+ \ell^-$ event reference samples: fractions of events in each sample and their classification. Events in a class $x$ are correctly classified if $P_x > P_y,P_z$, where $y$, $z$ are the other classes.}
\label{tab:lik-2Q0-Z}
\end{table}

The event reconstruction proceeds in the same way as in the trilepton channel (see section~\ref{sec:3l-Z-2}) but replacing the leptonic $W$ boson by a second $W$ decaying hadronically. We use all jets pairings with a maximum of 6 light jets to construct two $W$ bosons, and for events in class ($b$) we require at least 8 jets ($b$-tagged or not), otherwise the events are rejected.
The number of signal and background events after reconstruction cuts and their distribution in the three classes is given in Table~\ref{tab:nsnb-2Q0-Z-C}. Notice that the total number of events includes those which are later rejected in reconstruction. We also remark that this signal discrimination based on topology brings an important ``cleaning'' of the background for events classified as $\BB$, as we already have anticipated.

\begin{table}[t]
\begin{center}
\begin{tabular}{ccccccc}
               & Total & ($a$) & ($b$) & $(c)$ \\[1mm]
$\TT$ ($\Ts$)  & 137.3 & 50.4 & 56.5  & 21.7 \\
$\TT$ ($\TBd$) & 135.6 & 48.6 & 56.6  & 21.3 \\
$\TT$ ($\TBD$/$\XTd$) & 297.9 & 67.4 & 165.0 & 40.5 \\
$\XX$ ($\XTd$) & 45.0  & 14.2 & 18.6  & 2.8 \\
$\BB$ ($\Bs$)  & 220.4 & 65.4  & 31.1 & 113.5 \\
$\BB$ ($\TBd$) & 218.3 & 61.1  & 28.9 & 121.3 \\
$\BB$ ($\TBD$) & 38.8 & 11.3 & 15.3 & 3.3 \\
$\BB$ ($\BYd$) & 372.2 & 134.8 & 45.9 & 180.9 \\
$\YY$ ($\BYd$) & 5.3   & 1.9   & 0.3  & 2.7 \\
\hline
$t \bar t nj$    & 450 & 113  & 129   & 16 \\
$tW$             & 9   & 4    & 1     & 0  \\
$Z^*/\gamma^*nj$ & 181 & 32   & 86    & 12 \\
$Zb\bar b nj$  & 335   & 109   & 74   & 51 \\
$Zc \bar c nj$ & 61    & 24    & 14   & 8 \\
$Zt \bar t nj$ & 65    & 15    & 28   & 8 
\end{tabular}
\end{center}
\caption{Number of signal events in the $\ell^+ \ell^-$ ($Z$) final state at the selection level assigned to each event class. The luminosity is 30 fb$^{-1}$.}
\label{tab:nsnb-2Q0-Z-C}
\end{table}

\begin{figure}[t]
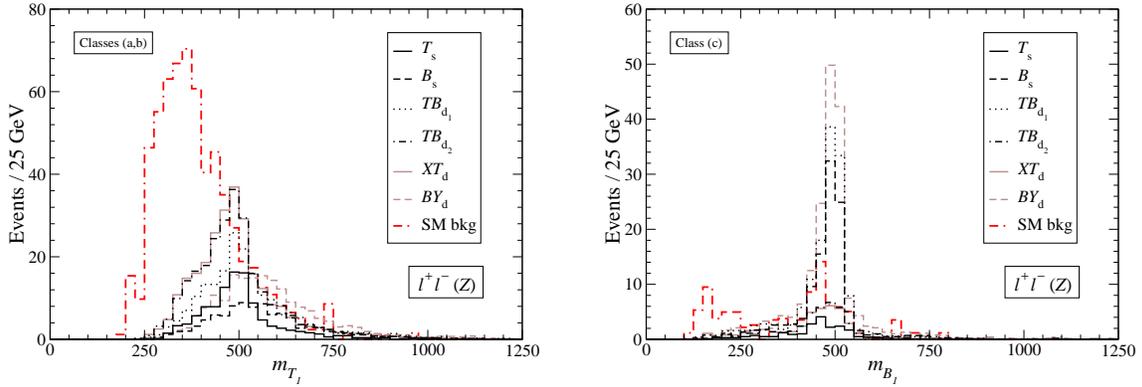

\begin{center}
\begin{tabular}{ccc}
\epsfig{file=Figs/mtZ-2Q0-Z.eps,height=5.1cm,clip=} & \quad &
\epsfig{file=Figs/mbZ-2Q0-Z.eps,height=5.1cm,clip=}
\end{tabular}
\caption{Reconstructed heavy quark masses in the $\ell^+ \ell^-$ ($Z$) final state.
The luminosity is 30 \fbin.}
\label{fig:mrec-2Q0-Z}
\end{center}
\end{figure}

We show in Fig.~\ref{fig:mrec-2Q0-Z} the two most interesting signal peaks, those involving the decays $T_1 \to Zt$, common to classes ($a,b$), and $B_1 \to Zb$ in class ($c$). These peaks are less biased by the reconstruction process although we can notice that, for example, the $B\to Zb$ singlet signals misidentified and reconstructed as $T \to Zt$ have a small bump around 500 GeV.
Notice that the $B_1$ peaks in the $Zb$ invariant mass distribution are very sharp and significative, resulting in an excellent discovery potential for $B$ quarks.
For the other heavy quarks $T_2$, $B_2$ with full hadronic decay the distributions are more biased, and the heavy quark peaks are almost equally well reconstructed for the signals as for the SM background: among the many jet combinations it is always possible,
especially when the background involves top quarks and $W$ bosons, to find one which is kinematically similar to the signal. These distributions are uninteresting and are not presented for brevity. Analogously, we do not present nor perform quality cuts on reconstructed $W$ and top masses, which are very similar for the signals and backgrounds.
We estimate the signal significance by performing a cut around the mass peaks,
\begin{equation}
400~\text{GeV} < m_{T_1},m_{B_1} < 600~\text{GeV} \,, 
\end{equation}
giving the numbers of signal and background events in Table~\ref{tab:nsnb-2Q0-Z-C2} for completeness.
\begin{table}[t]
\begin{center}
\begin{tabular}{cccccccccc}
 & $T_1$ ($a$,$b$) & $B_1$ $(c)$  & \quad & & $T_1$ ($a$,$b$) & $B_1$ $(c)$ \\[1mm]
$\TT$ ($\Ts$)   & 77.5  & 15.1   && $\BB$ ($\Bs$)  & 55.7   & 93.6  \\
$\TT$ ($\TBd$)  & 74.2  & 14.1   && $\BB$ ($\TBd$) & 52.9   & 103.7 \\
$\TT$ ($\TBD$/$\XTd$)  & 159.6 & 27.6   && $\BB$ ($\TBD$) & 13.2   & 2.2 \\
$\XX$ ($\XTd$)  & 18.7  & 1.3    && $\BB$ ($\BYd$) & 101.3  & 148.2 \\
                &       &        && $\YY$ ($\BYd$) & 1.4    & 1.7 \\
\hline
$t \bar t nj$   & 69    & 0      && $Zb\bar b nj$  & 78     & 28\\
$tW$            & 1     & 0      && $Zc \bar c nj$ & 15     & 4 \\
$Z^*/\gamma^*nj$& 22    & 8      && $Zt \bar t nj$ & 22     & 5 
\end{tabular}
\end{center}
\caption{Number of signal events in the $\ell^+ \ell^-$ ($Z$) final state at the $T_1$, $B_1$ heavy quark peaks. The luminosity is 30 fb$^{-1}$.}
\label{tab:nsnb-2Q0-Z-C2}
\end{table}
The discovery luminosities obtained summing all signal contributions within a given model and combining the significances for the $T_1$ and $B_1$ peaks are presented in Table~\ref{tab:sig-2Q0-Z}. We find that the amount of work necessary to build the likelihood function and discriminate the different signals pays off, and the discovery luminosities achieved are quite small in some cases.
The excellent result obtained for $\BB$ production in a $\BY$ doublet, where the $B$ quark only decays in $B \to Zb$, $B \to Hb$, deserves a special mention: in most final states examined up to now the discovery potential for this model was rather limited but the opposite-charge dilepton one constitutes a remarkable exception. The reconstructed peak in the $Zb$ invariant mass distribution shows the presence of a heavy $B$ quark with charge $-1/3$. The same can be said regarding the $Zt$ distribution and $T$ quarks, although in this case the background is much larger and the observation in the trilepton final state is easier and cleaner.

\begin{table}[t]
\begin{center}
\begin{tabular}{ccccccc}
       & $L$      & Rec. & \quad &        & $L$       & Rec. \\[1mm]
$\Ts$  & 22  \fbin  & $m_T$   &       & $\TBD$ & 4.4 \fbin & $m_T$ \\
$\Bs$  & 4.5 \fbin & $m_B$   &       & $\XTd$ & 4.4 \fbin & $m_T$ \\
$\TBd$ & 2.4 \fbin & $m_T$, $m_B$    &       & $\BYd$ & 1.8 \fbin    & $m_B$
\end{tabular}
\end{center}
\caption{Luminosity $L$ required to have a $5\sigma$ discovery in the $\ell^+ \ell^-$ ($Z$) final state. We also indicate whether a mass peak can be reconstructed in this final state.}
\label{tab:sig-2Q0-Z}
\end{table}

\subsubsection{Discovery of $B \to Hb$}

We now concentrate ourselves on the process $\BB \to ZbH \bar b$. As selection criteria we only ask (i) the presence of four $b$-tagged jets with $p_T > 20$ GeV, which is sufficient to practically eliminate all backgrounds, and (ii) less than four light jets, to remove the overlap between this final state and the previous one.\footnote{This is not strictly necessary as long as we do not intend to combine the statistical sensitivities of both samples, but we include it for simplicity and in order to be conservative. Dropping the requirement on light jets the signals in this sample are larger and the discovery luminosity for the $(B \, Y)$ doublet is reduced about a factor of two.} We give in Table~\ref{tab:nsnb-2Q0-Zbb} the numbers of events at pre-selection and selection for all signals and backgrounds.
\begin{table}[t]
\begin{center}
\begin{tabular}{cccccccccc}
                 & Pre.  & Sel.   & \quad &          & Pre.  & Sel. \\[1mm]
$\TT$ ($\Ts$)    & 179.6 & 3.1  & & $\BB$ ($\Bs$)  & 393.4 & 12.9  \\
$\TT$ ($\TBd$)   & 174.1 & 4.1  & & $\BB$ ($\TBd$) & 388.1 & 14.4 \\
$\TT$ ($\TBD$/$\XTd$)   & 341.6 & 7.7  & & $\BB$ ($\TBD$) & 55.4 & 0.4 \\
$\XX$ ($\XTd$)   & 60.7  & 0.2  & & $\BB$ ($\BYd$) & 780.3 & 38.8 \\
                 &       &      & & $\YY$ ($\BYd$) & 25.8  & 0.0 \\
\hline
$t \bar t nj$    & 7464  & 0    & & $Z^*/\gamma^* nj$   & 4875 & 0 \\
$tW$             & 212   & 0    & & $Zb\bar b nj$  & 9807  & 6 \\
$t\bar tb\bar b$ & 38    & 4    & & $Zc \bar c nj$ & 883   & 0 \\
$Wt \bar t nj$   & 4     & 0    & & $Zt \bar t nj$ & 88    & 0 
\end{tabular}
\end{center}
\caption{Number of events in the $\ell^+ \ell^-$ ($Z$) final state at pre-selection and with the selection requirement of four $b$-tagged jets. The luminosity is 30 fb$^{-1}$.}
\label{tab:nsnb-2Q0-Zbb}
\end{table}
This final state has an excellent discovery potential for the Higgs boson with a 
$\BY$ doublet (see Table~\ref{tab:sig-2Q0-Zbb}), and moderate for the $\TBd$ model. In our estimations for the sensitivity we include a 20\% systematic uncertainty in the background when necessary.

\begin{table}[t]
\begin{center}
\begin{tabular}{ccccccc}
       & $L$      & Rec. & \quad &        & $L$       & Rec. \\[1mm]
$\Ts$  & --  & no   &       & $\TBD$ & -- & no \\
$\Bs$  & --  & no   &       & $\XTd$ & -- & no \\
$\TBd$ & 30 \fbin & no   &       & $\BYd$ & 9.2 \fbin    & $m_B$, $M_H$ \\
\end{tabular}
\end{center}
\caption{Luminosity $L$ required to have a $5\sigma$ discovery in the $\ell^+ \ell^-$ ($Z$) final state with four $b$ tags. We also indicate whether a mass peak can be reconstructed in this final state.}
\label{tab:sig-2Q0-Zbb}
\end{table}

The Higgs boson mass can also be reconstructed when it results from a $B$ decay, doing as follows.
\begin{enumerate}
\item We select a $b$ jet to be paired with the $Z$ boson candidate and reconstruct the heavy quark $B_1$; the other heavy quark $B_2$ is reconstructed from the three remaining $b$ jets.
\item The combination minimising the mass difference $(m_{B_2}-m_{B_1})$ is chosen.
\item Among the three $b$ jets from $B_2$, we choose the two ones with minimum invariant mass to be the ones corresponding to the Higgs decay.
\end{enumerate} 

\begin{figure}[t]
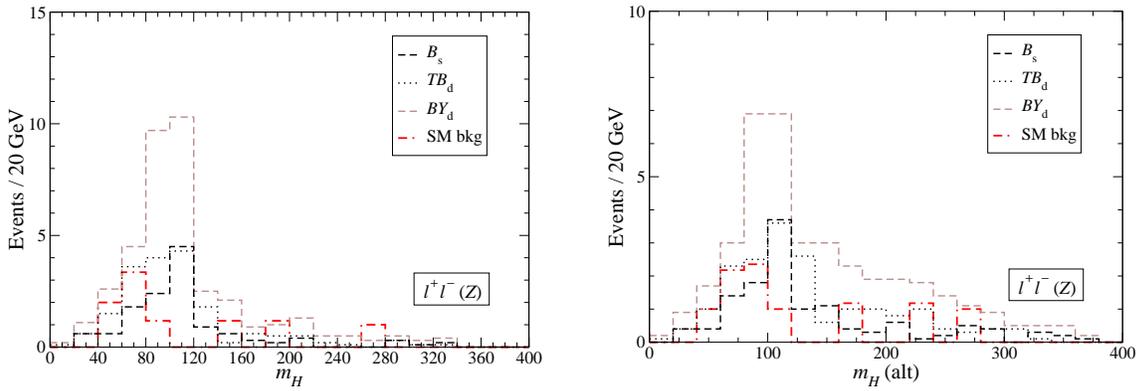

\begin{center}
\begin{tabular}{ccc}
\epsfig{file=Figs/mH-2Q0-Z.eps,height=5.1cm,clip=} & \quad &
\epsfig{file=Figs/mHalt-2Q0-Z.eps,height=5.1cm,clip=}
\end{tabular}
\caption{Reconstructed Higgs boson mass in the $\ell^+ \ell^-$ ($Z$) final state
with four $b$ tags. The luminosity is 30 fb$^{-1}$.}
\label{fig:mhrec-2Q0-Z}
\end{center}
\end{figure}

The resulting reconstructed Higgs mass is shown in Fig.~\ref{fig:mhrec-2Q0-Z} (left). As we already have mentioned, this final state is most interesting for the $\BY$ doublet which has a large signal and in which the Higgs mas peak can be clearly reconstructed with sufficient luminosity.
For $T \to Ht$ decays a small signal could be seen with different selection criteria but we do not address this here, since $T \to Ht$ signals are far more interesting in the single lepton channel.
Finally, one may wonder whether the ``Higgs'' peak results from the presence of a resonance or if it is merely a kinematical effect. To investigate this, we can use a different reconstruction by selecting the two $b$ jets with smallest $p_T$. The resulting distribution, shown in Fig.~\ref{fig:mhrec-2Q0-Z} (right), also displays a peak at the same place although the combinatorial background is larger in this case.

\subsection{Final state $\ell^+ \ell^-$ (no $Z$)}

In this final state the signals involve two $W$ boson decays from different heavy quarks in general, and hence the heavy mass peaks are difficult to reconstruct except for $\BB$ production. The detection of a signal must then rely on event counting, which requires an efficient background suppression. We perform here two analyses: first a generic one aiming to discover the new quark signals, and then a specific one to reconstruct the heavy $B$ quark mass. This mass reconstruction is useful for the $\TBD$ model, where the $B \to Zb$ decay does not take place.

\subsubsection{Discovery potential}

\begin{figure}[htb]
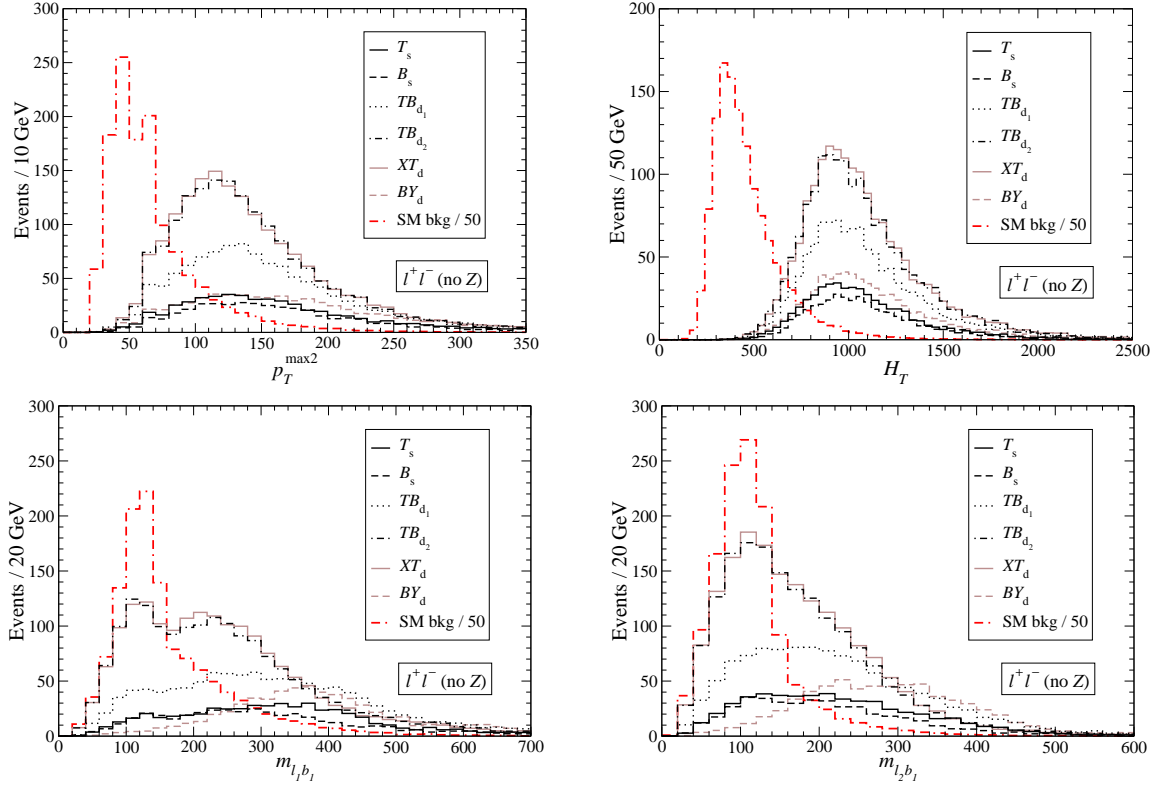

\begin{center}
\begin{tabular}{ccc}
\epsfig{file=Figs/ptmax2-2Q0-noZ.eps,height=5.1cm,clip=} & \quad &
\epsfig{file=Figs/HT-2Q0-noZ.eps,height=5.1cm,clip=} \\
\epsfig{file=Figs/ml1b1-2Q0-noZ.eps,height=5.1cm,clip=} & \quad &
\epsfig{file=Figs/ml2b1-2Q0-noZ.eps,height=5.1cm,clip=}
\end{tabular}
\caption{Kinematical distributions of variables used in selection criteria
for the $\ell^+ \ell^-$ (no $Z$) final state: transverse momentum of the second highest-$p_T$ jet, total transverse energy and invariant masses of the two leptons and the leading $b$ jet. The luminosity is 30 fb$^{-1}$.}
\label{fig:dist-2Q0-noZ}
\end{center}
\end{figure}

For event selection we demand
(i) transverse momentum $p_T > 100$ GeV for the sub-leading jet ($b$-tagged or not);
(ii) transverse energy $H_T > 750$ GeV;
(iii) the invariant mass of the highest-$p_T$ $b$ jet $b_1$ and each of the two leptons must be larger than the top mass, taken here as 175 GeV. The first two conditions reduce backgrounds in general, while the third one strongly suppresses $t \bar t nj$ production, where the $b$ quarks and charged leptons result from top decays. The kinematical distributions of these variables are presented in Fig.~\ref{fig:dist-2Q0-noZ}.
The number of events after these cuts can be read in Table~\ref{tab:nsnb-2Q0-noZ}, where we also include for better comparison the numbers of events at pre-selection.
\begin{table}[t]
\begin{center}
\begin{tabular}{ccccccc}
                 & Pre.   & Sel.  & \quad &          & Pre.  & Sel.  \\[1mm]
$\TT$ ($\Ts$)    & 536.3  & 236.6 & & $\BB$ ($\Bs$)  & 426.4 & 170.1 \\
$\TT$ ($\TBd$)   & 625.3  & 274.4 & & $\BB$ ($\TBd$) & 519.4 & 202.7 \\
$\TT$ ($\TBD$/$\XTd$)   & 666.1  & 175.6 & & $\BB$ ($\TBD$) & 1050.0 & 233.3 \\
$\XX$ ($\XTd$)   & 1086.7 & 240.5 & & $\BB$ ($\BYd$) & 122.2 & 89.1 \\
                 &        &       & & $\YY$ ($\BYd$) & 544.6 & 359.1 \\
\hline
$t \bar t nj$    & 61029  & 80    & & $Z^*/\gamma^* nj$    & 370 & 1 \\
$tW$             & 1923   & 14    & & $Zb\bar b nj$    & 325 & 6  \\
$t\bar tb\bar b$ & 309    & 22    & & $Zc \bar c nj$   & 48  & 1 \\
$W t \bar t nj$  & 59     & 3     & & $Zt \bar t nj$   & 18  & 6 
\end{tabular}
\end{center}
\caption{Number of events in the $\ell^+ \ell^-$ (no $Z$) final state (discovery analysis) for
the signals and main backgrounds with a luminosity of 30 fb$^{-1}$, at pre-selection and selection level.}
\label{tab:nsnb-2Q0-noZ}
\end{table}
We point out that among the opposite-sign $\BB$ events produced in the decay
\begin{align}
& B \bar B \to W^- t \, W^+ \bar t \to W^- W^+ b \, W^+ W^- \bar b
&& \quad 2W \to \ell \nu , 2W \to q \bar q' \,,
\label{ec:BBdec}
\end{align}
those surviving the $m_{\ell_1 b}$ and $m_{\ell_2 b}$ cuts mostly correspond to leptonic decay of the two opposite-sign $W$ bosons produced in $B \to W^- t$, $\bar B \to W^+ \bar t$ decays. Then, a mass reconstruction is difficult with these event selection criteria, which are anyway very efficient to reduce the $t \bar t$ background and observe a heavy quark signal.
The luminosities required for $5\sigma$ discovery are given in Table~\ref{tab:sig-2Q0-noZ}. Since in this final state the background is still relatively important, the uncertainty in its overall normalisation affects the significance of the signals. We then include a 20\% systematic uncertainty in our estimations in order to be more realistic.
\begin{table}[t]
\begin{center}
\begin{tabular}{ccccccc}
       & $L$        & Rec. & \quad &        & $L$       & Rec. \\[1mm]
$\Ts$  & 2.7 \fbin  & no   &       & $\TBD$ & 1.1 \fbin & $m_B$ \\
$\Bs$  & 9.3 \fbin  & $m_B$   &       & $\XTd$ & 1.1 \fbin & no \\
$\TBd$ & 0.83 \fbin & $m_B$   &       & $\BYd$ & 0.87 \fbin    & no \\
\end{tabular}
\end{center}
\caption{Luminosity $L$ required to have a $5\sigma$ discovery in the $\ell^+ \ell^-$ (no $Z$) final state. We also indicate whether a mass peak can be reconstructed in this final state.}
\label{tab:sig-2Q0-noZ}
\end{table}
We point out the excellent sensitivity of this final state to $\TT$, $\XX$ and $\YY$ production. The latter is specially important, because $Y$ pair production only gives signals in the opposite-sign dilepton and single lepton channels, and the discovery of charge $-4/3$ quarks must be done in one of them.

\subsubsection{Heavy quark reconstruction}

In order to reconstruct the heavy $B$ masses we drop from the selection criteria the $\ell b$ invariant mass requirements, to allow for top quark semileptonic decays in Eq.~(\ref{ec:BBdec}).
The selection 
criteria in this case are:
(i) the presence of four jets with $p_T > 20$ GeV;
(ii) transverse momentum $p_T > 100$ GeV for the sub-leading jet ($b$-tagged or not);
(iii) transverse energy $H_T > 750$ GeV.
The number of events are given in Table~\ref{tab:nsnb-2Q0-noZ-2}. We notice that the $t \bar t nj$ background is much larger here than in the previous discovery analysis (Table~\ref{tab:nsnb-2Q0-noZ}).

\begin{table}[t]
\begin{center}
\begin{tabular}{ccccccccc}
                      & Pre.   & Sel.  & Rec.  & \quad &          & Pre.   & Sel.  & Rec. \\[1mm]
$\TT$ ($\Ts$)         & 536.3  & 209.3 & 21.1  & & $\BB$ ($\Bs$)  & 426.4  & 206.4 & 24.6 \\
$\TT$ ($\TBd$)        & 625.3  & 248.7 & 27.0  & & $\BB$ ($\TBd$) & 519.4  & 249.4 & 36.2 \\
$\TT$ ($\TBD$/$\XTd$) & 666.1  & 395.7 & 42.4  & & $\BB$ ($\TBD$) & 1050.0 & 623.6 & 122.4 \\
$\XX$ ($\XTd$)        & 1086.7 & 661.1 & 127.2 & & $\BB$ ($\BYd$) & 122.2  & 53.6  & 4.6 \\
                      &        &       &       & & $\YY$ ($\BYd$) & 544.6  & 118.7 & 11.0 \\
\hline
$t \bar t nj$         & 61029  & 1419  & 139   & & $Z^*/\gamma^* nj$  & 370    & 2     & 0 \\
$tW$                  & 1923   & 18    & 0     & & $Zb\bar b nj$  & 325    & 4     & 0 \\
$t\bar tb\bar b$      & 309    & 28    & 2     & & $Zc \bar c nj$ & 48     & 0     & 0 \\
$W t \bar t nj$       & 59     & 6     & 0     & & $Zt \bar t nj$ & 18     & 8     & 2
\end{tabular}
\end{center}
\caption{Number of events in the $\ell^+ \ell^-$ (no $Z$) final state (reconstruction analysis) for
the signals and main backgrounds with a luminosity of 30 fb$^{-1}$, at pre-selection and selection level, and including reconstructed mass cuts.}
\label{tab:nsnb-2Q0-noZ-2}
\end{table}

\begin{figure}[htb]
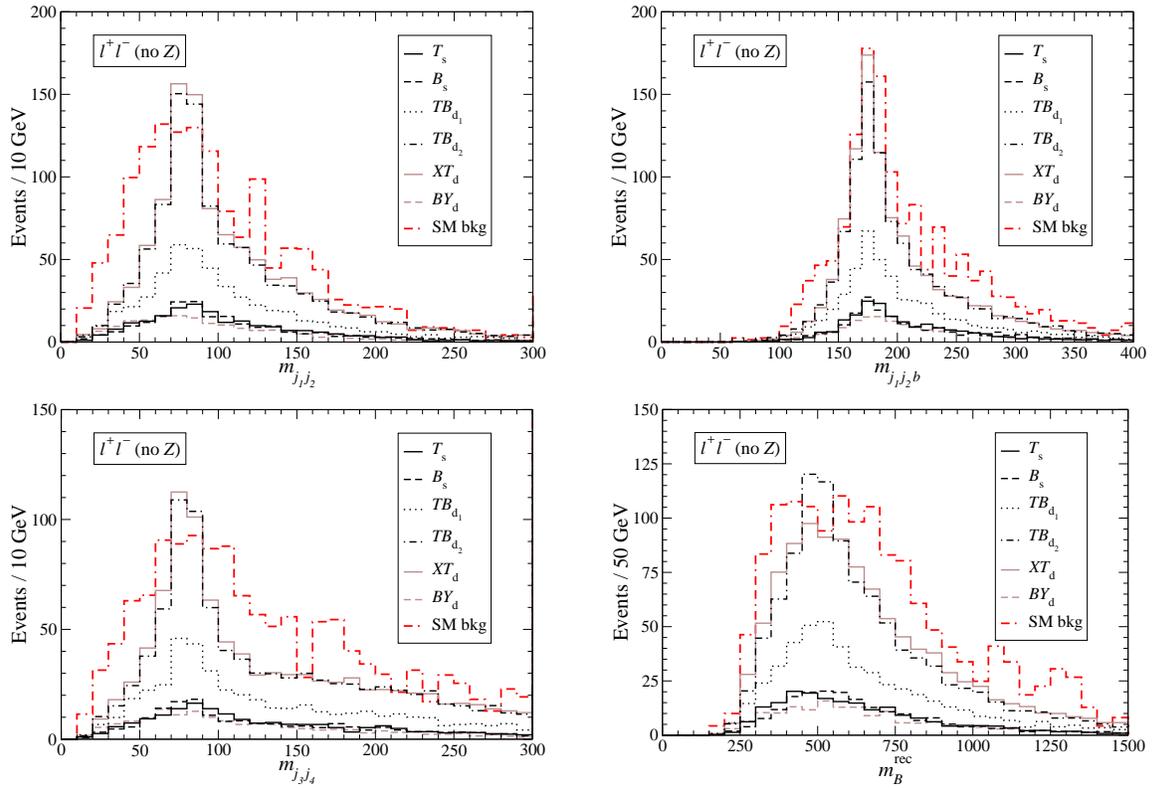

\begin{center}
\begin{tabular}{ccc}
\epsfig{file=Figs/mw1-2Q0-noZ.eps,height=5.1cm,clip=} & \quad &
\epsfig{file=Figs/mt-2Q0-noZ.eps,height=5.1cm,clip=} \\
\epsfig{file=Figs/mw2-2Q0-noZ.eps,height=5.1cm,clip=} & \quad &
\epsfig{file=Figs/mQ-2Q0-noZ.eps,height=5.1cm,clip=}
\end{tabular}
\caption{Reconstructed masses of the two $W$ bosons, the top and the heavy quark. The luminosity is 30 \fbin.}
\label{fig:mrec-2Q0}
\end{center}
\end{figure}

\begin{figure}[htb]
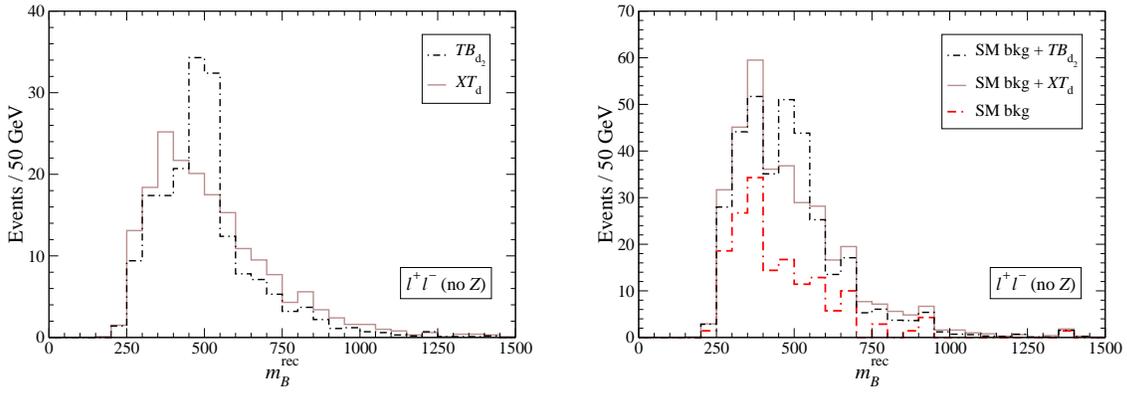

\begin{center}
\begin{tabular}{ccc}
\epsfig{file=Figs/mQ-2Q0-noZcut1.eps,height=5.1cm,clip=} & \quad &
\epsfig{file=Figs/mQ-2Q0-noZcut.eps,height=5.1cm,clip=}
\end{tabular}
\caption{Left: reconstructed mass of the heavy quark for the two largest signals. Right: the same, including the SM background. The luminosity is 30 \fbin.}
\label{fig:mrec-2Q0cut}
\end{center}
\end{figure}

For events in which the $W^+ W^-$ pair corresponds to the same heavy quark, the invariant mass of the $W$ bosons decaying hadronically plus one of the $b$ quarks will peak at $m_B$. The heavy quark mass reconstruction is done analogously as for the $X$ quark in the like-sign dilepton channel in section~\ref{sec:6.2}, and
the reconstructed heavy quark mass $m_B$ is then defined as the invariant mass of the $b$ quark and  four light jets selected. These distributions are presented in Fig.~\ref{fig:mrec-2Q0}.
We point out that, in contrast with the $X$ quark reconstruction in the like-sign dilepton channel,
half of the events in $\BB$ decays have opposite-sign leptons resulting from different heavy quark decays. The heavy $B$ quark mass peak is reasonably well reconstructed as it is shown in the last plot of Fig.~\ref{fig:mrec-2Q0} but the distribution is quite similar for a $X$ quark. Then, although the presence of a signal would be apparent, the observation of a clear peak and the discrimination
among these two possiblities is rather difficult, even more in the presence of a large $t \bar t$ background.\footnote{For higher masses the background suppression is more efficient via transverse energy requirments, and the $B$ quark peak may be easier to reconstruct~\cite{Skiba:2007fw}.} We then apply quality cuts to improve the reconstruction and reduce the background,
\begin{align}
60~\text{GeV} < m_{j_1 j_2} < 100~\text{GeV} \,, \notag \\
60~\text{GeV} < m_{j_3 j_4} < 100~\text{GeV} \,, \notag \\
125~\text{GeV} < m_{j_1 j_2 b} < 225~\text{GeV} \,.
\end{align}
The number of events after these cuts is given in Table~\ref{tab:nsnb-2Q0-noZ-2}. With these cuts, the reconstructed mass for the $\TBD$ model (displaying a peak at $m_B$) and the $\XT$ doublet (without peak) are quite different, as it can be seen in Fig.~\ref{fig:mrec-2Q0cut} (left). Both possibilities could also be distinguished in the presence of background, as shown in the right panel.
The reconstruction of a peak in the $Wt$ invariant mass distribution shows that the heavy quark has charge $-1/3$ or $5/3$. Unfortunately the transverse mass of the $B$ quark with leptonic $W$ decays does not display a clear endpoint, and the direct identification of the quark charge is not possible as for $X$ quarks in the like-sign dilepton channel.

\subsection{Summary}

Despite the {\em a priori} large opposite-sign dilepton backgrounds, this final state turns out to have an excellent sensitivity to heavy quark signals. These signals and the SM backgrounds result from either the leptonic decay of a $Z$ or of two $W$ bosons. Therefore, as it was done in other final states, it is advantageous to divide this final state into two subsamples, with or without a $Z$ candidate.

The dilepton sample with a $Z$ candidate has a huge background from $Znj$ production which can be practically removed by asking for the presence of two $b$-tagged jets and four light jets. Once that SM backgrounds are manageable, we have implemented a likelihood function which discriminates among the three processes in Eqs.~(\ref{ec:ch2Q0Z}), classifying events with a good efficiency. The event reconstruction is then performed according to this classification. For events identified as $\BB \to ZbWt$, the $Zb$ invariant mass distribution displays a sharp peak which signals the presence of a charge $-1/3$ quark. For $\TT$ production a peak can be observed in the $Zt$ invariant mass distribution but the background is larger in this case. The discovery luminosities are small,
as it can be seen in Table~\ref{tab:sig-2Q0-Z}. The case of $B$ quarks in the $\BY$ doublet (1.8 \fbin\ for 500 GeV quarks) deserves a special mention, since these quarks are much harder to see in other final states. For this model, the observation of the decay $B \to Hb$, with $H \to b \bar b$ is also possible if we concentrate on final states with four $b$ tags, and $5\sigma$ discovery could be possible with a luminosity below 10 \fbin\ for $m_B = 500$ GeV and $M_H = 115$ GeV. This decay is interesting not only to establish the identity of the $B$ quark but because it is a possible discovery channel for the Higgs boson if such doublets exist.

In the subsample without $Z$ candidates we have performed two different analyses, first a generic one which achieves the best signal significance and then a specific one to reconstruct the heavy $B$ quark mass in $B \to W^- t$ decays. The background is again important but the largest one, $t \bar t nj$, can be practically removed by requiring invariant masses $m_{\ell b} > m_t$, so that the charged leptons and $b$ quarks cannot result from a top quark decay (this requirement must be dropped in the reconstruction analysis). After background suppression, this final state also offers an excellent discovery potential for the pair production of $T$, $X$ and $Y$ quarks. For example, $5\sigma$ significance can be achieved with a luminosity around 1 \fbin\ for the four models with quark doublets.
The sensitivity to $Y$ quarks is especially important because they only have decays $Y \to W^- b$, and their detection can only be performed in the opposite-sign dilepton and single lepton final states.

Finally, we have performed the reconstruction of the $B$ quark mass in $B \to W^- t \to W^- W^+ b$ decays (or the charge conjugate), with both $W$ bosons decaying hadronically.
This is specially interesting for the $\TBD$ model where the $B \to Zb$ decay does not take place. 
In this analysis the $m_{\ell b} > m_t$ requirement must be dropped in order to keep the events which actually display a peak in the invariant mass of four light jets plus a $b$-tagged jet. With adequate reconstruction quality cuts a clear peak could be observed distinguishing $\BB$ and $\XX$ production, but the quark charge cannot be directly measured.

\section{Final state $\ell^\pm$}
\label{sec:1l}

Single lepton signals result from heavy quark pair decays when one of the $W$ bosons (up to four are present, depending on the channel) decays leptonically and the rest of $W$, $Z$ and Higgs bosons decay hadronically. Hence, single lepton signals benefit from a large branching ratio. This final state is fundamental to establish whether the $T \to Wb$ decay takes place (this can also be seen in the trilepton final state but needs about ten times more luminosity, see section~\ref{sec:3l-Z-2}). Due to the large size of the signals, one can also look for subsamples with high $b$ jet multiplicities to establish the decays $T \to Ht$, $B \to Hb$. Without being completely exhaustive, we will perform three analyses in this section. The first one, in the single lepton channel with exactly two $b$ tags, is devoted to the search of the $T \to Wb$ decay. The second one, in a sample with four $b$ tags, allows to search for $T \to Ht$ and $B \to Hb$. The third one, requiring six $b$ tags, is very useful to look for $T \to Ht$ in the models where this decay has an enhanced branching ratio. For event pre-selection we require (i) one charged lepton with $p_T > 30$ GeV; (ii) at least two $b$ jets with $p_T > 20$ GeV; (iii) at least two light jets also with $p_T > 20$ GeV. The total number of events for the signals and main backgrounds at pre-selection  are given in Table~\ref{tab:nsnb-1Q1}, as well as the numbers in the three subsamples.

\begin{table}[htb]
\begin{center}
\begin{tabular}{ccccc}
                      & Total   & $2b$   & $4b$   & $6b$  \\[1mm]
   $\TT$ ($\Ts$)      & 9415.3  & 5797.8 & 874.7  & 26.5 \\
   $\TT$ ($\TBd$)     & 10064.4 & 6172.9 & 931.8  & 29.0 \\
$\TT$ ($\TBD$/$\XTd$) & 11782.9 & 5294.3 & 1873.4 & 112.1 \\
   $\XX$ ($\XTd$)     & 9213.8  & 7506.3 & 172.9  & 0.4   \\
   $\BB$ ($\Bs$)      & 6535.6  & 4460.1 & 412.9  & 2.5 \\
   $\BB$ ($\TBd$)     & 7021.4  & 4802.3 & 434.2  & 2.8 \\
   $\BB$ ($\TBD$)     & 9193.4  & 7484.4 & 164.6  & 0.0 \\
   $\BB$ ($\BYd$)     & 2146.1  & 1399.6 & 150.7  & 0.7 \\
   $\YY$ ($\BYd$)     & 7444.1  & 6588.5 & 58.7   & 0.3 \\
\hline
   $t \bar tnj$       & 965051  & 902205 & 1629   & 0 \\  
   $tW$               & 31920   & 30280  & 38     & 0 \\
   $t\bar tb\bar b$   & 4355    & 2287   & 423    & 2 \\
   $t\bar tt\bar t$   & 27      & 12     & 1      & 0 \\
   $Wnj$              & 38185   & 37236  & 14     & 0 \\
   $W b \bar bnj$     & 20634   & 19920  & 16     & 0 \\
   $W t \bar tnj$     & 654     & 592    & 0      & 0 \\
   $Z/\gamma nj$      & 3397    & 3314   & 0      & 0 \\
   $Z b \bar bnj$     & 4874    & 4715   & 5      & 0 \\
\end{tabular}
\end{center}
\caption{Number of signal and background events in the $\ell^\pm$ final state at the pre-selection level, and in the four subsamples studied. The luminosity is 30 fb$^{-1}$.}
\label{tab:nsnb-1Q1}
\end{table}

\subsection{Final state $\ell^\pm$ ($2b$)}

\begin{figure}[b]
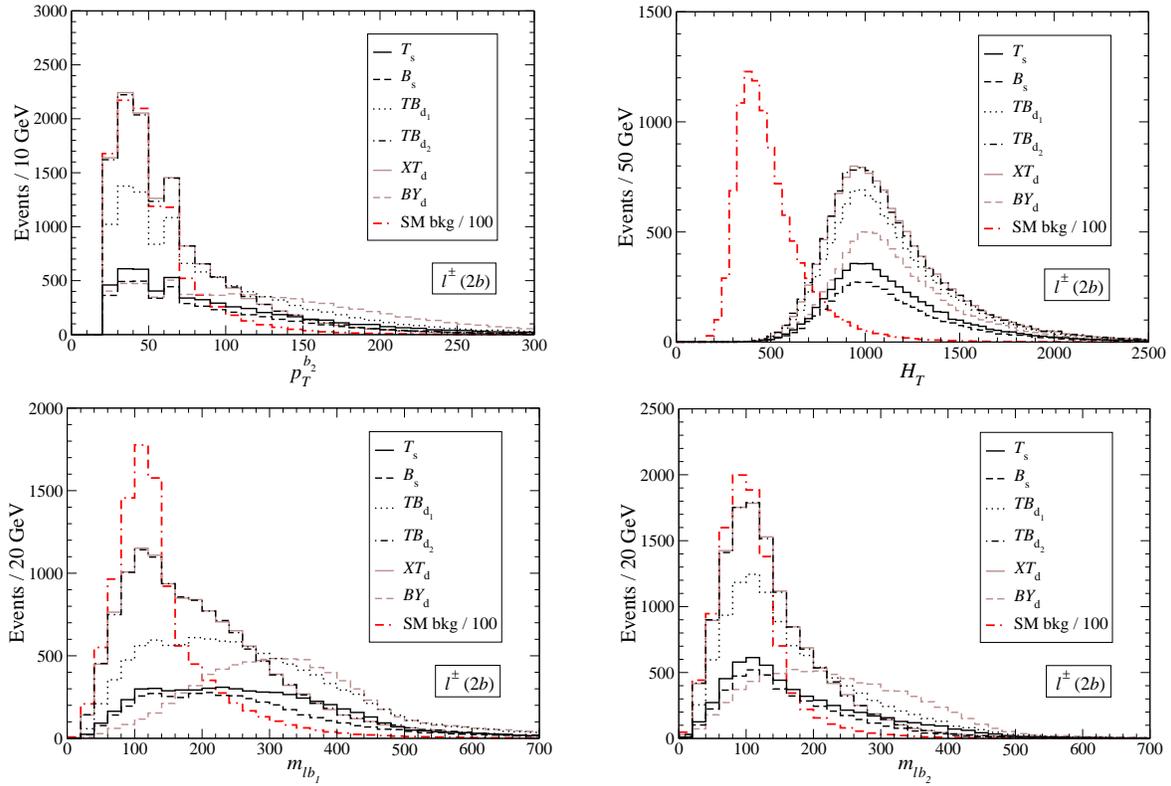

\begin{center}
\begin{tabular}{ccc}
\epsfig{file=Figs/ptbmax2-1Q1-2b.eps,height=5.1cm,clip=} & \quad &
\epsfig{file=Figs/HT-1Q1-2b.eps,height=5.1cm,clip=} \\
\epsfig{file=Figs/mlb1-1Q1-2b.eps,height=5.1cm,clip=} & \quad &
\epsfig{file=Figs/mlb2-1Q1-2b.eps,height=5.1cm,clip=} 
\end{tabular}
\caption{Kinematical distributions of variables used in selection criteria
for the $\ell^\pm$ ($2b$) final state: transverse momentum of the subleading $b$ jet, total transverse energy and invariant masses of the charged lepton and the two $b$ jets. The luminosity is 30 fb$^{-1}$.}
\label{fig:dist-1Q1-2b}
\end{center}
\end{figure}

In this channel, the processes we are mainly interested in are
\begin{align}
& T \bar T \to W^+ b \, W^- \bar b 
&& \quad WW \to \ell \nu q \bar q' \,, \nonumber \\
& Y \bar Y \to W^- b \, W^+ \bar b 
&& \quad WW \to \ell \nu q \bar q' \,.
\label{ec:ch1Q12b}
\end{align}
There are other decay channels for $T$ quarks, involving $T \to Zt$ and $T \to Ht$, which give the same final state, but they are suppressed by the selection criteria.
Some production and decay channels of $B$ and $X$ quarks also give single lepton signals, for example
\begin{align}
& B \bar B \to W^- t \, W^+ \bar t \to W^- W^+ b \, W^+ W^- \bar b
&& \quad 3W \to q \bar q' , 1W \to \ell \nu \,, \nonumber \\
& X \bar X \to W^+ t \, W^- \bar t \to W^+ W^+ b \, W^- W^- \bar b
&& \quad 3W \to q \bar q' , 1W \to \ell \nu \,,
\label{ec:ch1Q12b-2}
\end{align}
but they are rather difficult to see due to their greater similarity with $t \bar t$ production. Then, we concentrate our analysis on the channels in Eqs.~(\ref{ec:ch1Q12b}). The selection criteria applied for this are:
(i) transverse momentum $p_T > 150$ GeV for both $b$ jets; 
(ii) transverse energy $H_T > 750$ GeV;
(iii) the invariant mass of both $b$ jets and the charged lepton must be larger than the top mass, taken here as 175 GeV.
The distributions of the relevant variables are shown in Fig.~\ref{fig:dist-1Q1-2b}.
The first condition is inspired by the specific decays in Eqs.~(\ref{ec:ch1Q12b}).
(The dependence on the $p_T$ cut is not very strong, and we have choosen 150 GeV for simplicity.)
The transverse energy requirement is a general one to look for high mass states, not very optimised for these heavy quark masses.
The invariant mass requirements are extremely useful to reduce the $t \bar t nj$ background (still some events remain due to mistag of the charm or light jets), and allows to improve our results over previous analyses for $\TT$ production in this channel~\cite{AguilarSaavedra:2005pv,AguilarSaavedra:2006gv}.
However, it reduces the $\BB$ and $\XX$ signals for which some of the decay channels have the charged lepton and $b$ quarks both resulting from a top quark.
 We give in Table~\ref{tab:nsnb-1Q1-2b} the number of events at selection, also including the ones at pre-selection for better comparison.
\begin{table}[htb]
\begin{center}
\begin{tabular}{cccccccccccc}
                  & Pre.   & Sel.  & Peak  & \quad &    
                  & Pre.   & Sel.  & Peak
               \\[1mm]
   $\TT$ ($\Ts$)  & 5797.8 & 796.6 & 538.8 && $\BB$ ($\Bs$)  & 4460.1 & 391.7  & 217.7 \\
   $\TT$ ($\TBd$) & 6172.9 & 795.0 & 551.7 && $\BB$ ($\TBd$) & 4802.3 & 352.8  & 193.4 \\
   $\TT$ ($\TBD$/$\XTd$) & 5294.3 & 123.8 & 55.2  && $\BB$ ($\TBD$) & 7484.4 & 186.5  & 84.7 \\
   $\XX$ ($\XTd$) & 7506.3 & 165.5 & 74.8  && $\BB$ ($\BYd$) & 1399.6 & 307.0  & 198.8 \\ 
                  &        &       &       && $\YY$ ($\BYd$) & 6588.5 & 1974.3 & 1508.8\\
\hline
   $t \bar tnj$   & 902205 & 299   & 117   && $W b \bar bnj$ & 19920  & 125    & 52\\
   $tW$           & 30280  & 68    & 34    && $W t \bar tnj$ & 592    & 12     & 1 \\
   $t\bar tb\bar b$ & 2287 & 15    & 6     && $Z/\gamma nj$  & 3314   & 2      & 1 \\
   $t\bar tt\bar t$ & 12   & 1     & 0     && $Z b \bar bnj$ & 4715   & 24     & 9 \\
   $Wnj$          & 37236  & 49    & 22
\end{tabular}
\end{center}
\caption{Number of signal and background events in the $\ell^\pm$ ($2b$) final state at the pre-selection and selection level, and at the reconstructed mass peak. The luminosity is 30 fb$^{-1}$.}
\label{tab:nsnb-1Q1-2b}
\end{table}
We observe the excellent background reduction achieved with these simple selection criteria, especially with the $\ell b$ invariant mass cuts: the $t \bar t nj$ background is reduced by a factor of 3000, while the $\YY$ signal is kept at one third. The $\TT$ signal is reduced to one seventh because there are contributing channels other than $\TT \to W^+ b W^- \bar b$, and those are quite suppressed by the event selection necessary to reduce $t \bar t nj$. 

\begin{figure}[t]
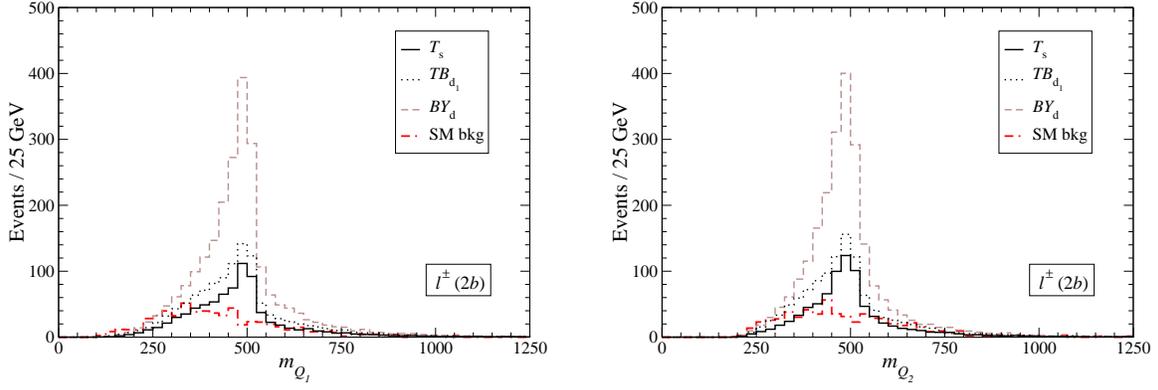

\begin{center}
\begin{tabular}{ccc}
\epsfig{file=Figs/mQ1-1Q1-2b.eps,height=5.1cm,clip=} & \quad &
\epsfig{file=Figs/mQ2-1Q1-2b.eps,height=5.1cm,clip=}
\end{tabular}
\caption{Reconstructed heavy quark masses at the selection level in the $\ell^\pm$ ($2b$) final state.
The luminosity is 30 \fbin.}
\label{fig:mrec-1Q1-2b}
\end{center}
\end{figure}

The $\TT$ and $\YY$ signals are reconstructed by choosing the best pairing between $b$ jets and reconstructed $W$ bosons:
\begin{enumerate}
\item The hadronic $W$ is obtained with the two jets (among the three ones with largest $p_T$) having an invariant mass closest to $M_W$.
\item The leptonic $W$ is obtained from the charged lepton and the missing energy with the usual method, keeping both solutions for the neutrino momentum.
\item The two heavy quarks $Q=T,Y$ are reconstructed with one of the $W$ bosons and one of the $b$ jets. We label them as $Q_{1,2}$, corresponding to the hadronic and leptonic $W$, respectively.
\item The combination minimising
\begin{small}
\begin{equation}
\frac{(m_{W_H}^\text{rec}-M_W)^2}{\sigma_W^2} + 
\frac{(m_{W_L}^\text{rec}-M_W)^2}{\sigma_W^2} + 
\frac{(m_{Q_1}^\text{rec}-m_{Q_2}^\text{rec})^2}{\sigma_Q^2}
\end{equation}
\end{small}%
is selected, with $\sigma_W = 10$ GeV, $\sigma_Q = 20$ GeV.
\end{enumerate}
We present the reconstructed mass distributions at the selection level in Fig.~\ref{fig:mrec-1Q1-2b}. 
With the criteria applied the peaks are very good even for the $\TT$ signals which involve several competing decay chains giving up to six $b$ quarks. In particular, the $p_T$ cut on the subleading $b$ jet suppresses the $T \to Ht$ decays selecting only $T \to Wb$, in which we are interested. The signal significance can be estimated by performing the invariant mass cuts
\begin{equation}
350~\text{GeV} < m_{Q_{1,2}}^\text{rec} < 650~\text{GeV} \,.
\end{equation}
The numbers of signal and background events at the peak, defined by the above mass windows, can be found in Table~\ref{tab:nsnb-1Q1-2b}. The high signal significance achieved for the case of $T$ and $Y$ quarks implies an excellent discovery potential, summarised in Table~\ref{tab:sig-1Q1-2b}.
We include a 20\% systematic uncertainty in the estimations in all cases. For $T$ quarks the discovery luminosities are rather small except for the $\TBD$ and $\XT$ doublet model, where the $T \to Wb$ decay does not take place. For $\YY$ production the discovery potential is even better, because the $Y \to Wb$ channel is the only one present.
For $B$ and $X$ quarks the signals are smaller, 
of the size of the background itself,
implying a signal significance below $5\sigma$ even for large luminosities due to the background normalisation uncertainty assumed. However, in case that a signal is detected in other final states it should be possible to detect also $\BB$ and $\XX$ signals in the single lepton channel with two $b$ tags using a dedicated and optimised analysis.
The heavy quark reconstruction as a peak in $Wb$ invariant mass distributions implies that it has either charge $2/3$ or $-4/3$. Both possibilities cannot be distinguished unless the $b$ jet charge is measured, which is very difficult. However, for the models considered in this paper a strong hint is offered by the signal size itself, which is much larger for $\YY$ production than for $\TT$, and
the observation of $T \to Zt$ in the opposite-sign dilepton and trilepton final states establishes the quark charge.

\begin{table}[t]
\begin{center}
\begin{tabular}{ccccccc}
       & $L$      & Rec. & \quad &        & $L$       & Rec. \\[1mm]
$\Ts$  & 1.1 \fbin  & $m_T$   &       & $\TBD$ & -- & no \\
$\Bs$  & -- & no   &       & $\XTd$ & -- & no \\
$\TBd$ & 0.60 \fbin & $m_T$   &       & $\BYd$ & 0.18 \fbin        & $m_Y$
\end{tabular}
\end{center}
\caption{Luminosity $L$ required to have a $5\sigma$ discovery in the $\ell^\pm$ ($2b$) final state. A dash indicates no signal or a luminosity larger than 100 \fbin.
We also indicate whether a mass peak can be reconstructed in this final state.}
\label{tab:sig-1Q1-2b}
\end{table}

Finally, it is worth mentioning that $T$ quarks singlets and those in the $\TBd$ model (which have the same decay channels) could in principle be distinguished by the $W$ helicity fractions~\cite{Kane:1991bg}, but for large $m_T$ the $W$ bosons in $T \to W^+ b$ are mainly longitudinal and the difference between a left- and right-handed $WTb$ coupling is washed out. For a 500 GeV $T$ singlet we have
$F_L \simeq  0.05$, $F_0 \simeq  0.95$, $F_R \simeq 0$, while for the $T$ quark in
a $\TB$ doublet $F_L \simeq  0$, $F_0 \simeq  0.95$, $F_R \simeq 0.05$. With 500 events for a 30 \fbin, the statistical error $\sim 1/\sqrt N$ in angular asymmetries, etc. is expected to be around 5\%, of the order of the difference between the two models. Systematic uncertainties in the measurement of helicity fractions can also be important~\cite{AguilarSaavedra:2007rs}.

\subsection{Final state $\ell^\pm$ ($4b$)}

\begin{figure}[t]
\begin{center}
\epsfig{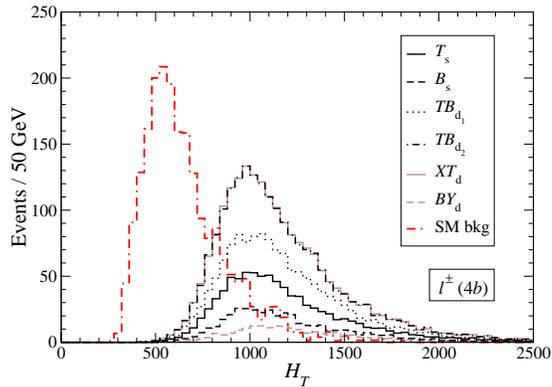}
\caption{Total transverse energy distribution for the signals and backgrounds in the
for the $\ell^\pm$ ($4b$) final state. The luminosity is 30 fb$^{-1}$.}
\label{fig:dist-1Q1-4b}
\end{center}
\end{figure}

\begin{figure}[p]
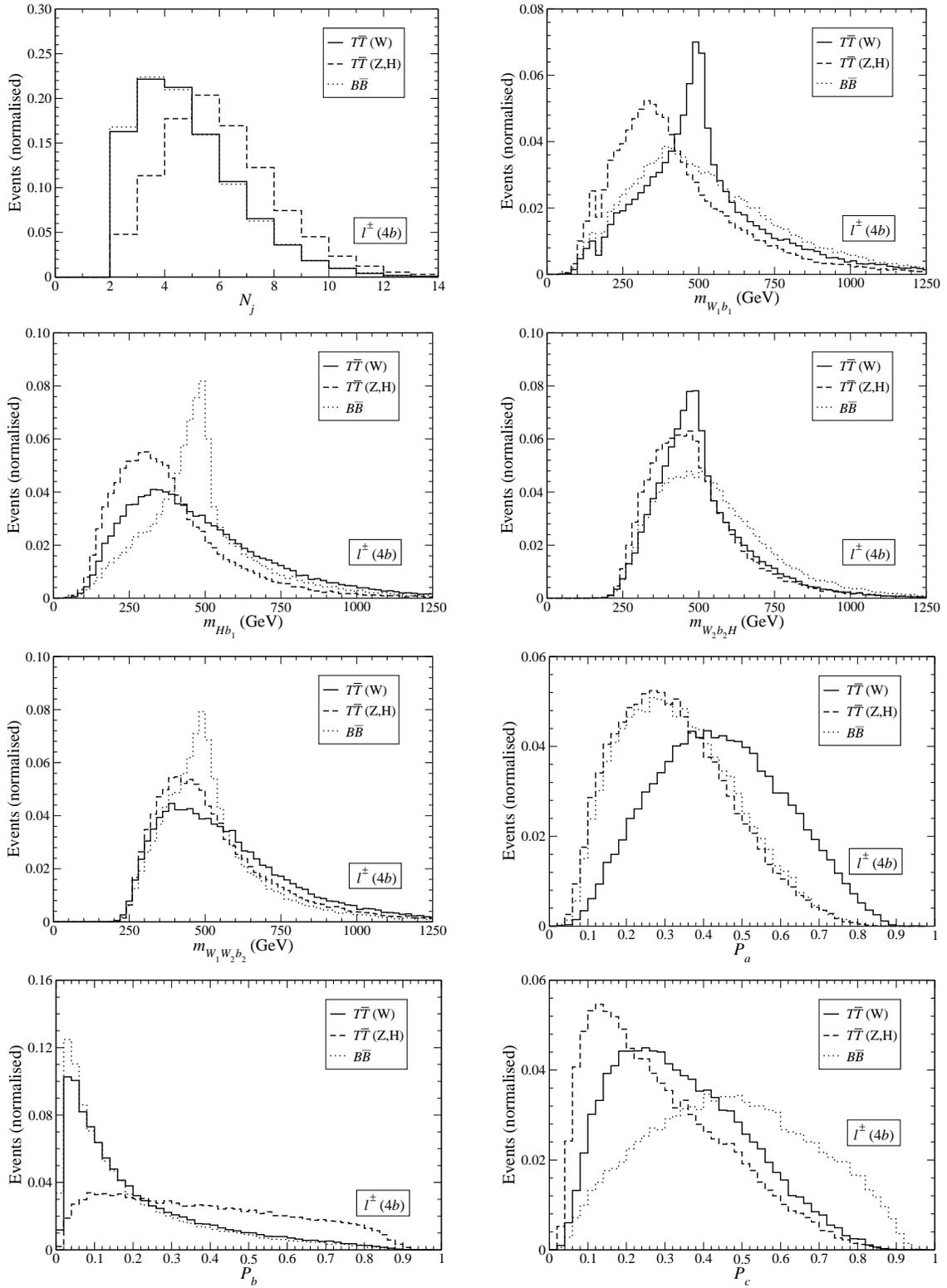

\begin{center}
\begin{tabular}{ccc}
\epsfig{file=Figs/D-mult-1Q1-4b.eps,height=5.0cm,clip=} & \quad &
\epsfig{file=Figs/D-mW1b1-1Q1-4b.eps,height=5.0cm,clip=} \\
\epsfig{file=Figs/D-mHb1-1Q1-4b.eps,height=5.0cm,clip=} & \quad &
\epsfig{file=Figs/D-mtH-1Q1-4b.eps,height=5.0cm,clip=} \\
\epsfig{file=Figs/D-mW1t-1Q1-4b.eps,height=5.0cm,clip=} & \quad &
\epsfig{file=Figs/D-Pa-1Q1-4b.eps,height=5.0cm,clip=} \\
\epsfig{file=Figs/D-Pb-1Q1-4b.eps,height=5.0cm,clip=} & \quad &
\epsfig{file=Figs/D-Pc-1Q1-4b.eps,height=5.0cm,clip=}
\end{tabular}
\caption{Kinematical variables used to classify the three heavy quark signals in the
$\ell^\pm$ (4$b$) final state, and the resulting probability distributions for events in the reference samples.}
\label{fig:lik-1Q1-4b}
\end{center}
\end{figure}

The main interest of this final state, in addition to the discovery of new quarks, lies in the observation of the decays $T \to Ht$, $B \to Hb$, which would also allow an early discovery of a light Higgs boson if new vector-like quarks exist~\cite{delAguila:1989ba,delAguila:1989rq,AguilarSaavedra:2006gw,Sultansoy:2006cw}. The relevant decay channels are
\begin{align}
& T \bar T \to Ht \, W^- \bar b \to H W^+b W^- \bar b
&& \quad H \to b \bar b , WW \to \ell \nu q \bar q' \,, \nonumber \\
& T \bar T \to Ht \, \HZ \bar t \to H W^+b \, \HZ W^- \bar b
&& \quad H \to b \bar b , WW \to \ell \nu q \bar q' , \HZ \to q \bar q/\nu \bar \nu \,, \nonumber \\
& B \bar B \to H b \, W^+ \bar t \to H b \, W^+ W^- \bar b
&& \quad H \to b \bar b , WW \to \ell \nu q \bar q' \,.
\label{ec:ch1Q14b}
\end{align}
The first and last channels in the above equation give exactly four $b$ quarks in the final state, while the second gives up to six $b$ quarks. For $\TT$ production alone, it has been shown~\cite{AguilarSaavedra:2006gw} that the discrimination between the first two decay chains is very involved in final states with only four $b$-tagged jets, because the signals are actually not very different. The situation is worsened if additional $B$ quarks exist, for example in models introducing a $\TB$ doublet.
Here we implement a discriminating method based on a likelihood analysis similar to the ones used in previous sections.
In this final state, however, discrimination is less efficient than for multi-lepton signals due to the combinatorics resulting from presence of four $b$-tagged jets and their association to the other particles present in the event.
We use high statistics reference samples for three event classes ($a,b,c$) corresponding to the three decay channels in Eqs.~(\ref{ec:ch1Q14b}).
As selection criteria for this analysis we demand (i) four $b$-tagged jets with $p_T > 20$ GeV; (ii) transverse energy $H_T > 750$ GeV. The distribution of this variable for the signals and the SM background can be found in Fig.~\ref{fig:dist-1Q1-4b}.
The variables used in the discrimination are obtained using a preliminary reconstruction of a top quark and the Higgs boson:
\begin{enumerate}
\item The hadronic $W$ is obtained with the two jets (among the three ones with largest $p_T$) having an invariant mass closest to $M_W$.
\item The leptonic $W$ is obtained from the charged lepton and the missing energy with the usual method, keeping both solutions for the neutrino momentum.
\item The top quark is reconstructed with one of the $W$ bosons and one of the four $b$ jets, selecting the ones which give an invariant mass closest to the nominal top mass. The $W$ boson and $b$ quark selected are labelled as $W_2$ and $b_2$.
\item The Higgs boson ``candidate'' is obtained from the two $b$ jets, among the three remaining ones, which have the minimum invariant mass.
\item The remaining $W$ boson and $b$ jet are labelled as $W_1$, $b_1$.
\end{enumerate}
The interesting variables for signal discrimination are:
\begin{itemize}
\item The light jet multiplicity.
\item The $W_1 b_1$ invariant mass , which peaks around $m_T$ in class ($a$).
\item The $H b_1$ invariant mass, which peaks at $m_B$ in class ($c$).
\item The $H W_2 b_2$ invariant mass. This corresponds to $m_T$ in class ($a$), but the distribution is very similar for the other decay channels.
\item The $W_1 W_2 b_2$ invariant mass, which is $m_B$ in class ($c$) but does not differ much for events in the other decay channels.
\end{itemize}
The normalised distributions of these variables for the reference event samples are presented in Fig.~\ref{fig:lik-1Q1-4b}, together with the resulting probability distributions $P_{a,b,c}$ that the events belong to a given class. Comparing with the trilepton and dilepton final states, we find that the separation among channels is much less clean. This is reflected in Table~\ref{tab:lik-1Q1-4b}, which collects the fractions of events correctly and incorrectly classified.
\begin{table}[htb]
\begin{center}
\begin{tabular}{cccc}
Class & $P_a > P_b,P_c$ & $P_b > P_a,P_c$ & $P_c > P_a,P_b$ \\
\hline
($a$) & 0.53 & 0.17 & 0.30 \\
($b$) & 0.28 & 0.47 & 0.25 \\
($c$) & 0.27 & 0.15 & 0.58
\end{tabular}
\end{center}
\caption{Performance of the likelihood function on the $\ell^\pm$ ($4b$) event reference samples: fractions of events in each sample and their classification. Events in a class $x$ are correctly classified if $P_x > P_y,P_z$, where $y$, $z$ are the other classes.}
\label{tab:lik-1Q1-4b}
\end{table}

We give in Table~\ref{tab:nsnb-1Q1-4b} the numbers of signal and background events at the selection level, and their classification (as in the previous sections, we select the class which has the highest probability).
\begin{table}[htb]
\begin{center}
\begin{tabular}{ccccc}
               & Total  & ($a$) & ($b$) & $(c)$ \\[1mm]
$\TT$ ($\Ts$)  & 836.8  & 342.5 & 260.4 & 233.9 \\
$\TT$ ($\TBd$) & 886.5  & 363.9 & 286.4 & 236.2 \\
$\TT$ ($\TBD$/$\XTd$) & 1780.7 & 509.9 & 841.9 & 428.9 \\
$\XX$ ($\XTd$) & 167.3  & 44.9  & 86.5  & 35.9  \\
$\BB$ ($\Bs$)  & 396.8 & 119.6 & 64.3 & 212.9 \\
$\BB$ ($\TBd$) & 416.4 & 119.7 & 67.2 & 229.5 \\
$\BB$ ($\TBD$) & 160.0 & 43.0  & 83.1 & 33.9 \\
$\BB$ ($\BYd$) & 146.1 & 62.0  & 10.5 & 73.6 \\
$\YY$ ($\BYd$) & 57.9  & 28.1  & 4.9  & 24.9 \\
\hline
$t \bar tnj$   & 404    & 122   & 228   & 54 \\
$tW$           & 5      & 3     & 0     & 2 \\
$t\bar tb\bar b$& 158   & 47    & 66    & 45 \\
$t\bar tt\bar t$& 1     & 0     & 0     & 1 \\
$Wnj$          & 1     & 0     & 1    & 0 \\
$Wb \bar b nj$ & 3     & 0     & 1    & 2 \\
$Zb \bar b nj$ & 1     & 0     & 0    & 1
\end{tabular}
\end{center}
\caption{Number of signal and background events in the $\ell^\pm$ ($4b$) final state at the selection level assigned to each event class. The luminosity is 30 fb$^{-1}$.}
\label{tab:nsnb-1Q1-4b}
\end{table}
An important remark here is the presence of $\XX$ and $\YY$ signals, as well as $\BB$ ones in the $\TBD$ model, which involve only two $b$ quarks at the partonic level. Events with four $b$-tagged jets result from the mistag of the charm quarks from $W$ decays, and light quarks to a lesser extent. The presence and size of these signals illustrates the relative importance of mistags in the processes we are interested in: for $\TT$ and $\BB$ events with four $b$ jets it may well happen that only three of them correspond to true $b$ quarks and one is a charm quark from a $W$ decay. This, added to the kinematical similarity of the signals and the several possibilities in $b$ jet assignments, makes the separation among the channels difficult.
The luminosities required for $5\sigma$ discovery are collected in Table~\ref{tab:sig-1Q1-4b}, summing all contributions in a given model and combining the significance of the three classes $(a,b,c)$. A systematic uncertainty of 20\% is included in the estimations.

\begin{table}[t]
\begin{center}
\begin{tabular}{ccccccc}
       & $L$        & Rec. & \quad &        & $L$       & Rec. \\[1mm]
$\Ts$  & 0.70  \fbin & $m_T$, $M_H$   &       & $\TBD$ & 0.16 \fbin & no \\
$\Bs$  & 1.9  \fbin & $m_B$, $M_H$   &       & $\XTd$ & 0.16 \fbin & no \\
$\TBd$ & 0.25 \fbin & $m_T$, $m_B$, $M_H$   &       & $\BYd$ & 6.2 \fbin & no
\end{tabular}
\end{center}
\caption{Luminosity $L$ required to have a $5\sigma$ discovery in the $\ell^\pm $ ($4b$) final state.
We also indicate whether a mass peak can be reconstructed in this final state.}
\label{tab:sig-1Q1-4b}
\end{table}

Comparing with Ref.~\cite{AguilarSaavedra:2006gw}, we observe that the discovery luminosity is significantly smaller than the values quoted there. The reasons for this difference are: (i) for consistency with the other channels  we are giving here the statistical significance of the signal (including the Higgs boson) compared to the ``only background'' hypothesis, while in Ref.~\cite{AguilarSaavedra:2006gw} we compared the ``Higgs'' and ``no Higgs'' hypotheses in the presence of new quarks, for which the significance is lower; (ii) multi-jet SM backgrounds in Ref.~\cite{AguilarSaavedra:2006gw} were quite pessimistically overestimated, and with our new evaluation with updated tools and improved matrix element-parton shower matching they turn out to be smaller; (iii) the new likelihood classification performed here, with the subsequent statistical combination of channels, also improves the significance.

We finally address the heavy quark and Higgs boson reconstruction, which depends on the decay channel in which events are classified. We only perform the reconstruction of events classified in the first and third classes, because those in the second class have six $b$ jets at the partonic level.

{\em Class} ($a$): $\TT \to Ht W \bar b \to H Wb Wb$. Events identified as resulting from this decay chain are reconstructed using this procedure:
\begin{enumerate}
\item Two light jets are selected to form the hadronic $W$, labelled as $W_H$.
If there are only two light jets these are automatically chosen; if there are more than two, only up to three (ordered by decreasing $p_T$) are considered. 
\item The leptonic $W$ (labelled as $W_L$) is obtained from the charged lepton $\ell$ and the missing energy, in the way explained in previous sections.
Both solutions for the neutrino momentum are kept, and the one giving best reconstructed masses is selected.
\item Two $b$ jets are selected among the ones present, to be paired with $W_H$ and $W_L$, respectively.
\item The top quark is reconstructed from one of the $Wb$ pairs, and its parent heavy quark $T_1$ from the top quark and the two remaining $b$ jets.
\item The other heavy quark $T_2$ is reconstructed from the remaining $Wb$ pair.
\item Among all choices for $b$ and light jets and all possible pairings, the combination minimising the quantity
\begin{small}
\begin{equation}
\frac{(m_{W_H}^\text{rec}-M_W)^2}{\sigma_W^2} + 
\frac{(m_{W_L}^\text{rec}-M_W)^2}{\sigma_W^2} + 
\frac{(m_t^\text{rec}-m_t)^2}{\sigma_t^2} +
\frac{(m_{T_1}^\text{rec}-m_{T_2}^\text{rec})^2}{\sigma_T^2}
\end{equation}
\end{small}%
is selected, with $\sigma_t = 14$ GeV, $\sigma_T = 20$ GeV.
After the final choice, the Higgs is reconstructed from the two $b$ jets not assigned to the $W$ bosons.
\end{enumerate}
\begin{figure}[t]
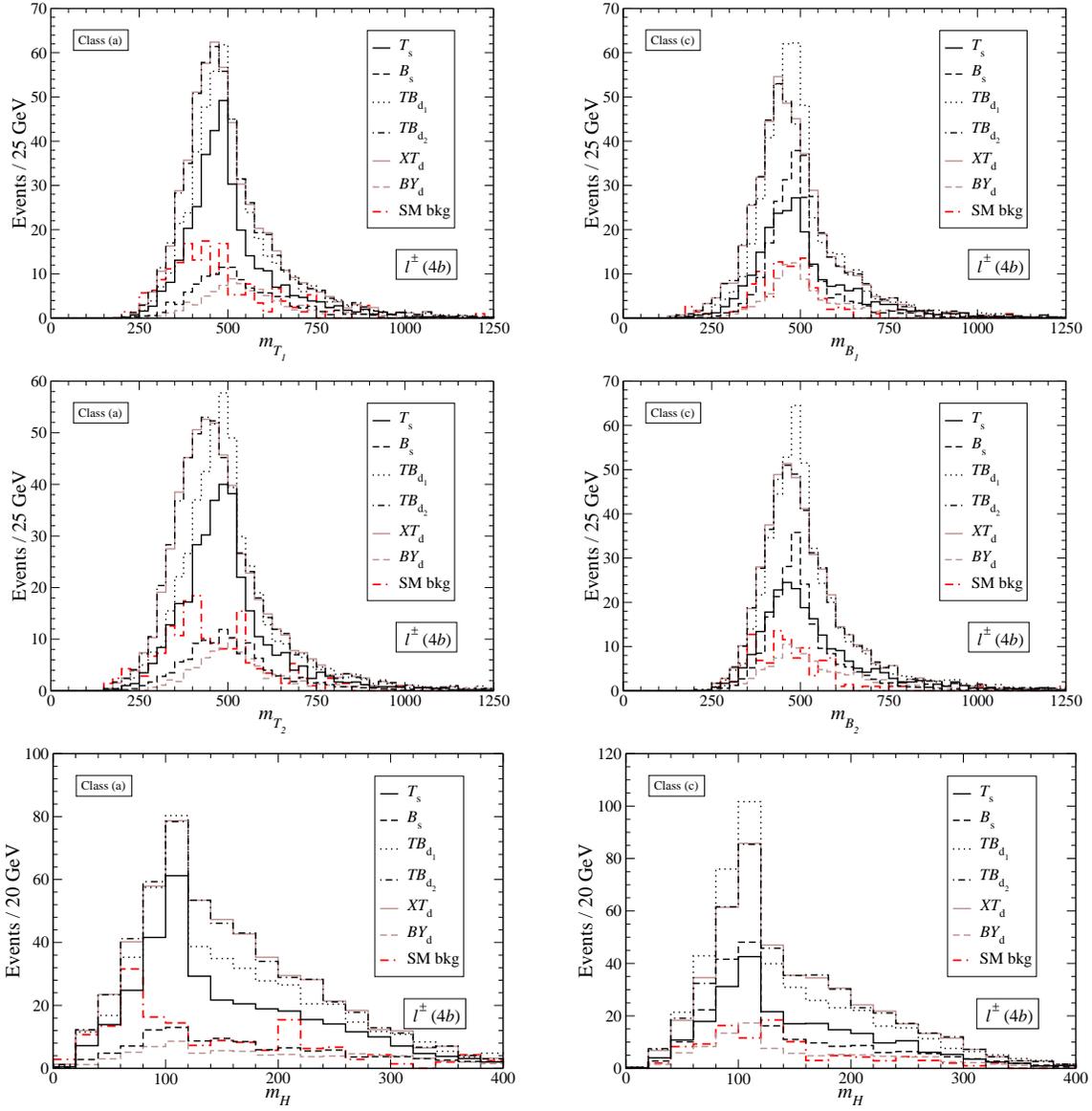

\begin{center}
\begin{tabular}{ccc}
\epsfig{file=Figs/mtH-1Q1-4b.eps,height=5.0cm,clip=} & \quad &
\epsfig{file=Figs/mbH-1Q1-4b.eps,height=5.0cm,clip=} \\
\epsfig{file=Figs/mbW-1Q1-4b.eps,height=5.0cm,clip=} & \quad &
\epsfig{file=Figs/mtW-1Q1-4b.eps,height=5.0cm,clip=} \\
\epsfig{file=Figs/mH-1Q1-4b-a.eps,height=5.0cm,clip=} & \quad &
\epsfig{file=Figs/mH-1Q1-4b-c.eps,height=5.0cm,clip=} \\
\end{tabular}
\caption{Reconstructed heavy quark and Higgs masses in the $\ell^\pm$ ($4b$) final state.}
\label{fig:mrec-1Q1-4b}
\end{center}
\end{figure}

{\em Class} ($c$): $\BB \to Hb Wt \to Hb WWb$. The reconstruction of this channel proceeds through  
the same steps $1-2$ as in the previous two channels, and then:
\begin{enumerate}\setcounter{enumi}{2}
\item One $b$ jet is selected and paired with one of the two $W$ bosons to form a top quark, and with the other $W$ to form the heavy quark $B_2$.
\item The three remaining $b$ jets then reconstruct the heavy quark $B_1$.
\item The combination minimising
\begin{small}
\begin{equation}
\frac{(m_{W_H}^\text{rec}-M_W)^2}{\sigma_W^2} + 
\frac{(m_{W_L}^\text{rec}-M_W)^2}{\sigma_W^2} + 
\frac{(m_{t}^\text{rec}-m_t)^2}{\sigma_t^2} +
\frac{(m_{B_1}^\text{rec}-m_{B_2}^\text{rec})^2}{\sigma_B^2} \,,
\end{equation}
\end{small}%
with $\sigma_B = 20$ GeV,
is finally selected. Among the three $b$ jets corresponding to $B_1$, the two with the minimum invariant mass are chosen to reconstruct the Higgs boson.
\end{enumerate}

The results are presented in Fig.~\ref{fig:mrec-1Q1-4b}. For brevity we do not include the reconstructed $W$ boson and top quark distributions, which have good peaks by construction.
\begin{figure}[b]
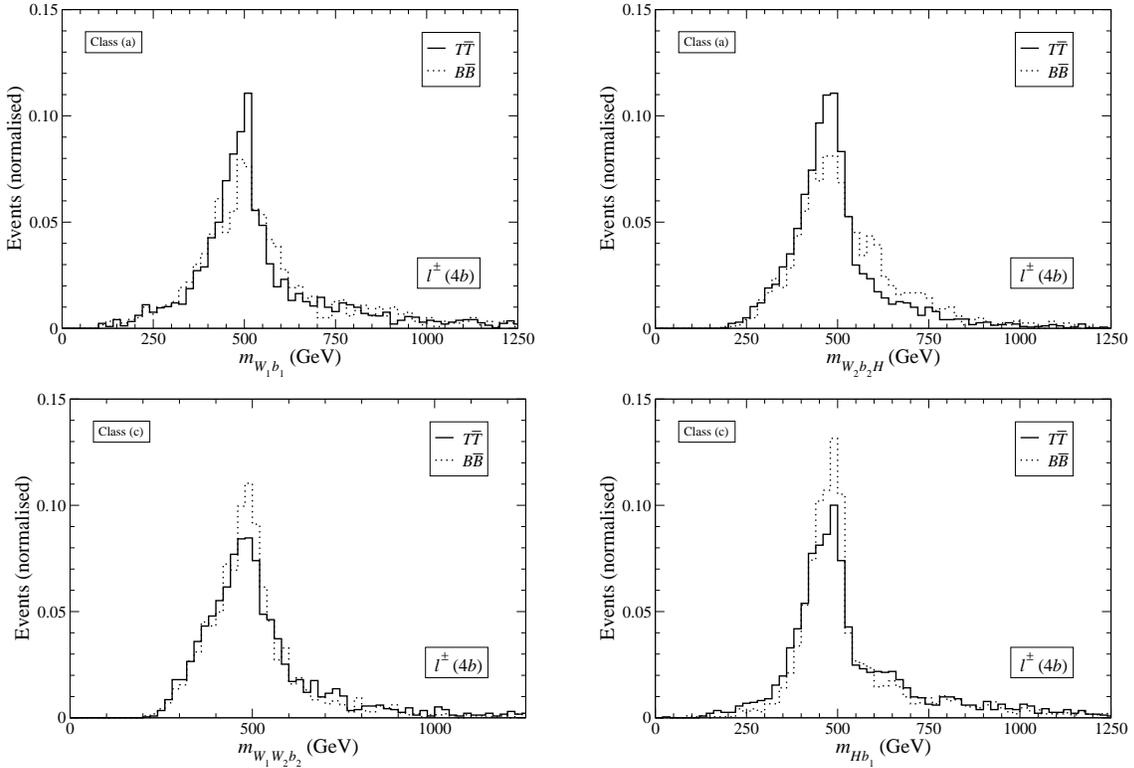

\begin{center}
\begin{tabular}{ccc}
\epsfig{file=Figs/W-mW1b1-1Q1-4b.eps,height=5.0cm,clip=} & \quad &
\epsfig{file=Figs/W-mtH-1Q1-4b.eps,height=5.0cm,clip=} \\
\epsfig{file=Figs/W-mW1t-1Q1-4b.eps,height=5.0cm,clip=} & \quad &
\epsfig{file=Figs/W-mHb1-1Q1-4b.eps,height=5.0cm,clip=} 
\end{tabular}
\caption{Comparison between kinematical distributions for correctly and wrongly classified events (see the text).}
\label{fig:wrong-1Q1-4b}
\end{center}
\end{figure}
We observe that for $\TT$ signals in class $(a)$ the $T_{1,2}$ peaks are well reconstructed, as the Higgs boson peak. The same happens for $\BB$ signals in class $(c)$: the $B_{1,2}$ peaks are clear and the Higgs boson peak is observable. However, the intriguing fact is that, for $\TT$ and $\BB$ signals included in the ``wrong'' class (respectively, $(c)$ and $(a)$) the reconstruction procedure produces peaks which are as sharp as those for the signals ``correctly'' classified, except for the Higgs peaks. This point deserves a detailed discussion.
Clearly, if an event is assigned to a given decay channel in Eqs.~(\ref{ec:ch1Q14b}) based on its likelihood, it is because its kinematics is quite compatible with that decay channel. Then, it is 
not so surprising that, for example, if a $\BB$ event is classified as $\TT$ based on its topology, when it is reconstructed as $\TT$ it rather looks as a $\TT$ event.
For the reader's illustration we present in Fig.~\ref{fig:wrong-1Q1-4b} several distributions 
for the production of $\TT$ and $\BB$ events in the case of singlets in all decay channels.
In the upper part we show the normalised $W_1 b_1$ and $W_2 b_2 H$ invariant mass distributions assigned to class $(a)$ For $\BB$ events (incorrectly classified) the distributions display peaks very similar to the ones for $\TT$ events correctly included in this class. In the lower part of this figure we plot the $W_1 W_2 b_2$ and $H b_1$ distributions for $\TT$ and $\BB$ events in class ($c$). The peaks are quite similar for $\TT$ (wrong classification) and $\BB$ (correct).
Therefore, we can conclude that distinguishing $\TT$ and $\BB$ signals in this channel is a more demanding task, and the multi-leptonic channels are much more appropriate for that. Fortunately,
all these difficulties in signal discrimination do not affect the discovery potential, which is excellent for this final state.

\subsection{Final state $\ell^\pm$ ($6b$)}

The single lepton final state with six $b$ jets allows a clean reconstruction of the decay
\begin{align}
& T \bar T \to Ht \, H \bar t \to H W^+b \, H W^- \bar b
&& \quad H \to b \bar b , WW \to \ell \nu q \bar q' \,,
\label{ec:ch1Q16b}
\end{align}
with $H \to b \bar b$, which seems impossible if only four jets are tagged. This final state is most interesting for the models in which the decay $T \to Ht$ is enhanced and the $6b$ signal is larger. We do not impose any further selection criteria apart from having six $b$-tagged jets with $p_T > 20$ GeV, which defines the sample studied. The number of background events is given in Table~\ref{tab:nsnb-1Q1-6b}.
\begin{table}[htb]
\begin{center}
\begin{tabular}{cccccccccc}
                 & Sel.  & \quad &        & Sel. \\[1mm]
$\TT$ ($\Ts$)    & 26.5  & & $\BB$ ($\Bs$)  & 2.5 \\
$\TT$ ($\TBd$)   & 29.0  & & $\BB$ ($\TBd$) & 2.8 \\
$\TT$ ($\TBD$/$\XTd$)   & 112.1 & & $\BB$ ($\TBD$) & 0.0 \\
$\XX$ ($\XTd$)   & 0.4   & & $\BB$ ($\BYd$) & 0.7 \\ 
                 &       & & $\YY$ ($\BYd$) & 0.3 \\
\hline
$t\bar tb\bar b$ & 2
\end{tabular}
\end{center}
\caption{Number of events in the $\ell^\pm$ ($6b$) final state at selection level. The luminosity is 30 fb$^{-1}$.}
\label{tab:nsnb-1Q1-6b}
\end{table}
This final state is extremely clean, and discovery could be made merely by an event counting. The $5\sigma$ discovery potential for the different models is given in Table~\ref{tab:sig-1Q1-6b}, summing all signal contributions. The discovery potential (in models with $T$ quarks, when a signal is produced) is determined by the requirement of having at least 10 signal events. The background normalisation in this case has little effect on the significance, because for the discovery luminosities it is rather small.

\begin{table}[ht]
\begin{center}
\begin{tabular}{ccccccc}
       & $L$        & Rec. & \quad &        & $L$       & Rec. \\[1mm]
$\Ts$  & 11  \fbin & $m_T$, $M_H$   &       & $\TBD$ & 2.7 \fbin & $m_T$, $M_H$ \\
$\Bs$  & --   & no   &       & $\XTd$ & 2.7 \fbin & $m_T$, $M_H$ \\
$\TBd$ & 9.4 \fbin & $m_T$, $M_H$   &       & $\BYd$ & --  & no
\end{tabular}
\end{center}
\caption{Luminosity $L$ required to have a $5\sigma$ discovery in the $\ell^\pm $ ($6b$) final state.
A dash indicates no signal or a luminosity larger than 100 \fbin.
We also indicate whether a mass peak can be reconstructed in this final state.}
\label{tab:sig-1Q1-6b}
\end{table}

\begin{figure}[t]
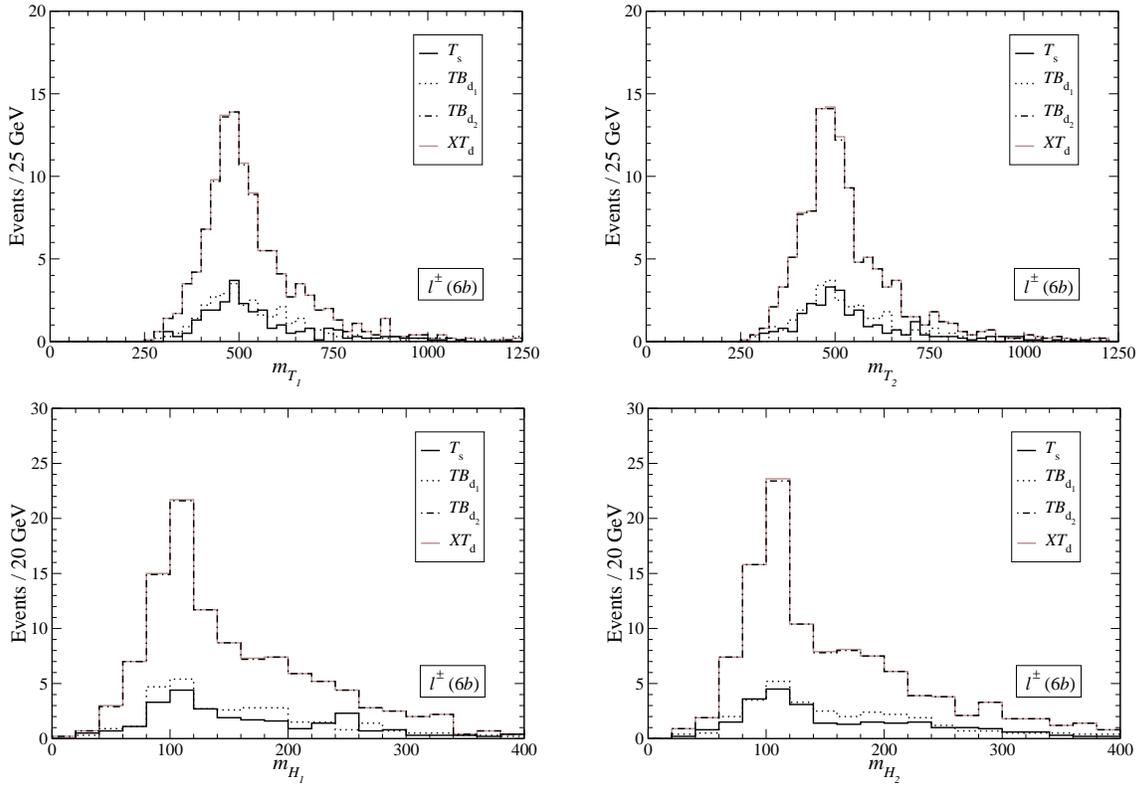

\begin{center}
\begin{tabular}{ccc}
\epsfig{file=Figs/mT1-1Q1-6b.eps,height=5.1cm,clip=} & \quad &
\epsfig{file=Figs/mT2-1Q1-6b.eps,height=5.1cm,clip=} \\
\epsfig{file=Figs/mH1-1Q1-6b.eps,height=5.1cm,clip=} & \quad &
\epsfig{file=Figs/mH2-1Q1-6b.eps,height=5.1cm,clip=}
\end{tabular}
\caption{Reconstructed heavy quark and Higgs masses in the $\ell^\pm$ ($6b$) final state.}
\label{fig:mrec-1Q1-6b}
\end{center}
\end{figure}

The event reconstruction can be easily done despite the large combinatorics from the six $b$ jets. The procedure is similar to the ones used in other final states:
\begin{enumerate}
\item Two light jets (among the three ones with largest $p_T$) are selected to form the hadronic $W$, labelled as $W_H$.
\item The leptonic $W$, labelled as $W_L$, is obtained from the charged lepton $\ell$ and the missing energy.
\item Two $b$ jets are selected among the ones present, to be paired with the two $W$ bosons to reconstruct the top quarks decaying hadronically and semileptonically ($t_H$ and $t_L$).
\item The four remaining $b$ jets are grouped in pairs to reconstruct the two Higgs bosons, $H_1$ and $H_2$.
\item The two heavy quarks $T_1$ (corresponding to $W_H$) and $T_2$ (with $W_L$) are reconstructed from a top quark plus a Higgs boson.
\item Among all choices for $b$ and light jets and all possible pairings, the combination minimising the quantity
\begin{small}
\begin{align}
& \frac{(m_{W_H}^\text{rec}-M_W)^2}{\sigma_W^2} + 
\frac{(m_{W_L}^\text{rec}-M_W)^2}{\sigma_W^2} + 
\frac{(m_{t_H}^\text{rec}-m_t)^2}{\sigma_t^2} +
\frac{(m_{t_L}^\text{rec}-m_t)^2}{\sigma_t^2} \notag \\
& + \frac{(m_{T_1}^\text{rec}-m_{T_2}^\text{rec})^2}{\sigma_T^2} +
\frac{(M_{H_1}^\text{rec}-M_{H_2}^\text{rec})^2}{\sigma_H^2}
\end{align}
\end{small}%
is selected, with $\sigma_H = 20$ GeV.
\end{enumerate}
The results are presented in Fig.~\ref{fig:mrec-1Q1-6b}.
We omit for brevity the $W$ boson and top quark reconstructed masses, which good have peaks at $M_W$ and $m_t$ by construction. It is seen that for the $\TBD$ and $\XT$ models the heavy quark and Higgs peaks are quite good and, with a moderate luminosity, they would give evidence for the $T \to Ht$ decay (and, in particular, for the production of a Higgs boson). For $T$ singlets and the $\TBd$ model the signals are smaller and they would require more luminosity, not only to be discovered but also to reconstruct the peaks.
We point out that an important difference with the $4b$ final state is that six $b$ jets can only be produced (up to mistags of charm quarks) from the decay in Eq.~(\ref{ec:ch1Q16b}). Hence, the model identification is cleaner here. We also note that for $\TT$ production within the $\TBD$ and $\XT$ models the signal in this final state is almost four times larger than in the $\ell^+ \ell^-$ ($Z$) one with four $b$ tags, so this final state is best suited to detect $T \to Ht$.

\subsection{Summary}

The single lepton final state offers the best heavy quark discovery potential for all the models studied, due to the large signal branching ratios. To achieve this result, an efficient reduction of the large backgrounds from $t \bar t nj$, $Wnj$ and $W b \bar b nj$ is necessary. We have concentrated on three different subsamples in this final state, with exactly two, four and six $b$ jets. Clearly, the discovery potential will improve further including the samples with three and five $b$ jets, which have not been considered here for brevity.

The final state with two $b$-tagged jets is the best suited for the discovery of $\YY$ production, which only requires 0.18 \fbin\ for a 500 GeV quark. It is also very good for $T$ singlets (1.1 \fbin\ for the same mass) and $\TB$ doublets in scenario 1 (0.60 \fbin). In this final state the $T$ and $Y$ masses can be reconstructed as peaks in the $Wb$ invariant mass distributions. The identity of the quarks cannot be established unless the $b$ jet charge is measured: the decays $T \to W^+ b$ and $\bar Y \to W^+ \bar b$ both give a $W^+$ boson plus a $b$-tagged jet, and the jet charge measurement is necessary to discriminate both possibilities. It is interesting to point out that all kinematical distributions are the same for $T$ and $Y$ quarks, including various angular asymmetries which can be built in the $W$ and top quark rest frames (see for example Ref.~\cite{AguilarSaavedra:2006fy}). The only discrimination between both possibilities comes either indirectly, from the cross section measurement (about three times larger for $\YY$ in this final state, after including efficiencies and cuts) or directly, via the observation of $T \to Zt$ in the dilepton or trilepton final states.

The final state with four $b$ jets is the best one for the discovery of $T$ quarks in either of the models considered. For heavy quark masses of 500 GeV, $5\sigma$ discovery
only requires 0.7 \fbin\ for singlets, 0.25 and 0.16 \fbin\ for the $\TBd$ and
$\TBD$ models, respectively, and 0.16 \fbin\ if the $T$ quark belongs to a $\XT$ doublet. For $B$ singlets the discovery potential is also the best one, with 1.9 \fbin.
 (For these results we have assumed a light Higgs, as suggested by precise electroweak data, taking $M_H = 115$ GeV.) This process could also be a discovery channel for the Higgs boson in the presence of $T$ or $B$ quarks. We have gone beyond the signal observation and studied the discrimination among $\TT$ and $\BB$ signals in this final state, which makes sense because they are both present in general for the case of the $\TB$ doublet. The separation is very difficult, for several reasons: (i) the combinatorics from the presence of four $b$ jets;
 (ii) two different decay chains contribute in the case of $\TT$; (iii) the signals are  kinematically not very different; (iv) the possibility of charm quark mistags. We have implemented a likelihood method to separate $\TT$ and $\BB$ signals, which has a reasonable efficiency if we bear in mind all these difficulties. After the $\TT$ and $\BB$ events are classified, the kinematics can be reconstructed according to the decay channel expected in each case, and sharp peaks are obtained in all cases, although the rate of ``wrong'' classifications is sizeable and a better discrimination between $T$ and $B$ quarks can be achieved in the trilepton final state. 
 
It is also interesting to remark the excellent discovery potential for $\XT$ doublets in this final state: only 0.16 \fbin\ for heavy quark masses of 500 GeV. The discovery potential in this channel is similar but better than in the like-sign dilepton and trilepton channels (0.23 and 0.25 \fbin, respectively) altough in those final states the main signal contribution comes from the charge $5/3$ quark $X$ and here it is the $T$ quark which gives most of the four-$b$ signal. The same results are obtained for the $\TBD$ model which includes a $T$ quark with the same decay modes: $5\sigma$ discovery of $T$ is possible with 0.16 \fbin\ in the single lepton final state with four $b$ tags, while the $B$ quark can be discovered in the like-sign dilepton and trilepton final states with 0.23 and 0.25 \fbin, respectively.

Finally, the sample with six $b$ jets has also been studied. In this final state the decay $\TT \to Ht \, H \bar t$ can be cleanly determined and peaks reconstructed without contamination of other decay modes. The signals are small, however, except for the $\TBD$ and $\XT$ models, for which $T \to Wb$ does not take place and thus
$T \to Ht$ has a larger branching ratio. The discovery potential is rather good for these models, 2.7 \fbin\ for $m_T = 500$ GeV.

\section{The roadmap to top partner identification}
\label{sec:summ}

We summarise in Table~\ref{tab:summ} the discovery luminosities for the six models in the different final states examined. The comparison among them clearly shows that the single lepton channel (either with two or four $b$ jets) offers the best discovery potential for new quarks. In the case of doublets, the signals may correspond to one or both members, as it is explained in detail in the summary at the end of each section, where it is also indicated whether heavy quark masses can be reconstructed. We now discuss case by case how the different models would be discovered and identified.

\begin{table}[ht]
\begin{center}
\begin{small}
\begin{tabular}{lcccccc}
       & $\Ts$      & $\Bs$     & $\TBd$     & $\TBD$     & $\XTd$     & $\BYd$ \\
$\ell^+ \ell^+ \ell^- \ell^-$ ($ZZ$)
       & --         & 24 \fbin  & 18 \fbin   & 23 \fbin   & 23 \fbin   & 10 \fbin \\ 
$\ell^+ \ell^+ \ell^- \ell^-$ ($Z$) 
       & 11 \fbin   & 14 \fbin  & 5.7 \fbin  & 3.4 \fbin  & 3.3 \fbin  & 50 \fbin \\ 
$\ell^+ \ell^+ \ell^- \ell^-$ (no $Z$) 
       & 35 \fbin   & 25 \fbin  & 11 \fbin   & 3.3 \fbin  & 3.5 \fbin  & -- \\ 
$\ell^\pm \ell^\pm \ell^\mp$ ($Z$) 
       & 3.4 \fbin  & 3.4 \fbin & 1.1 \fbin  & 0.73 \fbin & 0.72 \fbin & 26 \fbin \\ 
$\ell^\pm \ell^\pm \ell^\mp$ (no $Z$) 
       & 11 \fbin   & 3.5 \fbin & 1.1 \fbin  & 0.25 \fbin & 0.25 \fbin & -- \\ 
$\ell^\pm \ell^\pm$ 
       & 17 \fbin   & 4.1 \fbin & 1.5 \fbin  & 0.23 \fbin & 0.23 \fbin & -- \\ 
$\ell^+ \ell^-$ ($Z$) 
       & 22 \fbin   & 4.5 \fbin & 2.4 \fbin  & 4.4 \fbin  & 4.4 \fbin  & 1.8 \fbin \\ 
$\ell^+ \ell^-$ ($Z$, $4b$) 
       & --         & --        & 30  \fbin  & --         & --         & 9.2 \fbin \\ 
$\ell^+ \ell^-$ (no $Z$) 
       & 2.7 \fbin  & 9.3 \fbin & 0.83 \fbin & 1.1 \fbin  & 1.1 \fbin  & 0.87 \fbin \\ 
$\ell^\pm$ ($2b$) 
       & 1.1 \fbin  & --        & 0.60 \fbin & --         & --         & 0.18 \fbin \\ 
$\ell^\pm$ ($4b$) 
       & 0.70 \fbin & 1.9 \fbin & 0.25 \fbin & 0.16 \fbin & 0.16 \fbin & 6.2 \fbin \\ 
$\ell^\pm$ ($6b$) 
       & 11  \fbin  & --        & 9.4 \fbin  & 2.7 \fbin  & 2.7  \fbin & -- \\ 
\end{tabular}
\end{small}
\end{center}
\caption{Luminosity required to have a $5\sigma$ discovery in all final states studied.}
\label{tab:summ}
\end{table}

A $T$ singlet or the $T$ quark in the $\TBd$ model: They would be discovered in the single lepton final state with two or four $b$ jets. In the $2b$ final state, the peaks in the $Wb$ invariant mass distributions give evidence of the charged current decay but do not identify completely the new quark (it could be a charge $-4/3$ quark $Y$, although the cross section would not be consistent with that hypothesis). In the $4b$ final state, the peak in the $Ht$ distribution, with $H$ reconstructed from two $b$ jets exhibiting a peak at $M_H$, gives quite strong hints of the $T \to Ht$ decay, although $B$ quarks also give signals not very different. (The $6b$ final state does not have this ambiguity but the observation of $T \to Ht$ requires much larger luminosity.)
The best confirmation of its nature comes with a little more luminosity in the $\ell^\pm \ell^\pm \ell^\mp$ ($Z$) final state, with the observation of a peak in the $Zt$ invariant mass distribution. This peak also establishes that the quark has charge $2/3$. The analysis of the charged lepton distribution in the top quark rest frame for the subset of events in which the top quark decays semileptonically can discriminate between a $T$ singlet with a left-handed $WTb$ coupling and the $T$ quark in a $\TB$ doublet for which the coupling is right-handed. With 30 \fbin\ the differences found in the forward-backward asymmetry would amount to $2.4\sigma$, and a better sensitivity is expected by using a more sophisticated analysis with a fit to the complete distribution.

A $T$ quark in the $\TBD$ model or in a $\XT$ doublet: in the single lepton final state with two $b$ quarks it does not exhibit peaks in the $Wb$ invariant mass distribution because the decay $T \to W^+ b$ does not take place and probably the signal is very difficult to separate from the $t \bar t nj$ background. In the $4b$ sample, however, the signal is very large and clean, and the quark is seen in the decay $T \to Ht$. With small luminosity, a signal should be also visible in the $6b$ sample. This quark also has enhanced decays $T \to Zt$, from which the quark charge is determined, and the signals in the trilepton and opposite-sign dilepton final states with a $Z$ candidate are $2-3$ times larger than expected for a $T$ singlet.

A $B$ singlet or the $B$ quark in the $\TBd$ model: they would be discovered in the single lepton final state with four $b$ jets. However, its discrimination from a $T$ quark might not be very clear due to combinatorics and the signal similarities. With practically the same luminosity, the $B$ quark would appear as a sharp peak in a $Zb$ invariant mass distribution in the $\ell^\pm \ell^\pm \ell^\mp$ ($Z$) final state. This would determine the quark charge, and would be confirmed by an opposite-sign dilepton signal. The evidence for the $B \to W^- t$ decay comes from the same trilepton final state, and the
charged lepton distribution in the top semileptonic decays would in principle probe the chirality of the $WtB$ coupling, but the statistics is smaller than for $T$ quarks.
Indirectly, evidence for the $B \to W^- t$ decay results from the presence of $\ell^\pm \ell^\pm \ell^\mp$ (no $Z$) and $\ell^\pm \ell^\pm$ signals, also observable with small luminosity. 

A $B$ quark in the $\TBD$ model: it would be discovered in the like-sign dilepton and $\ell^\pm \ell^\pm \ell^\mp$ (no $Z$) final states with similar luminosities. An indirect indication of the quark identity, in comparison with a charge $5/3$ quark $X$, would be given by the absence of the reconstructed mass peaks and endpoints which are present for $\XX$ production. Signals in the $\ell^+ \ell^-$ (no $Z$) final state are also interesting, not only because of the good discovery potential but also because the mass reconstruction is possible in the hadronic decay $B \to W^- t \to W^- W^+ b$ (or the charge conjugate) with a moderate luminosity. This mass reconstruction is important because in this model the $B$ quark does not have decays $B \to Zb$ and thus trilepton and opposite-sign dilepton signals with a $Z$ candidate are absent (see the paragraph above). Single lepton signals with four $b$-tagged jets, which are very significant for other models, are also absent for the $B$ quark in this model but can be produced by its $T$ partner and are kinematically not very different.

A $B$ quark in a $\BY$ doublet: it does not give trilepton signals without a $Z$ candidate
nor like-sign dilepton ones. On the other hand, it gives large opposite-sign dilepton signals with a $Z$ candidate with a sharp peak in the $Zb$ invariant mass distribution, from which the quark charge is determined. With five times more luminosity, this is also done in the four lepton final state with two $Z$ candidates. The decay $B \to Hb$ can be seen in the $\ell^+ \ell^-$ ($Z$, $4b$) final state, also with larger luminosity.

A charge $5/3$ quark $X$: it would be simultaneously discovered in the like-sign dilepton and $\ell^\pm \ell^\pm \ell^\mp$ (no $Z$) final states. In the former, the invariant mass can be reconstructed and the quark identity ({\em i.e.} that it has charge $5/3$) can be established under reasonable assumptions. The mass could also be determined from the trilepton final state with an endpoint analysis to confirm the quark identity. A signal due to this quark should also be visible in the four lepton final state without $Z$ candidates.

A charge $-4/3$ quark $Y$: it would be discovered in the single lepton final state with two $b$ jets. The peaks in the $Wb$ invariant mass distributions would give evidence of the charged current decay and indirect evidence of its nature: the signal is three times larger than for a $T$ singlet, for example. A clean signal in the $\ell^+ \ell^-$ (no $Z$) final state would also be visible, larger than for a $T$ quark. On the other hand, all the signals characteristic of a $T \to Zt$ decay would be absent, in the trilepton and dilepton final states with a $Z$ candidate, for example.

In more complicated scenarios with several singlets or doublets the signals would add up but it still would be possible to identify the new quarks with a thorough examination of all final states. For example, with a $T$ singlet plus a $\XT$ doublet in which both charge $2/3$ quarks have similar masses (which cannot be experimentally resolved), the $T \to W^+ b$ decay would be seen in the single lepton final state, and the
dilepton and trilepton signals involving $T \to Zt$ would be much larger than the ones corresponding to just one $T$ quark. Of course, the $X$ member of the doublet would also be detected. The same argument applies to a combination of a $B$ singlet and a $\BY$ doublet.
The simplest discrimination between $T$, $B$ singlets and the $T$, $B$ quarks in the $\TBd$ model is by the presence of two partners almost degenerate in mass. Still, one may imagine a situation in which a $T$ and a $B$ singlet almost degenerate were present. This scenario would be distinguished from the $\TBd$ model with the analysis of angular distributions, especially for $T \to Zt$ decays.

The discrimination of vector-like singlets and doublets from a fourth sequential generation with new quarks $t'$, $b'$
is also easy, because the latter would give different signatures (see Refs.~\cite{delAguila:2008iz,Holdom:2009rf} for recent reviews). Oblique corrections prefer a mass difference $m_{t'} > m_{b'} \sim 60$ GeV~\cite{Kribs:2007nz}, so the $t'$ quark would decay either $t' \to W^+ b$ (as a $T$ or $Y$) or $t' \to W^+ b'$ if the mass difference is larger. In the first case, the absence of $t' \to Zt$ would distinguish it from a $T$ singlet, and the absence of a degenerate $B$ partner from a $\BY$ doublet. In the second case, the decay between new heavy quarks would prove the non-singlet nature of both.
Regarding the $b'$ quark, for $m_{b'} > m_{t} + M_W$ the decay $b' \to W^- t$ would dominate, distinguishing this quark from a $B$ singlet in either model. The discovery potential for fourth generation quarks~\cite{Ozcan:2008zz,Burdman:2008qh} is similar to the models studied here.

To summarise, the analyses carried out in this paper show that the single lepton final state offers the best discovery potential, and is the one in which new vector-like quark signals would be first seen. Searches in the dilepton and trilepton channels would soon confirm a possible discovery, and with a luminosity around five times larger all the decay modes of the new quarks would be observed in these channels, establishing the nature of the new quarks. 
In some models four lepton signals could be sizeable and detectable as well and, in any case, these should be investigated as a further test of the models. 

\section{Conclusions}
\label{sec:concl}

In this work we have investigated in detail the LHC discovery potential for pair production of new vector-like quarks in five models: $T$ or $B$ singlets of charge $2/3$, $-1/3$ respectively,
and $\TB$, $\XT$, $\BY$ doublets of hypercharge $1/6$, $7/6$, $-5/6$, restricting ourselves to the case that new quarks mainly couple to the third generation, as it is expected from the SM quark mass hierarchy. In the case of $\TB$ doublets we have distinguished two scenarios: that both heavy quarks have similar mixing with the top and bottom quark (model $\TBd$) and that the mixing of the top with its heavy partner is much larger than for the bottom quark (model $\TBD$), as expected from the mass hierarchy $m_t \gg m_b$ and from indirect precision data.
Using a dedicated Monte Carlo generator {\tt Protos} \cite{AguilarSaavedra:2008gt} we have computed all signal contrubutions involving all heavy quark, gauge and Higgs boson decay channels. With a fast detector simulation of signals and backgrounds we have examined twelve final states which would give evidence of the presence of new quarks, with one to four charged leptons in different kinematical regions and several $b$ jet multiplicities.

We have identified the final state with one charged lepton plus two or four $b$ jets as the most sensitive one for new quark searches. Nevertheless, model discrimination requires the observation or exclusion of the different heavy quark decay channels. To achieve this goal, the dilepton and trilepton final states are essential. These final states have also good sensitivity to heavy quark signals, and with a luminosity at most five times larger than in the single lepton channel the $5\sigma$ observation would be possible and the heavy quarks might be identified. The reconstruction of mass peaks would also be possible when a sufficient number of events is collected.
In our simulations we have taken heavy quark masses of 500 GeV, focusing on early discoveries at LHC. We have obtained an excellent discovery potential for all models: 
0.70 and 1.9 \fbin\ for $T$ and $B$ singlets, respectively; 0.25 and 0.16 \fbin\ for the $\TBd$ and $\TBD$ models; and 0.16, 0.18 \fbin\ for the $\XT$ and $\BY$ doublets.
It is also interesting to know the mass reach for higher integrated luminosities. With a simple rescaling it can be seen that in the single lepton channel
alone and a luminosity of 100 \fbin\ heavy $T$, $B$ singlets with masses up to 800 and 720 GeV respectively can be discovered with $5\sigma$ significance, while for the doublets the reach is higher: 850 GeV and 900 GeV for the $\TB$ doublet in the two scenarios considered, 900 GeV for $\XT$ and 820 GeV for $\BY$. For higher masses the experimental detection of heavy quarks can also be done using jet mass measurements~\cite{Skiba:2007fw} but model discrimination would follow similar strategies as outlined here.

We have also obtained an excellent potential for the discovery of the new quarks in decay channels containing a Higgs boson, especially in the final state with one charged lepton and four $b$-tagged jets. For heavy quark masses of 500 GeV, the discovery luminosities are
0.16 \fbin\ for the $\TBD$ and $\XT$ models, 0.25 \fbin\ for $\TBd$ and 0.70, 1.9 \fbin\ for $T$ and $B$ singlets, respectively. These luminosities are much smaller than the ones required for a light Higgs discovery in the SM. Indeed, it is well known since some time~\cite{delAguila:1989ba,delAguila:1989rq} that vector-like quark production can be a copious source of Higgs bosons and, if such quarks exist and the Higgs boson is light, its discovery would possibly happen in one of these channels. For a heavier Higgs with different decay modes the analyses presented here (relying on the leading decay $H \to b \bar b$) must be modified accordingly. Nevertheless, the determination of the other modes  like $T \to W^+b$, $T \to Zt$, etc. would still be done in the same way as presented here, with few modifications.

In the summaries given at the end of each section we have compared the multi-lepton signals produced by new quarks with those arising from heavy leptons. Both possibilities for new fermions are easily distinguished by the different reconstructed mass peaks and the common presence of $b$ jets for quarks, which in lepton pair production only result from $H \to b \bar b$, $Z \to b \bar b$ decays.
Interestingly, a more general difference among models introducing new quarks and leptons is that the latter give signals which are more ``multi-leptonic'': for heavy leptons the trilepton signatures are usually the ones with the highest significance, while for heavy quarks the single lepton one is the most sensitive. This is not unexpected, since heavy lepton decays give SM leptons plus a gauge or Higgs boson, while heavy quarks give SM quarks instead. In the minimal supersymmetric standard model where squark and gluino pair production is large, it is also found~\cite{Aad:2009wy} that, although multi-lepton signatures are important, the final state with best discovery potential is the one with a charged lepton or large missing energy plus jets. 

Finally, it is worth pointing out that, although in this work we have restricted ourselves to heavy quark pair production, electroweak single production can also have a large cross section depending on the mass and couplings of the new quarks. These interesting processes are the only ones in which the heavy quark mixing with the SM sector can be measured because the pair production cross section is determined by the heavy quark mass alone and the heavy quark total width is likely to be very difficult to measure. Further model discrimination is also possible in single production, in particular from the study of cross sections and angular asymmetries, and it will be addressed elsewhere.

\section*{Acknowledgements}

I thank F. del \'Aguila, N. Castro, R. Contino and J. Santiago for useful discussions.
This work has been supported by a MEC Ram\'on y Cajal contract, MEC project FPA2006-05294 and
Junta de Andaluc{\'\i}a projects FQM 101 and FQM 437.

\appendix
\section{Feynman rules}
\label{sec:a}

We give in Tables~\ref{tab:F-Ts}--\ref{tab:F-BYd} the Feynman rules used for our matrix element evaluations for heavy quark pair production $\TT$, $\BB$, $\XX$, $\YY$ and single production $T\bar b j$,
$B \bar b j$, $X \bar t j$, $T \bar t j$, $Y \bar b j$. The rest of vertices are the same as in the SM, including QCD interactions of the new quarks.

\begin{table}[h]
\begin{center}
\begin{footnotesize}
\begin{tabular}{clccl}
\raisebox{-11mm}{\epsfig{file=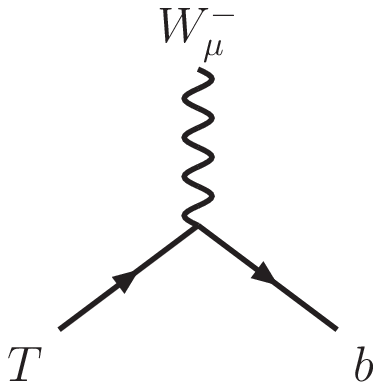,height=22mm,clip=}}
  & $\displaystyle -i \frac{g}{\sqrt 2} \mathrm{V}_{Tb}^* \gamma^\mu P_L$ & \quad &
\raisebox{-11mm}{\epsfig{file=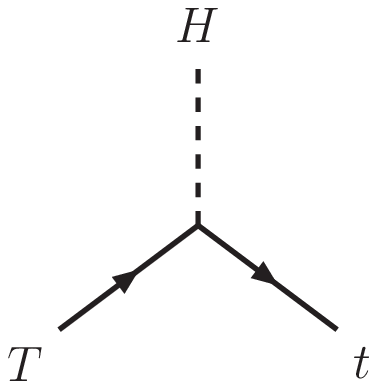,height=22mm,clip=}}
  & $\displaystyle -i\frac{g}{2 M_W} \mathrm{X}_{Tt}^* (m_t P_L + m_T P_R)$ \\ \\
\raisebox{-11mm}{\epsfig{file=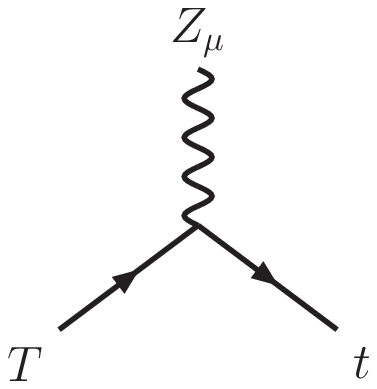,height=22mm,clip=}}
  & $\displaystyle -i\frac{g}{2 c_W} \mathrm{X}_{Tt}^* \gamma^\mu P_L$ & \quad &
\raisebox{-11mm}{\epsfig{file=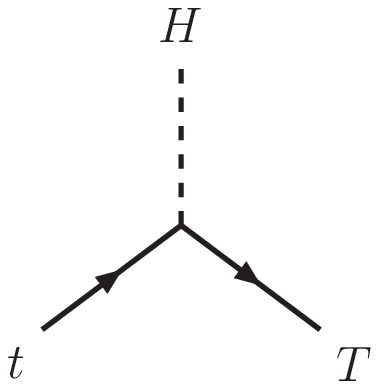,height=22mm,clip=}}
  & $\displaystyle -i\frac{g}{2 M_W} \mathrm{X}_{Tt} (m_T P_L + m_t P_R)$ \\ \\
\end{tabular}
\end{footnotesize}
\caption{Feynman rules for $T$ singlet electroweak and scalar interactions with the third generation.}
\label{tab:F-Ts}
\end{center}
\end{table}

\begin{table}[h]
\begin{center}
\begin{footnotesize}
\begin{tabular}{clccl}
\raisebox{-11mm}{\epsfig{file=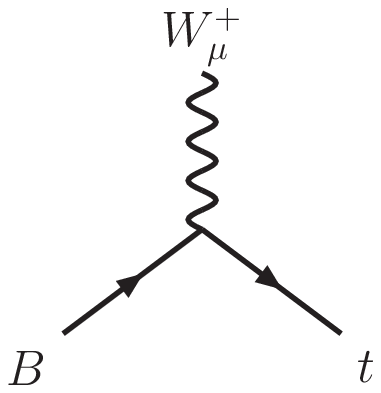,height=22mm,clip=}}
  & $\displaystyle -i \frac{g}{\sqrt 2} \mathrm{V}_{tB} \gamma^\mu P_L$ & \quad &
\raisebox{-11mm}{\epsfig{file=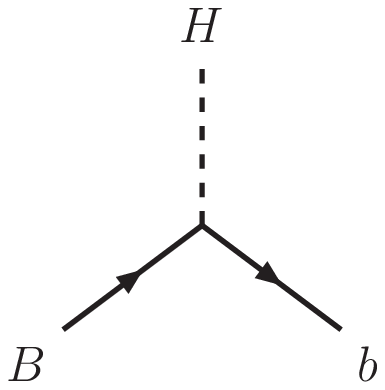,height=22mm,clip=}}
  & $\displaystyle -i\frac{g}{2 M_W} \mathrm{X}_{bB} (m_b P_L + m_B P_R)$ \\ \\
\raisebox{-11mm}{\epsfig{file=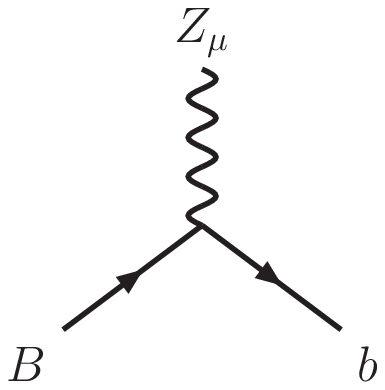,height=22mm,clip=}}
  & $\displaystyle i\frac{g}{2 c_W} \mathrm{X}_{bB} \gamma^\mu P_L$ & \quad &
\raisebox{-11mm}{\epsfig{file=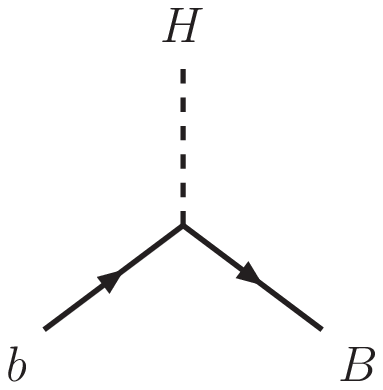,height=22mm,clip=}}
  & $\displaystyle -i\frac{g}{2 M_W} \mathrm{X}_{bB}^* (m_B P_L + m_b P_R)$ \\ \\
\end{tabular}
\end{footnotesize}
\caption{Feynman rules for $B$ singlet electroweak and scalar interactions with the third generation.}
\end{center}
\end{table}

\clearpage

\begin{table}[p]
\begin{center}
\begin{footnotesize}
\begin{tabular}{clccl}
\raisebox{-11mm}{\epsfig{file=Diags/WTb.eps,height=22mm,clip=}}
  & $\displaystyle -i \frac{g}{\sqrt 2} \mathrm{V}_{Tb}^{R*} \gamma^\mu P_R$ & \quad &
\raisebox{-11mm}{\epsfig{file=Diags/HTt.eps,height=22mm,clip=}}
  & $\displaystyle i\frac{g}{2 M_W} \mathrm{X}_{Tt}^{u*} (m_T P_L + m_t P_R)$ \\ \\
\raisebox{-11mm}{\epsfig{file=Diags/ZTt.eps,height=22mm,clip=}}
  & $\displaystyle -i\frac{g}{2 c_W} \mathrm{X}_{Tt}^{u*} \gamma^\mu P_R$ & \quad &
\raisebox{-11mm}{\epsfig{file=Diags/HtT.eps,height=22mm,clip=}}
  & $\displaystyle i\frac{g}{2 M_W} \mathrm{X}_{Tt}^u (m_t P_L + m_T P_R)$ \\ \\
\raisebox{-11mm}{\epsfig{file=Diags/WtB.eps,height=22mm,clip=}}
  & $\displaystyle -i \frac{g}{\sqrt 2} \mathrm{V}_{tB}^R \gamma^\mu P_R$ & \quad &
\raisebox{-11mm}{\epsfig{file=Diags/HBb.eps,height=22mm,clip=}}
  & $\displaystyle i\frac{g}{2 M_W} \mathrm{X}_{bB}^d (m_B P_L + m_b P_R)$ \\ \\
\raisebox{-11mm}{\epsfig{file=Diags/ZBb.eps,height=22mm,clip=}}
  & $\displaystyle i\frac{g}{2 c_W} \mathrm{X}_{bB}^d \gamma^\mu P_R$ & \quad &
\raisebox{-11mm}{\epsfig{file=Diags/HbB.eps,height=22mm,clip=}}
  & $\displaystyle i\frac{g}{2 M_W} \mathrm{X}_{bB}^{d*} (m_b P_L + m_B P_R)$ \\ \\
\end{tabular}
\end{footnotesize}
\caption{Feynman rules for $\TB$ doublet electroweak and scalar interactions with the third generation.}
\end{center}
\end{table}

\begin{table}[p]
\begin{center}
\begin{footnotesize}
\begin{tabular}{clccl}
\raisebox{-11mm}{\epsfig{file=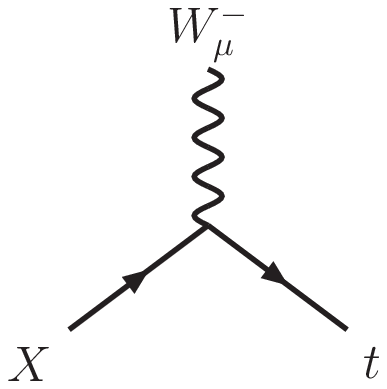,height=22mm,clip=}}
  & $\displaystyle -i \frac{g}{\sqrt 2} \mathrm{V}_{Xt}^{R*} \gamma^\mu P_R$ & \quad &
\raisebox{-11mm}{\epsfig{file=Diags/HTt.eps,height=22mm,clip=}}
  & $\displaystyle i\frac{g}{2 M_W} \mathrm{X}_{Tt}^* (m_T P_L + m_t P_R)$ \\ \\
\raisebox{-11mm}{\epsfig{file=Diags/ZTt.eps,height=22mm,clip=}}
  & $\displaystyle i\frac{g}{2 c_W} \mathrm{X}_{Tt}^* \gamma^\mu P_R$ & \quad &
\raisebox{-11mm}{\epsfig{file=Diags/HtT.eps,height=22mm,clip=}}
  & $\displaystyle i\frac{g}{2 M_W} \mathrm{X}_{Tt} (m_t P_L + m_T P_R)$ \\ \\
\end{tabular}
\end{footnotesize}
\caption{Feynman rules for $\XT$ doublet electroweak and scalar interactions with the third generation.}
\end{center}
\end{table}

\clearpage

\begin{table}[t]
\begin{center}
\begin{footnotesize}
\begin{tabular}{clccl}
\raisebox{-11mm}{\epsfig{file=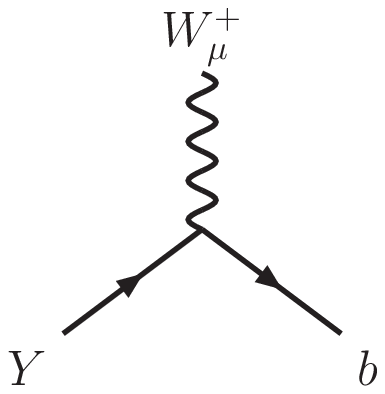,height=22mm,clip=}}
  & $\displaystyle -i \frac{g}{\sqrt 2} \mathrm{V}_{bY}^R \gamma^\mu P_R$ & \quad &
\raisebox{-11mm}{\epsfig{file=Diags/HBb.eps,height=22mm,clip=}}
  & $\displaystyle i\frac{g}{2 M_W} \mathrm{X}_{bB} (m_B P_L + m_b P_R)$ \\ \\
\raisebox{-11mm}{\epsfig{file=Diags/ZBb.eps,height=22mm,clip=}}
  & $\displaystyle -i\frac{g}{2 c_W} \mathrm{X}_{bB} \gamma^\mu P_R$ & \quad &
\raisebox{-11mm}{\epsfig{file=Diags/HbB.eps,height=22mm,clip=}}
  & $\displaystyle i\frac{g}{2 M_W} \mathrm{X}_{bB}^* (m_b P_L + m_B P_R)$ \\ \\
\end{tabular}
\end{footnotesize}
\caption{Feynman rules for $\BY$ doublet electroweak and scalar interactions with the third generation.}
\label{tab:F-BYd}
\end{center}
\end{table}

\end{document}